\documentclass[a4paper,11pt]{article}
\pdfoutput=1 % if your are submitting a pdflatex (i.e. if you have
\usepackage{jheppub} % for details on the use of the package, please
\usepackage[T1]{fontenc} % if needed

\usepackage{comment}
\usepackage{amsmath}    % need for subequations
\usepackage{amssymb}    % 
\usepackage{graphicx}   % need for figures
\usepackage{verbatim}   % useful for program listings
\usepackage{color}      % use if color is used in text
\usepackage{subfigure}  % use for side-by-side figures
\usepackage{hyperref}
\usepackage{multirow}
\usepackage{comment}
\usepackage{pdflscape}
\usepackage{rotating}
\usepackage{booktabs,tabularx}
\usepackage{float}	% Put {figure}[H] to force the position of a figure
\usepackage{color}
\definecolor{rosso}{RGB}{210,0,0}
\renewcommand\arraystretch{1.1}

\newcommand{\nn}{\nonumber}

\newcommand{\LOQCDt}{LO$_{\textrm{\tiny{QCD}}}$}
\newcommand{\NLOQCDt}{NLO$_{\textrm{\tiny{QCD}}}$}

\newcommand{\nNLOQCDt}{nNLO$_{\textrm{\tiny{QCD}}}$}

\newcommand{\cS}{\check{\Sigma}}

\def\LO{{\rm LO}}
\def\NLO{{\rm NLO}}

\def\NLOQCD{{\rm NLO}_{\rm QCD}}
\def\NLOEW{{\rm NLO}_{\rm EW}}
\def\LOQCD{{\rm LO}_{\rm QCD}}
\def\nLOQCD{{\rm nLO}_{\rm QCD}}
\def\nNLOQCD{{\rm nNLO}_{\rm QCD}}

\def\MttW{m(t\bar t W^+)}
\def\MttH{m(t\bar t H)}
\def\MttZ{m(t\bar t Z)}
\def\MttV{m(t\bar t V)}
\def\pt{p_{T}}
\def\pttt{p_{T}(t\bar t)}

\def\alphas{\alpha_s}

\def\ttwm{t \bar t W^-}
\def\ttwp{t \bar t W^+}
\def\ttw{t \bar t W^{\pm}}
\def\tth{t \bar t H}
\def\ttz{t \bar t Z}
\def\ttv{t \bar t V}

\def\ttt{t \bar t}

\def\beq{\begin{equation}}
\def\beqn{\begin{eqnarray}}
\def\eeq{\end{equation}}
\def\eeqn{\end{eqnarray}}
\def\beal{\begin{align}}
\def\endal{\end{align}}

\newcommand\mf{{\sc\small MadFKS}}
\newcommand\ml{{\sc\small MadLoop}}
\newcommand\ct{{\sc\small CutTools}}
\newcommand\nin{{\sc\small Ninja}}
\newcommand\coll{{\sc\small Collier}}
\newcommand\ol{{\sc\small Openloops}}

\newcommand{\MSb}{\overline{\rm MS}}

 \newcommand{\ord}[1]{\mathcal{O}(#1)}
 
 \newcommand{\ff}{f\hspace{-0.3cm}f}

\title{Top-quark pair hadroproduction in association with a heavy boson at NLO+NNLL including EW corrections}

\author[a,b]{Alessandro Broggio,}
\author[c,d]{Andrea Ferroglia,}
\author[e]{Rikkert Frederix,}
\author[f]{Davide Pagani,}
\author[g]{Benjamin D. Pecjak,}
\author[f]{Ioannis Tsinikos}

\emailAdd{alessandro.broggio@unimib.it}
\emailAdd{aferroglia@citytech.cuny.edu}
\emailAdd{rikkert.frederix@thep.lu.se}
\emailAdd{ben.pecjak@durham.ac.uk}
\emailAdd{davide.pagani@tum.de}
\emailAdd{ioannis.tsinikos@tum.de}

\affiliation[a]{Universit\`{a} degli Studi di Milano-Bicocca, Piazza della Scienza~3, I-20126 Milano, Italy}

\affiliation[b]{INFN, Sezione di Milano-Bicocca, Piazza della Scienza~3, I-20126 Milano, Italy}

\affiliation[c]{Physics Department, New York City College of Technology, The City
	University of New York, 
	300 Jay Street, Brooklyn, NY 11201 USA}

\affiliation[d]{The Graduate School and University Center,
	The City University of New York, 365 Fifth Avenue,
	New York, NY 10016  USA}
	
\affiliation[e]{Theoretical Particle Physics, Department of Astronomy and Theoretical Physics, Lund University, S\"olvegatan 14A, SE-223 62 Lund, Sweden}

\affiliation[f]{Technische Universit\"{a}t M\"{u}nchen, James-Franck-Str.~1, D-85748 Garching, Germany}

\affiliation[g]{Institute for Particle Physics Phenomenology, Ogden Centre for Fundamental Physics,
	Department of Physics, University of Durham, Science Laboratories,
	South Rd, Durham DH1 3LE, United Kingdom}

\note{Preprint:  TUM-HEP-1208/19, LU-TP 19-30,
	IPPP/19/57}

\abstract{This work studies the associated production of a top-quark pair with a
$W$, $Z$, or Higgs boson at the LHC.   Predictions for the total cross sections as well as for several differential distributions of the massive particles in the final state are provided. These predictions, valid for the LHC operating at $13$ TeV, include
without any approximation all the NLO
electroweak and QCD contributions of $\mathcal{O}(\alpha_s^{i} \alpha^{j+1})$ with $i+j=2,3$. In addition, the predictions presented here improve upon the NLO QCD results by adding  the effects of soft gluon emission corrections resummed  to next-to-next-to-leading logarithmic accuracy. The residual dependence of the predictions on scale and PDF choices is analyzed. }

\begin{document} 
\maketitle
\flushbottom

\section{Introduction}
\label{sec:introduction}

The Run-II at the LHC, with a center of mass energy of $13$~TeV  and a higher instant luminosity w.r.t.~Run-I, made this collider a fully operational top-quark factory. Indeed, the heaviest of the Standard Model (SM) particles can be produced via different channels, many of which have been observed at the LHC. To date, not only top-quark pair \cite{Abe:1995hr,D0:1995jca,Aad:2010ey, Khachatryan:2010ez} and single top-production \cite{Aaltonen:2009jj,Abazov:2009ii,Chatrchyan:2011vp,Aad:2012ux,Chatrchyan:2014tua} modes, but also the  production of a top-quark pair in association with a heavy electroweak (EW) boson have been measured. The latter class involves $\ttw$ \cite{Khachatryan:2015sha,Aad:2015eua, Aaboud:2016xve, Sirunyan:2017uzs}, $\ttz$ \cite{Khachatryan:2015sha, Aad:2015eua, Aaboud:2016xve,Sirunyan:2017uzs, Sirunyan:2017nbr} and $\tth$ \cite{Aaboud:2018urx, Sirunyan:2018hoz} production processes. These three processes are  extremely important in the searches for beyond-the-SM (BSM) effects, both as components of the background and of the signal itself. For example, $\ttw$ and $\ttz$ production constitute 
the main backgrounds in the measurement of the leptonic signatures emerging from $\tth$ production \cite{Aaboud:2017jvq, Sirunyan:2018shy, Maltoni:2015ena}, which in turn enables the direct measurement of the coupling of the top quark to the Higgs boson. Analogously, the $\ttz$ production process can be employed for the measurement of the coupling of the top quark to the $Z$ boson \cite{Sirunyan:2017uzs, Bylund:2016phk}. Finally, it is worth noting that very recently also single-top plus $Z$  associated production was observed \cite{Sirunyan:2018zgs}.

For a correct interpretation of current and future measurements and the possible identification of BSM effects, precise predictions for these processes, and, consequently, the study of their  radiative corrections, are of paramount relevance. For top-quark pair and single-top  production next-to-next-to-leading (NNLO) QCD corrections were computed in~\cite{Czakon:2013goa,Brucherseifer:2014ama, Berger:2016oht, Liu:2018gxa, Catani:2019hip}.  For top pair production, also next-to-leading-order  (NLO) electroweak (EW) corrections \cite{Czakon:2017wor, Czakon:2017lgo} and/or next-to-next-to-leading-logarithmic (NNLL) accuracy resummation of threshold and small-mass logarithms \cite{Czakon:2018nun, Czakon:2019txp}) were accounted for. This level of accuracy is not yet achievable for processes with three massive particles (two of which are colored) in the final state, nor is it 
expected in the near future. Still, it is desirable to have the best possible current predictions, {\it i.e.} those which include all corrections of QCD and EW origin that can be calculated with current technology. 
In addition, it is necessary to thoroughly study the phenomenological impact of these predictions at the differential level.   

In this paper we provide state-of-the-art SM predictions for top-quark pair hadroproduction in association with an EW heavy boson; we calculate the complete-NLO predictions for
$t\bar{t}W^{\pm}$, $t\bar{t}Z$ and $t\bar{t}H$  in
proton--proton collisions at 13 TeV and we resum soft gluon emission effects at NNLL accuracy in QCD. All the EW and QCD 
 contributions of $\mathcal{O}(\alpha_s^{i} \alpha^{j+1})$ with $i+j=2,3$
 are  evaluated without any approximation. In addition, in Mellin space, the resummation procedure accounts for terms proportional  $\alpha_s^{2+n}\alpha \ln^k \bar{N}$ with 
 $\max\{0,2n -3\} \le k \le 2 n$ at all orders ($n\geq 1$) in $\alpha_s$, where $\bar{N} = N e^{\gamma_{\rm E}}$ with $N$ the Mellin parameter, and $\bar{N} \to \infty$ is the soft emission limit.
  
 The calculation of the complete-NLO corrections to $\ttw$ production is based on the work in~\cite{Frederix:2017wme} and has been carried out with the new public version of {\sc\small MadGraph5\_aMC@NLO} \cite{Frederix:2018nkq}. This code was also used to obtain complete-NLO corrections to $\tth$ and $\ttz$ production. The calculations of soft gluon effects to NNLL accuracy in QCD for $\ttw$, $\ttz$ and $\tth$ are based on the work in~\cite{Broggio:2015lya, Broggio:2016zgg, Broggio:2016lfj, Broggio:2017kzi, Broggio:2017oyu} and on the in-house parton level Monte Carlo code that was developed for those papers. The resummation of soft emission effects was also studied in~\cite{Kulesza:2017ukk, Kulesza:2018tqz}, where a resummation framework different from the one considered in~\cite{Broggio:2015lya, Broggio:2016zgg, Broggio:2016lfj, Broggio:2017kzi} was employed. Very recently, in~\cite{Ju:2019lwp} also the resummation of Coulomb effects for $\tth$ production was studied. 
  
The paper is organized as follows. In Section~\ref{sec:calcframe} we briefly summarize the salient features of the calculational framework used in order to evaluate the various corrections. Section~\ref{sec:inputparam} includes a description of the input parameters and PDF sets employed in the calculation, as well a discussion of the values chosen for the factorization and resummation scales. Predictions for the total cross section and differential distributions for the processes considered in this study  are collected in Section~\ref{sec:results}. Finally, Section~\ref{sec:conclusions} contains our conclusions.

\section{Calculational framework \label{sec:calcframe}}

In this section we describe the calculational framework on which the phenomenological predictions presented in Section~\ref{sec:results} are based. In Sections \ref{sec:completeNLO} and \ref{sec:resummation} we briefly summarize the calculation of the complete-NLO corrections of QCD and EW origin \cite{Frederix:2017wme,Frederix:2018nkq} and the resummation of soft-gluon effects at NNLL  accuracy \cite{Broggio:2015lya, Broggio:2016zgg, Broggio:2016lfj, Broggio:2017kzi}, respectively.  In Section \ref{sec:matching} we explain how the combination and matching of complete-NLO and resummation of soft-gluon effects is carried out. We will denote the class of processes considered in this work as $\ttv$, where $V$ can be $W^{+},W^{-},Z$ or $H$. In Section \ref{sec:phenointro} we recall the most relevant phenomenological features of the different contributions entering the complete-NLO calculation, and comment on the implications for soft gluon resummation.

\subsection{Complete-NLO}
\label{sec:completeNLO}

The fixed order expansion of  a generic observable $\Sigma$ for the processes $pp \to \ttv(+X)$ (where $X$ indicates that the process is inclusive over extra QCD and QED radiation) in powers of $\alpha_s$ and $\alpha$ can be expressed as
\noindent
\begin{align}
\Sigma^{\ttv}(\alpha_s,\alpha) = \sum_{m+n\geq 2} \alpha_s^{m} \alpha^{n+1} \Sigma_{m+n+1,n}^{\ttv} \label{expansion:wp}\, ,
\end{align}
\noindent
 with  $m$ and $n$   positive integers.  LO contributions consist of $\Sigma_{m+n+1,n}^{\ttv}$  terms with $m+n=2$ and involve  tree-level diagrams only. NLO corrections correspond to the terms with $m+n=3$ and are induced by the interference among all the possible one-loop and tree-level Born diagrams  as well among all the possible  tree-level diagrams involving one additional quark, gluon or photon in the final state. 
 
In this work, ``complete-NLO'' is used to indicate the quantity $\Sigma^{\ttv}(\alpha_s,\alpha)$, in which all terms $\Sigma_{m+n+1,n}^{\ttv}$  with $m+n=2,3$ are included. On the other hand, a more user-friendly notation can be used to refer to any individual term in Eq.~(\ref{expansion:wp}).
We denote $\ttv$  observables  at LO as $\Sigma^{\ttv}_{\rm LO}$ and further redefine the individual perturbative orders as
\begin{align}
\Sigma^{\ttv}_{\rm LO}(\alpha_s,\alpha) &= \alpha_s^2 \alpha \Sigma_{3,0}^{\ttv} + \alpha_s \alpha^2 \Sigma_{3,1}^{\ttv} + \alpha^3 \Sigma_{3,2}^{\ttv} \nonumber\\
 &\equiv \Sigma_{\rm LO_1} + \Sigma_{\rm LO_2} + \Sigma_{\rm LO_3}\, .\label{eq:blobsttwLO} 
\end{align}
\noindent
Similarly, NLO corrections and their individual perturbative orders can be defined as
\begin{align}
\cS^{\ttv}_{\rm NLO}(\alpha_s,\alpha) &= \alpha_s^3 \alpha \Sigma_{4,0}^{\ttv} + \alpha_s^2 \alpha^2 \Sigma_{4,1}^{\ttv} + \alpha_s \alpha^3 \Sigma_{4,2}^{\ttv} + \alpha^4 \Sigma_{4,3}^{\ttv} \nonumber\\
 &\equiv \cS_{\rm NLO_1} + \cS_{\rm NLO_2} + \cS_{\rm NLO_3} + \cS_{\rm NLO_4}\, .
\label{eq:blobsttwNLO}
\end{align}
\noindent

In contrast to the notation used in previous works~\cite{Frixione:2014qaa, Frixione:2015zaa, Pagani:2016caq, Frederix:2016ost, Czakon:2017wor, Frederix:2017wme, Frederix:2018nkq}, here and in the rest of the text $\cS$ indicates a quantity that does not include any LO contribution, while $\Sigma$ indicates a quantity that does include LO contributions. In particular, all the $\LO_i$ are included for predictions beyond the LO, unless the subscript ``QCD'' is present; in this case only the $\Sigma_{\LO_1}$ is included. Consequently, with this convention an observable $\Sigma$ evaluated at complete-NLO accuracy can be written as
\begin{equation}
\Sigma_{\rm NLO} = \Sigma_{\rm LO} + \cS_{\rm NLO} \, .
\label{eq:NLOvsc}
\end{equation}
Observe that the quantities $\Sigma$ and $\cS$ are in general defined in such a way that they do include the appropriate multiplicative factor of powers of $\alpha$ and $\alpha_s$, as shown in Eqs.~\eqref{eq:blobsttwLO} and \eqref{eq:blobsttwNLO}.
We  use the symbols $\Sigma_{\LO_i}(\cS_{\NLO_i})$ or interchangeably their shortened aliases $\LO_i(\NLO_i)$ to indicate individual terms in the l.h.s.~of Eqs.~\eqref{eq:blobsttwLO} and \eqref{eq:blobsttwNLO}.
It is important to remember that in the literature the term ``LO'' usually 
refers only to $\LO_1$, which instead here is  denoted by $\LOQCD$. Therefore, with this notation one has
\begin{equation}
\Sigma_{\rm NLO_{QCD}} \equiv  \Sigma_{\rm LO_1} + \cS_{\rm NLO_1}  \, . 
\end{equation}
NLO EW corrections, which are of $\ord{\alpha}$ w.r.t.~the  $\LO_1$ observable, correspond to the $\Sigma_{\rm NLO_2}$ terms, so we  also  denote them as $\NLOEW$. On occasion, we also   refer to the set of $\Sigma_{\rm LO_i}$ and $\cS_{\rm NLO_i}$ corrections with $i\geq2$ as ``electroweak corrections'' (not to be confused with $\NLOEW$ corrections just defined).  The prediction at complete-NLO accuracy, which is the sum of all the $\LO_i$ and $\NLO_ i$ ($i \ge 1$) terms, is denoted as ``$ \NLO$'' \footnote{In Ref.~\cite{Frederix:2017wme} a slightly different notation has been used. Therein NLO and $\NLOQCD$ predictions refer to the corrections only, without including LO contributions. On the other hand, note that for the case of EW corrections also here $\NLOEW\equiv\NLO_2$.}. Consistently with the notation introduced above,  the sum of the ${\rm NLO}_i$ corrections without the LO is indicated by $\cS_{\rm NLO}$ (see Eq.~(\ref{eq:blobsttwNLO})).

It is important to point out that for all the $\ttv$ processes we do
not include the (finite) contributions from the real-emission of heavy
particles ($W^\pm$, $Z$ and $H$ bosons), also denoted in the literature as  heavy-boson-radiation (HBR) contributions. Although
they are formally  part of the inclusive predictions
at complete-NLO accuracy, these finite contributions in general lead to very different collider signatures and are typically small. For $\ttv$ processes, the HBR contributions to $\NLO_2$  were evaluated in~\cite{Frixione:2015zaa}.

The calculation of the complete-NLO predictions is carried out by employing  the latest version of {\sc\small MadGraph5\_aMC@NLO} \cite{Frederix:2018nkq}, which is now public. In {\sc\small MadGraph5\_aMC@NLO}, the FKS method~\cite{Frixione:1995ms,
Frixione:1997np} (automated in the module \mf~\cite{Frederix:2009yq,
Frederix:2016rdc}) is used in order to deal  with infrared singularities. One-loop amplitudes are evaluated by dynamically switching  among
different  kinds of techniques for integral reduction, namely, the OPP method~\cite{Ossola:2006us},
the Laurent-series expansion~\cite{Mastrolia:2012bu},
and the tensor integral reduction~\cite{Passarino:1978jh,Davydychev:1991va,Denner:2005nn}.
These techniques are automated in the module \ml~\cite{Hirschi:2011pa}, which is used for generating  the amplitudes. We remind the reader that
\ml \, employs \ct~\cite{Ossola:2007ax}, \nin~\cite{Peraro:2014cba,
Hirschi:2016mdz} and \coll~\cite{Denner:2016kdg}, and includes an in-house implementation of the {\sc OpenLoops} optimization~\cite{Cascioli:2011va}.

\subsection{Resummation}
\label{sec:resummation}

The resummation of the soft-gluon emission corrections to the $\ttv$ production processes is carried out as described in detail in  \cite{Broggio:2016zgg,Broggio:2016lfj,Broggio:2017kzi}, with techniques based on Soft Collinear Effective Theory\footnote{For an introductory review of SCET, see \cite{Becher:2014oda}} (SCET) \cite{Bauer:2000yr, Bauer:2001yt, Beneke:2002ph}  and renormalization-group-improved perturbation theory.  We summarize here the salient features of the resummation procedure. In $\ttv$ production, the underlying partonic processes  are of the form
\begin{equation}
i(p_1) + j(p_2) \longrightarrow t(p_3) + \bar{t}(p_4) + V(p_5)+ X \, ,
\end{equation}
where $X$ indicates unobserved final-state light-quark and/or gluon radiation. The incoming partons $i,j$ which enter the production process depend on the boson $V$ under consideration. At lowest order in QCD, if $V = W^\pm$ then $i,j \in {q\bar{q}',\bar{q}'q}$, where $q'$ indicates the isospin partner of the quark $q$.  If $V = H,Z$ instead,   both the quark-annihilation channel and the gluon-fusion channel contribute to the process, so that  $i,j \in {q\bar{q},\bar{q}q,gg}$. 

One can then define the invariants\footnote{In Refs.~\cite{Broggio:2015lya,Broggio:2016lfj,Broggio:2016zgg,Broggio:2017kzi}, as well as in a number of papers on top-quark pair production (see for example \cite{Ahrens:2009uz,Ahrens:2010zv,Broggio:2013uba,Broggio:2014yca}),  the invariant mass of the massive particles in the final states is indicated by $M$, as it is done in this section. However, in Section~\ref{sec:results} we discuss simultaneously results for $\ttw,\ttz$ and $\tth$ production. In that section, in order to avoid any possible source of confusion, we differentiate the three different processes considered by indicating the invariant mass of each one of them as $m(t\bar{t}W),m(\ttz),m(\tth)$, respectively.}
\begin{equation}
\hat{s} \equiv (p_1+p_2)^2 = 2 p_1\cdot p_2 \, , \qquad M^2 \equiv (p_3+p_4+p_5)^2 \, , 
\end{equation}
and starting from these quantities one can define the parameter
\begin{equation}
z \equiv \frac{M^2}{\hat{s}} \, .
\end{equation}
At lowest order in QCD, $z = 1$, while beyond leading order $z \leq 1$. We define the ``soft'' or ``partonic threshold'' limit as the limit $z \to 1$, since in this limit the final state radiation $X$ must be soft.

In the partonic threshold limit the $\ttv$ production cross section factorizes as follows:
\begin{align}
\sigma(s,m_t,m_V) =& \frac{1}{2 s} \int_{\tau_{{\footnotesize \text{min}}}}^{1} d \tau
\int^1_{\tau} \frac{dz}{z} \sum_{i,j} \ff_{ij}\left(\frac{\tau}{z},\mu \right) \nn \\
&\times \int d\text{PS}_{\ttv} \text{Tr} \left[\mathbf{H}_{ij}\left(\{p\},\mu\right) \mathbf{S}_{ij}\left(\frac{M (1-z)}{\sqrt{z}},\{p\},\mu\right) \right] \, . \label{eq:factorization}
\end{align}
In Eq.~(\ref{eq:factorization}), $s$ indicates the  square of the hadronic center of mass energy, the symbol $\{p\}$ is used to indicate the list of momenta $p_1, \cdots, p_5$,   while
\begin{equation}
\tau \equiv \frac{M^2}{s} \, , \qquad \text{and} \quad \tau_{\text{min}} \equiv \frac{\left(2 m_t + m_V\right)^2}{s} \, .
\end{equation}
The functions $\mathbf{H},\mathbf{S}$ and $\ff$ are the hard function, the soft function and the parton luminosity function, respectively. 
These functions are channel dependent and therefore they appear in Eq.~(\ref{eq:factorization}) with an $ij$ subscript.
The trace of the product of the hard and soft function is integrated over the $\ttv$ phase space, whose integration measure is indicated by $d\text{PS}_{\ttv}$. 
The hard and soft functions are matrices in color space.  Only partonic channels that are open at $\LOQCD$ contribute to the cross section in the partonic threshold limit. In the quark-annihilation channel, which contributes to $\ttw, \tth$ and $\ttz$, the hard and soft functions are two-by-two matrices in color space, while in the gluon-fusion channel, which contributes only to $\tth$ and $\ttz$, the hard and soft functions are three-by-three matrices.
Details on the definition of the hard, soft and luminosity functions as well as on the final state phase space can be found in Refs.~\cite{Broggio:2015lya,Broggio:2016zgg,Broggio:2016lfj,Broggio:2017kzi}. 

It is important to observe that the soft functions are singular in the partonic threshold limit $z \to 1$. They contain delta functions and plus distributions of the form
\begin{equation}
P_k(z) \equiv \left[ \frac{\ln^k (1-z)}{1-z} \right]_+ \, .
\end{equation}
The plus distributions are defined in such a way that they can be integrated up to $z = 1$; if $f(z)$ represents a smooth test function that is not singular in the $z \to 1$ limit, then one has
\begin{equation}
\int_0^1 f(z) P_k(z) dz \equiv \int_0^1 \frac{\ln^k(1-z)}{1-z}\left[f(z) - f(1)\right] dz \, .
\end{equation}
At each fixed order in perturbation theory, the soft function involves terms proportional to $\alpha_s^n P_k(z)$, where $n$ indicates the order of QCD corrections and  $0 \le k \le 2 n-1$. For example, NLO QCD corrections include $P_1$ and $P_0$ distributions, NNLO QCD corrections include $P_3,P_2,P_1$ and $P_0$ distributions, etc. These terms arise from soft gluon emission corrections and provide numerically large contributions to the hadronic cross section and differential distributions. In a sense, the purpose of resummation is to account for some of the  terms proportional to the plus distributions  to  all orders in perturbation theory. One convenient way of achieving this goal is to derive and solve the renormalization group equations satisfied by the hard and soft functions. 
 The renormalization group equations  are regulated by anomalous dimensions, which were computed to two loops 
 in Refs.~\cite{Ferroglia:2009ep,Ferroglia:2009ii}.
 
The hard functions and soft functions are free from large logarithmic corrections at appropriately chosen (and different) scales $\mu_h$ and $\mu_s$. At those scales, the hard and soft functions are well behaved in fixed-order perturbation theory. In order to achieve NNLL accuracy, one needs to evaluate the hard and soft function up to NLO. The soft functions are process independent. The soft function for the quark-annihilation channel in $\ttw$ is identical to the quark-annihilation channel soft function for $\ttz$ or $\tth$, up to a trivial replacement of the mass of the heavy boson $m_W \to m_Z$ or $m_W \to m_H$, respectively. Similarly, the soft functions for the gluon-fusion channel in $\tth$ and $\ttz$ production are also identical. The NLO hard functions are instead process dependent. They receive contributions only from one-loop QCD corrections to the production channels that are already open at tree level in QCD: quark annihilation channel for $\ttw$ production, quark-annihilation  and gluon-fusion channels for $\tth$ and $\ttz$ production. The hard functions needed for this work were evaluated by means of a customized version of the code~\ol    \, \cite{Cascioli:2011va} run in combination with the tensor reduction library \coll \, \cite{Denner:2016kdg}.  

The resummation of the soft emission corrections is carried out in Mellin space, where the integral form of the cross section becomes
\begin{align}
\sigma(s,m_t,m_V) = & \frac{1}{2 s} \int_{\tau_{\text{min}}}^{1} \frac{d \tau}{\tau} \frac{1}{2 \pi i} \int_{c - i \infty}^{c + i \infty} dN \tau^{-N} \sum_{ij} \widetilde{\ff}_{ij}\left(N, \mu \right) \int d \text{PS}_{t \bar{t} V} \, \widetilde{c}_{ij} \left(N, \{p\},\mu\right) \, .
\label{eq:Mellinfac}
\end{align}
The Mellin parameter is indicated by $N$ and the threshold limit $z \to 1$ corresponds to the limit $N \to \infty$ in Mellin space.
The functions $\widetilde{\ff}$ and $\widetilde{c}$ are the Mellin transforms of the luminosity function $\ff$ and of the trace of the product of the hard and soft function, respectively. The plus distributions found in the soft function in momentum space are mapped into logarithms of the Mellin parameter in Mellin space, such that in Mellin space the QCD corrections contain terms of the form $\alpha_s^{2+n} \ln^k N$, with $ 0 \leq k \leq 2n $.
Terms suppressed by inverse powers of $N$ in the partonic cross section in Mellin space are neglected in Eq.~(\ref{eq:Mellinfac}).

While the hard and soft functions included in $\widetilde{c}$ are evaluated in fixed order perturbation theory at the scales $\mu_h$ and $\mu_s$, their product is evolved to a common scale $\mu_f$ by solving the renormalization group equations satisfied by the functions. The scale $\mu_f$ is the scale which enters in the PDFs and, consequently, in the parton luminosity function $\tilde{\ff}$.  Ultimately, the resummed hard scattering kernels $\widetilde{c}$ have the following structure
\begin{align}
\widetilde{c}_{ij}(N, \{p\},\mu_f) =  
\mbox{Tr} \Bigg[&\widetilde{\mathbf{U}}_{ij}(\!\bar{N},\{p\},\mu_f,\mu_h,\mu_s) \, \mathbf{H}_{ij}( \{p\},\mu_h) \, \widetilde{\mathbf{U}}_{ij}^{\dagger}(\!\bar{N},\{p\},\mu_f,\mu_h,\mu_s)
\nn \\
& \times \widetilde{\mathbf{s}}_{ij}\left(\ln\frac{M^2}{\bar{N}^2 \mu_s^2},\{p\},\mu_s\right)\Bigg] \,  ,
\label{eq:Mellinresum}
\end{align}
with $\bar{N}=N e^{\gamma_E}$. The evolution factors $\widetilde{\mathbf{U}}$ include the full dependence on potentially large logarithms of the ratios $\mu_h/\mu_s, \mu_h/\mu_f, \mu_f/\mu_s$ and are, like the hard and soft functions, channel-dependent matrices in color space. The explicit expression for the evolution factors in terms of the anomalous dimensions regulating the renormalization group equations can be found in Eq.~(3.7) in reference \cite{Broggio:2016zgg}  for the $\ttw$ case. The evolution factors are  identical also for the $\tth$ and $\ttz$ cases, provided that one accounts for the fact that the explicit expressions of the anomalous dimensions are different for the quark-annihilation and gluon-fusion channels.

If all of the factors in the r.h.s.~of Eq.~(\ref{eq:Mellinresum}) were known at all orders in perturbation theory, the l.h.s.~of the equation would not depend on $\mu_h$ nor on $\mu_s$. However, since the hard function, the soft function, and the anomalous dimensions entering in the evolution factor are all evaluated up to a certain order in perturbation theory, a residual numerical dependence on the choice of $\mu_s$ and $\mu_h$ remains in the predictions presented in Section~\ref{sec:results}. This residual dependence on the scale choices is used, as usual in QCD, to estimate the theoretical error induced by the truncation of the perturbative series in the calculation of the various elements in the resummation formula. As discussed above, one should choose the hard and soft scales $\mu_h$ and $\mu_s$ in such a way that the hard and soft functions are free from large logarithmic corrections and are therefore calculable, at their characteristic scales, in fixed-order perturbation theory. Reasonable choices for these scales are, {\it e.g.}, $\mu_h \sim M, \mu_s \sim M/\bar{N}$  or $\mu_h \sim H_T, \mu_s \sim H_T/\bar{N}$, where $M$ is the invariant mass of the $t\bar{t}V$ final state and $H_T$ is the sum of the transverse mass of the top quark, antitop quark and heavy vector boson:
\begin{equation}
H_T = \sqrt{m_t^2 + p_{T,t}^2} + 
\sqrt{m_t^2 + p_{T,\bar{t}}^2} +
\sqrt{m_V^2 + p_{T,V}^2} \, . \label{eq:HT}
\end{equation}
The issue of scale choices is discussed in Section~\ref{sec:scales}. However, at this stage, it is important to observe that, in order to eliminate large logarithms from the soft function in Mellin space, the soft scale must depend on the Mellin parameter $\bar{N}$. This fact gives rise to a branch cut in $\widetilde{c}$ for large values of $\bar{N}$, which in turn is related to the Landau pole in $\alpha_s$. The integration path in the complex 
$\bar{N}$ plane is chosen according to the {\em Minimal Prescription} \cite{Catani:1996yz}. Notice that the ratio $\mu_h/\mu_s \sim \bar{N}$. An alternative to this approach is to perform the resummation directly in momentum space, fixing the soft scale at the hadronic level through a fitting procedure, see for example \cite{Becher:2007ty,Ahrens:2008nc, Ahrens:2010zv,Broggio:2011bd,Broggio:2013cia}.
  When resummation is carried out up to NNLL accuracy, as it is the case in this work, one is accounting for terms proportional to $\alpha_s^{2+n} \ln^k \bar{N}$ with $ 2 n \ge k \ge \max\{0,2n -3\}$ to all orders in $\alpha_s$ in the partonic cross section in Mellin space. 
 Finally, the parton luminosity functions  in Mellin space, $\tilde{\ff}$, that appear in Eq.~(\ref{eq:Mellinfac}) can be obtained using techniques described in Refs.~\cite{Bonvini:2012sh, Bonvini:2014joa}.
 
NNLL corrections to differential distributions such as the top-quark transverse momentum distribution, the vector boson transverse momentum distribution, the top-pair invariant mass, the $\ttv$ system invariant mass etc., can be obtained by evaluating Eq.~(\ref{eq:Mellinfac}) by means of the in-house Monte Carlo code developed for \cite{Broggio:2016zgg,Broggio:2016lfj,Broggio:2017kzi}. The code evaluates the total cross section while simultaneously binning events w.r.t.~variables which can be built out of the $\ttv$ momenta, such as the ones listed above. 
However, it must be pointed out that, in its current implementation, the code calculates the Mellin transform of the luminosity function in Eq.~(\ref{eq:Mellinfac}) and loses the information about the $x$ values at which the PDFs are evaluated. Hence it cannot be employed to evaluate rapidity distributions to NNLL accuracy in the laboratory frame\footnote{Note that this is not a matter of principle, and indeed NNLL resummation for rapidity distributions was recently carried out in \cite{Pecjak:2018lif}.}.

Nevertheless, the NNLL resummation formula can also be employed to obtain approximate $\rm NNLO_{QCD}$ results, which are indicated by $\rm nNLO_{QCD}$ in this work. The $\rm nNLO_{QCD}$ cross section can be obtained by solving the renormalization group equation satisfied by the NLO soft function. The $\rm nNLO_{QCD}$ predictions discussed in Section~\ref{sec:results} include, on top of the complete NLO,  all of the terms of order $\alpha_s^4 P_k (z)$ ($3 \ge k \ge 0$) in the partonic QCD cross section in momentum space, as well as part of the terms proportional to  $\alpha_s^4 \delta(1-z)$. A detailed description of the terms of the latter class that are included in the  $\rm nNLO_{QCD}$ calculations can be found in Section~3 in~\cite{Broggio:2015lya}. These calculations depend on a single scale $\mu_f$, in contrast with resummed calculations, which have a residual dependence on the scales $\mu_s,\mu_f,\mu_h$. In the context of this work, $\rm nNLO_{QCD}$ calculations allow us to obtain predictions also for rapidity distributions.

We conclude this section by returning to a point briefly mentioned in the discussion of the hard function. The resummation carried out in Refs.~\cite{Broggio:2015lya,Broggio:2016lfj,Broggio:2016zgg,Broggio:2017kzi} and in this work deals with QCD corrections only, meaning that the resummation formulas are linear in the fine structure constant $\alpha$. While it is in principle possible to consider the resummation of soft-gluon emission corrections to contributions that are proportional to higher powers of $\alpha$, their implementation is not trivial. However, the contribution of these corrections is expected to be numerically smaller than the contribution of the soft emission to the QCD process. In addition, one can gain some rough sense of the size of neglected higher order mixed QCD-electroweak corrections by comparing  the multiplicative and additive approaches to the matching of NLO and NNLL calculations, discussed in the next section. Results given in Section~\ref{sec:results} indicate that the difference between the matched results in the additive approach and in multiplicative approach is, with few exceptions, a small effect.

\subsection{Matching procedure}
\label{sec:matching}

The main goal of this paper is to match the NLO QCD and electroweak corrections to $\ttv$ production ({\it i.e.}~the complete-NLO corrections) to the resummation of soft gluon emissions to NNLL accuracy in QCD. In order to achieve this goal, it is necessary to avoid the double counting of terms that are included in both the NLO QCD corrections and the NNLL resummation formula. The method that allows one to avoid such a double counting is well understood and goes under the name of {\em matching procedure}.

In order to understand the details of the matching procedure it is necessary to identify terms in the NLO QCD partonic cross section that are included in the resummation formula in Eq.~(\ref{eq:Mellinresum}). If one sets $\mu_s = \mu_h = \mu_f$ in that equation, the evolution factors $\widetilde{\mathbf{U}}$ become identity matrices in color space. In that situation, the trace of the hard function and Mellin-space soft function (both evaluated to NLO)  includes terms proportional to $\alpha \alpha_s^3 \ln^2{\bar{N}}$ and $\alpha \alpha_s^3 \ln{\bar{N}}$, as well as terms that do not depend on the Mellin parameter $\bar{N}$. The latter class of terms  still depends on the Mandelstam invariants; nevertheless, those $\bar{N}$ independent terms are referred to as ``constant'' terms. Terms proportional to inverse powers of the Mellin parameter, which are present in the full QCD partonic cross section at NLO in Mellin space, cannot be reconstructed starting from the NNLL resummation formula. The trace of the hard function and soft function at NLO in Mellin space, including the terms discussed above, can be  inserted in Eq.~(\ref{eq:Mellinfac}) to obtain what is referred to as the {\em approximate} NLO QCD cross section, denoted here with the subscript ${\rm nLO_{ QCD}}$. The ${\rm nLO_{ QCD}}$ cross section contains the contribution of all of the terms proportional to $\alpha \alpha_s^3 P_k(z)$ ($k=0,1$) and $\delta(1-z)$ in the partonic cross section in momentum space. In analogy with the notation introduced in Section~\ref{sec:completeNLO}, we indicate the terms of $\ord{\alpha \alpha_s^3}$ included in the ${\rm nLO_{ QCD}}$ corrections to a given observable with $\cS_{\rm nLO_{ QCD}}$.  Consequently, we  define the observable $\Sigma$ evaluated to ${\rm nLO_{ QCD}}$ as
\begin{equation}
\Sigma_{\rm nLO_{ QCD}} \equiv \Sigma_{\rm LO_{QCD}} + \cS_{\rm nLO_{ QCD}} \, .
\end{equation}

Once the QCD and EW complete-NLO corrections (whose sum   will simply be referred to as NLO), the NNLL corrections, and the ${\rm nLO_{ QCD}}$ predictions for a given observable $\Sigma$ are available, it is straightforward to combine them into an NLO+NNLL prediction by using the matching formula
\begin{equation}
\Sigma_{\text{NLO+NNLL}} \equiv \Sigma_{\text{NLO}} + \left[\Sigma_{\rm NNLL} - \Sigma_{\rm nLO_{QCD}}\right] \, . \label{eq:matching}
\end{equation}

The symbol $\Sigma_{\rm NNLL}$ indicates the numerical value of the resummed total cross section in Eq.~(\ref{eq:Mellinfac}) or, in the case of differential distributions, the value of that resummed cross section in a specific bin of the distribution.
The terms included in square brackets in Eq.~(\ref{eq:matching}) are of  $\ord{\alpha \alpha_s^4}$ and higher, and represent the NNLL corrections to be added to the NLO result.
The quantity $\Sigma_{\text{NLO+NNLL}}$ is defined in such a way as to include all of the corrections to the observable $\Sigma$ considered in this work. In discussing the results of this study, it is also useful to match the resummed formulas to the QCD cross section only, by excluding all the EW corrections. 
In that case, Eq.~\eqref{eq:matching} must be modified by replacing $\NLO \rightarrow \NLOQCD$ in the first term in the r.h.s.~of the equation. The predictions obtained in this way include only QCD effects and are indicated by the $\NLOQCD+{\rm NNLL}$ subscript:  
\begin{equation}
\Sigma_{\rm NLO_{QCD}+NNLL} \equiv \Sigma_{\rm NLO_{QCD}} + \left[\Sigma_{\rm NNLL} - \Sigma_{\rm nLO_{QCD}}\right] \, . \label{eq:matchingNLO}
\end{equation}
Calculations at  $\NLOQCD+{\rm NNLL}$ accuracy correspond to the results presented in Refs.~\cite{Broggio:2016lfj,Broggio:2016zgg,Broggio:2017kzi}.

Analogously, it is possible to match $\rm nNLO_{QCD}$ predictions discussed at the end of Section~\ref{sec:resummation} to the complete-NLO prediction. A given observable $\Sigma$ can be evaluated to $\rm nNLO$ by calculating the quantity 
\begin{equation}
\Sigma_{\text{nNLO}} \equiv \Sigma_{\text{NLO}} + \left[\Sigma_{\rm nNLO_{QCD}} - \Sigma_{\rm NLO_{QCD}} \right] \, . \label{eq:matchingnNLO}
\end{equation}
In Eq.~(\ref{eq:matchingnNLO}), $\Sigma_{\rm NLO_{QCD}}$ includes ${\rm LO_{QCD}}$ terms of $\ord{\alpha \alpha_s^2}$ and ${\rm NLO_{QCD}}$ terms of $\ord{\alpha \alpha_s^3}$.  $\Sigma_{\rm nNLO_{QCD}}$ contains terms of $\ord{\alpha \alpha_s^n}$ ($2 \le n \le 4$), including the complete $\Sigma_{\rm NLO_{QCD}}$ cross section. Consequently, the square bracket in Eq.~(\ref{eq:matchingnNLO}) includes only the terms of  $\ord{\alpha \alpha_s^4}$ that must be added to the complete-NLO calculation in order to evaluate the observable to  nNLO.
Finally, one can exclude the  EW corrections from Eq.~\eqref{eq:matchingnNLO} by replacing  $\NLO \rightarrow \NLOQCD$ in the first term on the r.h.s.: in this way one obtains approximate NNLO corrections to the  QCD process, which are indicated with $\rm nNLO_{QCD}$.

Eqs.~\eqref{eq:matching} and~\eqref{eq:matchingnNLO} combine NLO to NNLL QCD  or approximate NNLO QCD
corrections in an additive approach, which is well defined in perturbation theory. However, it is possible to combine these contributions within a multiplicative approach,  which is often employed in combining NLO QCD  and NLO EW  corrections, denoted in this work by $\NLO_1$ and $\NLO_2$, respectively. While in the additive approach $\NLO_1$ and $\NLO_2$ are simply summed so that
\begin{equation}
\cS_{\rm NLO_{QCD+EW}}= \cS_{\rm NLO_1} + \cS_{\rm NLO_2}\, , \label{eq:additive}
\end{equation}
 in the multiplicative approach these two corrections are combined via the prescription 
\begin{equation}
\cS_{\rm NLO_{QCD \times EW}}=\cS_{\rm NLO_1} + \cS_{\rm NLO_2} \left(\frac{\Sigma_{\rm NLO_{QCD}}}{\Sigma_{\rm LO_{QCD}}} \right) . \label{eq:multiplicative}
\end{equation}
By comparing Eqs.~(\ref{eq:additive}) and (\ref{eq:multiplicative}), it is possible to see that differences between the
two approaches only enters at the level of mixed QCD-EW NNLO corrections of $\ord{\alpha_s\alpha}$ relative to $\LO_1$, {\it i.e.}, in the case of  $t \bar t V$  cross sections at $\ord{\alpha_s^3\alpha^2}$, which is beyond the accuracy of the calculations presented in this work.
 However, there are  specific configurations where the
multiplicative approach is well-motivated and expected to provide improved predictions. The typical case is 
when the $\NLO_1$
contribution is dominated by soft-QCD physics, and the $\NLO_2$ correction by large EW Sudakov logarithms. Indeed, these two
classes of corrections factorize, and therefore the entire
mixed QCD-EW NNLO corrections  of $\ord{\alpha_s\alpha}$ relative to $\LO_1$ are expected to be  well approximated by the difference between Eq.~(\ref{eq:multiplicative}) and Eq.~(\ref{eq:additive}), namely
\begin{equation}
\cS_{\rm NLO_{QCD \times EW}}  - \cS_{\rm NLO_{QCD + EW}} = \frac{\cS_{\rm NLO_2}\cS_{\rm NLO_1}}{\Sigma_{\rm LO_{QCD}}} \, .
\end{equation}

The resummation procedure allows one to account for   soft emission corrections  at all orders in $\alpha_s$. In particular, the NNLL resummation discussed in this work accounts for  terms in the partonic cross section in Mellin space that are proportional to $\alpha \alpha_s^{2+n} \ln^k \bar{N}$ with $2n \ge k \ge \max\{0,2n -3\}$,  where the soft configuration corresponds to the limit $\bar N \to \infty$. Consequently, one can generalize the multiplicative approach to approximate not only the mixed QCD-EW NNLO corrections of $\ord{\alpha_s\alpha}$ relative to the $\LOQCD$ observables, but also the corrections to  the Mellin space partonic cross section   proportional to  $\alpha^2 \alpha_s^{1+n} \ln^k \bar{N}$ with $2 n  \ge k \ge 2n -3$, for all orders in $\alpha_s$.
A resummed observable $\Sigma$ can then be evaluated  in the multiplicative approach at ${\rm NLO \times NNLL}$ accuracy as follows:
\begin{equation}
\Sigma_{\rm NLO \times NNLL} = \Sigma_{\text{NLO+NNLL}} + \cS_{\rm NLO_2} \left(\frac{\Sigma_{\NLOQCD+{\rm NNLL}}}{\Sigma_{\LOQCD}} -1\right)\, . \label{eq:matchingmult}
\end{equation}
Similarly, it is also possible to combine $\rm nNLO_{QCD}$ predictions to the complete-NLO ones in the multiplicative approach by using the matching relation
\begin{equation}
\Sigma_{\rm nNLO_{mult}} = \Sigma_{\text{nNLO}} + \cS_{\rm NLO_2} \left(\frac{\Sigma_{\rm nNLO_{QCD}}}{\Sigma_{\LOQCD}}-1\right)\, . \label{eq:matchingnNLOmult}
\end{equation}

In the tail of the differential distributions for $\tth$ and $\ttz$ productions, where Sudakov logarithms are large and QCD radiation is typically soft, ${\rm NLO \times NNLL}$ predictions can be considered as an improvement w.r.t.~those at ${\rm NLO + NNLL}$ accuracy. In the rest of the phase space this is not necessarily true. Therefore, the difference between the two approximations can be considered as an estimate of the impact of missing higher-order QCD-EW terms. The same argument holds for the comparison between ${\rm nNLO}$ and ${\rm nNLO_{mult}}$ predictions.

The situation is completely different in the case of $\ttw$, where the $\NLO_1$ contribution is dominated by hard radiation, as discussed in Section~\ref{sec:phenointro}. In addition, the Sudakov logarithms present in $\NLO_2$ are proportional to  the $\LO_1$ contribution, which arises from a $q\bar q '$ initial state, while the dominant $\NLO_1$ contributions arise from quark radiation in $qg$ initiated processes.  
Thus, in the case of $\ttw$ production, the multiplicative approach cannot be motivated by sound theoretical arguments. This is particularly relevant in the tail of the distributions, where both the $\NLO_1$ and $\NLO_2$ corrections  are large, the latter due to the presence of Sudakov logarithms. Therefore the multiplicative approach  can lead to uncontrolled  NNLO terms. Moreover, since in $\ttw$ production the $\NLO_3$ correction is numerically much larger than the $\NLO_2$ contribution, even if the multiplicative approach as defined in Eqs.~\eqref{eq:matchingmult} and \eqref{eq:matchingnNLOmult} were justified, it would probably not account for the dominant mixed QCD-EW NNLO contributions, which are expected to be those of $\ord{\alpha_s^2 \alpha^3}$ relative to the ${\rm LO}_1$ cross section.
For consistency, in Section~\ref{sec:results}  results in the additive and multiplicative approaches are shown and compared also for the $\ttw$ process. However, one should bear in mind that only in the case of $\tth$ and $\ttz$ production can the multiplicative approach be expected to improve the predictions.

\subsection{Structure of the fixed-order  corrections}
\label{sec:phenointro}

\begin{figure}[t]
	\centering
	\includegraphics[width=0.32\textwidth]{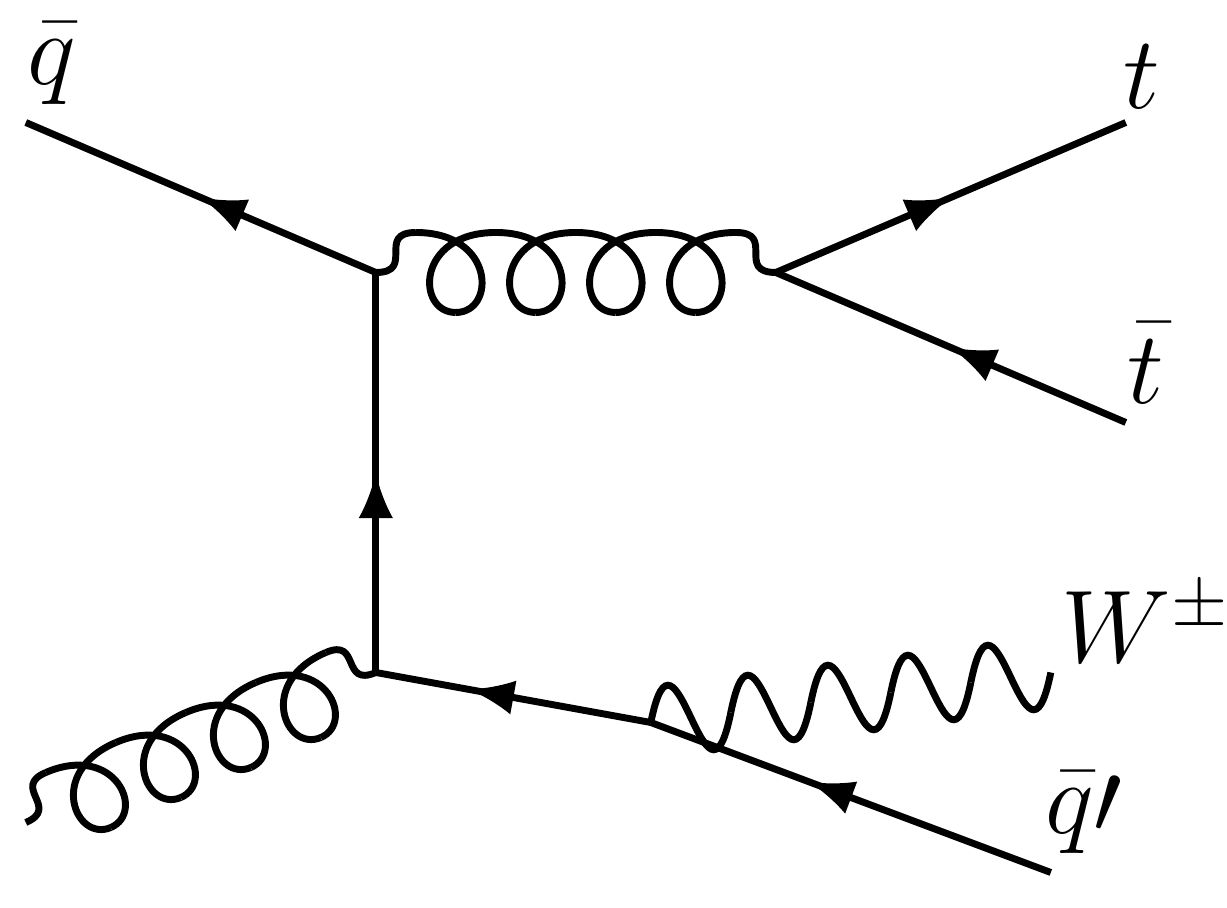}
	\hspace*{2.cm}
	\includegraphics[width=0.32\textwidth]{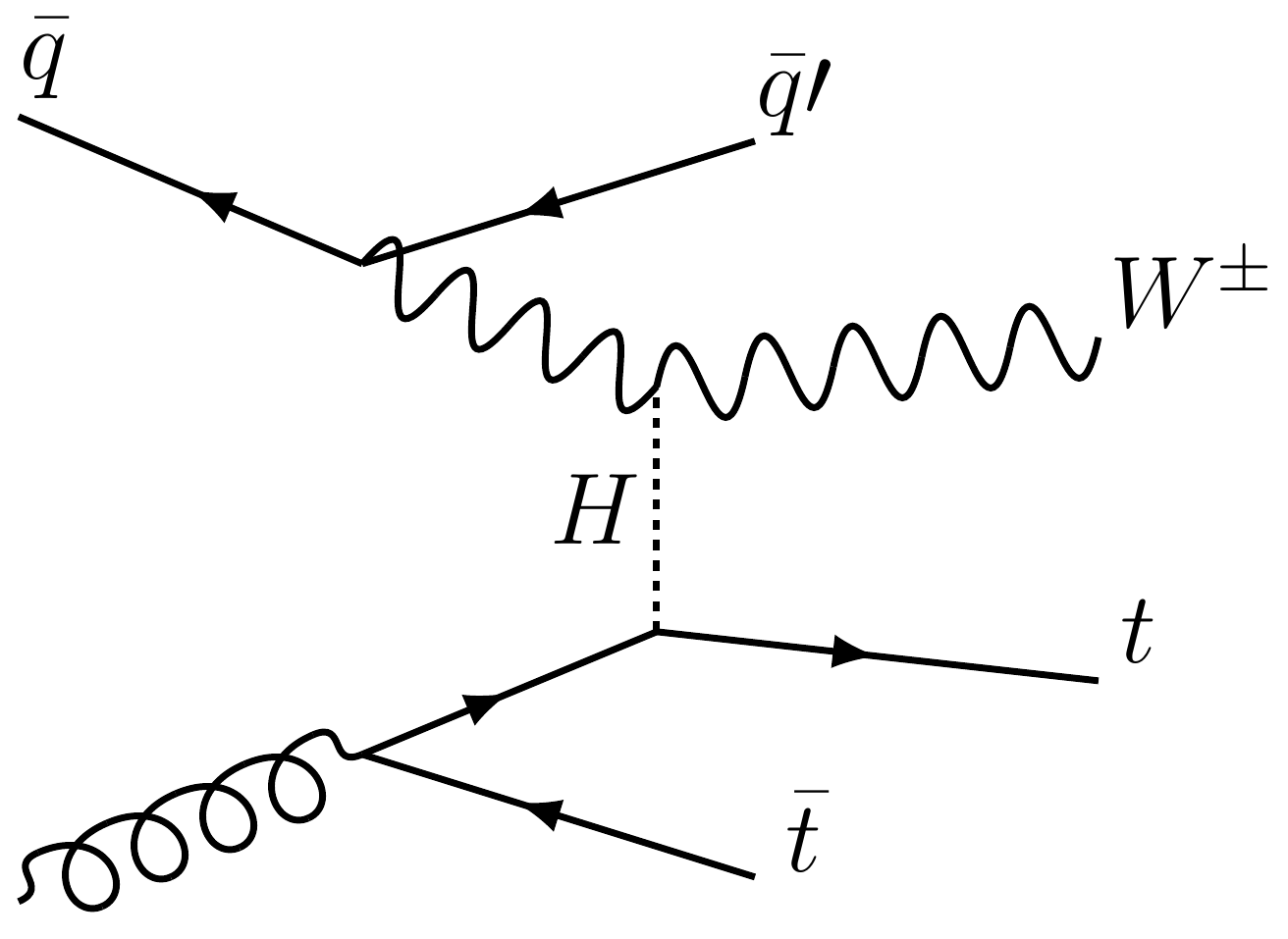}
	\caption{Representative diagrams for the  $\bar qg\to\ttw \bar q'$ real-emission amplitudes. The left diagram  leads to $\log^2(\pt^2(\ttt)/m_W^2)$ terms in the $\NLO_1$ contribution. The right diagram  shows an example of $t W \to t W$ scattering  and contributes to the $\NLO_3$. }
	\label{fig:realttw}
\end{figure}

This section describes the structures underlying the complete NLO corrections to $\ttv$  production.  We start by reviewing the most important features of  $\ttwp$ and $\ttwm$ production, which are discussed in detail in~\cite{Frederix:2017wme}. Subsequently, we consider $\tth$ and $\ttz$ production. 

In $\ttwp$ ($\ttwm$) production at LO only $u \bar d$ ($\bar{u} d$) initial states contribute, where $u$ and $d$ are a generic up- and down-type quarks. The $W^+ (W^-)$ boson is radiated from the $u\, (d)$ quark,  while the $\ttt $ pair is produced either via a gluon or a photon/$Z$ boson. The gluon mediated diagrams contribute to the $\LO_1$ cross section, while the diagrams involving a photon/$Z$ boson contribute to the   $\LO_3$ cross section.  The interference between these two classes of diagrams
vanishes after summing over colors, so that the $\LO_2$
cross section also vanishes.
On the contrary, all of the $\NLO_i$ contributions are non-vanishing. 

The $\NLO_1$ contribution to the $\ttw$ production process is in general large. It  was calculated in~\cite{Hirschi:2011pa,Garzelli:2012bn,Campbell:2012dh, Maltoni:2014zpa} and studied in detail in~\cite{Maltoni:2015ena}. Large QCD corrections are mainly induced  by the opening of the $gq \to \ttw q'$ channel, which depends on the gluon luminosity and  therefore is enhanced in high-energy proton--proton collisions. Moreover, the radiation of quarks in $gq \to \ttw q'$ is typically hard and in particular very large $K$-factors are present in the tail of the $\pttt$ distribution, which receives an additional $\log^2(\pt^2(\ttt)/m_W^2)$ enhancement on top of the one due to the $gq$ luminosity  (see left diagram in Figure~\ref{fig:realttw} and~\cite{Maltoni:2015ena} for a detailed discussion).  The impact of multiple soft-gluon emissions for this process  is scale sensitive and  non-negligible ~\cite{Li:2014ula,Broggio:2016zgg, Kulesza:2018tqz}; the predictions contained in Section~\ref{sec:results}  account for soft emission up to NNLL accuracy. However, it is important to observe that a  large component of $\NLO_1$ corrections, and therefore the associated scale uncertainties, originates from hard radiation in the  $gq \to \ttw q'$ channel. Therefore, the threshold resummation in the $q\bar q' \to \ttw $ channels is not expected to drastically reduce the total scale uncertainty. A detailed discussion of the size of the various corrections can be found in Section~\ref{sec:results}.

For what concerns  the EW contributions to $\ttw$ production, the $\NLO_2$ corrections were calculated for the first time in~\cite{Frixione:2015zaa} and further phenomenological studies were provided in~\cite{deFlorian:2016spz}. In a boosted regime, due to Sudakov logarithms,  the $\NLO_2$ corrections can be as large as the NLO QCD scale uncertainty.
The $\NLO_3$ contribution is sizable \cite{Frederix:2017wme} since it contains $gq \to \ttw q'$ real-emission channel that involves EW $t W \to t W$  scattering \cite{Dror:2015nkp} (see right diagram in Figure~\ref{fig:realttw} and~\cite{Frederix:2017wme} for a detailed discussion). Similarly to what happens in the case of the $\NLO_1$ corrections, this channel becomes even more relevant as the LHC center-of-mass energy grows,  due to the presence of an initial-state gluon. Although $t W \to t W$ scattering is present also in the $\NLO_4$ corrections, in that case it is induced by a $\gamma q$ initial state. It is therefore suppressed w.r.t.~$t W \to t W$ scattering contributing to $\NLO_3$ by the smaller luminosity of the photon and also by a factor $\alpha/\alpha_s$. Similarly,  all of the other $\NLO_4$ terms are negligible since they are  
of $\ord{\alpha^4}$.

\medskip
In contrast to the case of $\ttw$ production, tree-level Born diagrams for $\tth$ and $\ttz$ production are induced by both $gg$ and $q\bar q$ initial states. In particular, the gluon-fusion channel contributes only to the $\LO_1$ term and, due to the partonic luminosity, yields the largest part of the LO cross section. The $q\bar q$ initial states contribute also to $\LO_3$ via squared diagrams featuring $\ttt$ pairs stemming from a photon or $Z$ propagator. Similarly to the $\ttw$ case, their interference with diagrams contributing to $\LO_1$ vanishes due to color. However, the $\LO_2$ contribution to the cross section is non-vanishing for these two processes. Indeed, some of the $b \bar b$ initial-state diagrams feature a $t$-channel $W$-boson that leads to non-vanishing interference contributions. Moreover, the $\gamma g$ initial-state processes contribute to $\LO_2$ via squared diagrams. As shown in~\cite{Frederix:2018nkq}, the $\LO_2$ and $\LO_3$ contributions to the cross section are numerically negligible.

All of the $\NLO_i$ contributions are non-vanishing.  The $\NLO_1$ correction is in general large; it  was calculated in~\cite{Beenakker:2001rj, Beenakker:2002nc, Dawson:2002tg, Dawson:2003zu} for $\tth$ and in~\cite{Hirschi:2011pa,Lazopoulos:2008de,Garzelli:2011is,Kardos:2011na,Garzelli:2012bn} for $\ttz$. In addition, the $\NLO_1$ correction was studied in detail in~\cite{Maltoni:2015ena}, where, as in the case of $\ttw$ production, large $K$-factors for the $\pttt$ differential distribution were found. On the other hand, in presence of LO contributions involving two gluons in the initial state, the $qg$ luminosity is not providing a significant enhancement of the cross section. Furthermore, in contrast to the case of $\ttw$ production,  the QCD emissions in the $\NLO_1$ corrections are {\it not} typically hard. Also, the largest contribution from QCD emissions arises from the $gg$ initial state, which is in general the dominant partonic channel both for the $\LO_1$ and the $\NLO_1$. This also applies to the corrections to all the $\tth$ and $\ttz$ differential distributions considered in this work. For this reason, by resumming soft emission corrections to NNLL accuracy one observes, as expected, a sizable reduction of the residual scale uncertainty affecting the total cross section and differential distributions. Also this feature will be quantified in detail in Section~\ref{sec:results}.

For what concerns the EW corrections to $\tth$ and $\ttz$ production, the $\NLO_2$ corrections were calculated for the first time in~\cite{Frixione:2015zaa}  and further phenomenological studies were carried out in~\cite{deFlorian:2016spz}.
For the total cross section, the relative size of the $\NLO_2$ corrections is   smaller than in $\ttw$ production. However, in the tail of the differential distributions the $\NLO_2$ contribution can be non-negligible in comparison to the NLO QCD scale uncertainty.

For $\tth$ and $\ttz$ production, the $\NLO_3$ and $\NLO_4$ corrections  were calculated in~\cite{Frederix:2018nkq}; a phenomenological study involving these contributions to the cross section is presented for the first time in this work. Compared to $\ttw$ production, not only the $\NLO_4$ but also the $\NLO_3$ correction is small. At this order,  $\ttt Z$ ($\ttt H$) production involves $t Z \to t Z$ ($t H \to t H$) scattering in $gq \to \ttt Z q$ ($gq \to \ttt H q$) real-emission channels. However, as discussed for the case of the $\pttt$ enhancement, the $qg$ luminosity is not providing a significant enhancement and therefore the relative size of $\NLO_3$ correction in $\ttz$ and $\tth$ production is smaller than in $\ttw$ production.

The $\tth$ and $\ttz$ processes share several features at the diagrammatic level and therefore also at the phenomenological level. This similarity, which is present also at the level of $\NLO_1$ and $\NLO_2$ corrections to these processes, was advocated as a possible proxy to be used to reduce the theoretical uncertainties in the measurements of the top-quark Yukawa coupling \cite{Plehn:2015cta}. The main kinematic difference between these two processes concerns the rapidity of the $V$ bosons \cite{Maltoni:2015ena}, because the $Z$ bosons can be emitted both from initial-state quarks and final-state top-quarks, while the $H$ boson can be emitted only from the latter. This situation is markedly different from the case of  $\ttw$ production, where the $W^{\pm}$ bosons are emitted only from the initial state light quarks. Moreover, as discussed before, $\NLO_1$ and $\NLO_3$ corrections have a very different impact on predictions for $\ttw$ production compared to $\tth$ and $\ttz$ production. Similarly, the impact of soft gluon resummation is different for $\ttw$ production and in $\tth$ or $\ttz$ production.

\section{Input parameters, scales and PDFs}
\label{sec:inputparam}

The predictions presented in Section~\ref{sec:results} depend on the numerical values of physical input parameters, on the PDFs employed in the calculations, and on the choice of the unphysical scales that enter in fixed-order and resummed calculations. The choices made in this work are listed and discussed in this section.

\subsection{Masses and couplings}

The masses of the heavy SM particles are set equal to
\begin{equation}
m_t = 173.34 \text{ GeV}\, , \quad m_W = 80.385 \text{ GeV} \, , \quad m_Z = 91.1876 \text{ GeV}\, , \quad m_H = 125 \text{ GeV} \,,
\end{equation}
whereas all the other masses are set equal to zero. The decay widths of all particles are also set to zero. In addition, we use the on-shell renormalization scheme for all masses. The strong coupling $\alphas$ is renormalized in the $\MSb$-scheme with five active flavors, while the EW input parameters and the renormalization condition for $\alpha$ are in the $G_\mu$-scheme, with
\begin{equation}
G_\mu = 1.16639 \times 10^{-5} \text{ GeV}^{-2} \,.
\end{equation}
The CKM matrix  is set equal to the $3 \times 3$ unity matrix.

\subsection{Scale choices and uncertainties}
\label{sec:scales}
In the literature, two choices for the (functional form of the) central values for the factorization and renormalization scale
entering these processes are commonly adopted. In particular, in~\cite{Frederix:2018nkq} it was argued that
$H_T /2$, with $H_T$ defined as in Eq.~(\ref{eq:HT}),
is a reasonable choice for the factorization and renormalization scale. In~\cite{Broggio:2016lfj,Broggio:2016zgg,Broggio:2017kzi}, on the other hand,  scales based on the top-antitop-heavy-boson invariant mass, $M \equiv m(\ttv)$, were used. The former work is based on fixed-order perturbation theory, while the latter also considers the resummation of soft emission corrections. An additional study of the different scale choices was carried out in~\cite{Maltoni:2015ena}.

Since the factorization/renormalization scale and the hard and soft scales are unphysical, it is acceptable and even recommendable to explore different scale choices. Numerical differences among values of the same observables evaluated for different scale choices can be used as an estimate of the uncertainty associated to the truncation of the perturbative series.
In this work, we consider both  $H_T$ and  $m(t\bar{t}V)$ based scale choices.
In particular, when relating the central value of the three scales $\mu_f,\mu_h,\mu_s$ involved in the calculations  to the value of $m(\ttv)$, we choose
\begin{equation}
\mu_f^0 = \frac{m(\ttv)}{2} \, , \qquad \mu_h^0 = m(\ttv)\, , \qquad \mu_s^0 = \frac{m(\ttv)}{\bar{N}} \, . \label{scale1}
\end{equation}
When we relate the scales $\mu_f,\mu_h,\mu_s$ to $H_T$ we set instead
\begin{equation}
\mu_f^0 = \frac{H_T}{2} \, , \qquad \mu_h^0 = \frac{H_T}{2}\, , \qquad \mu_s^0 = \frac{H_T}{\bar{N}} \, . \label{scale2}
 \end{equation}

The uncertainty associated to missing higher-order corrections can be estimated by considering the dependence of the predictions for a given observable on the non-physical scales that
enter the calculation. 
At fixed order, this is done by varying  the
renormalization and factorization scales  in the range 
$\mu_i \in \{\mu^0_i/2, 2 \mu_i^0\}$ ($i = r,f$). The uncertainty estimate is then given by the bin-by-bin envelope of the 9 predictions obtained in this way.
For the resummed results, the hard, soft and factorization scales are varied  in the range $\mu_i \in \{\mu^0_i/2, 2 \mu_i^0\}$ ($i = s,h,f$). In particular, by introducing the notation $\kappa_i \equiv \mu_i /\mu_i^0$ ($i\in\{f,h,s\}$), one can rewrite  Eq.~(\ref{eq:matching}) by making explicit the dependence of each element on $\kappa_i$:
\begin{equation}
\Sigma_{\rm NLO+NNLL} \left(\kappa_f,\kappa_h,\kappa_s\right) = \Sigma_{\rm NLO}\left(\kappa_f\right) + \left[ \Sigma_{\rm NNLL} \left(\kappa_f,\kappa_h,\kappa_s\right) - \Sigma_{\nLOQCD} \left(\kappa_f\right)\right]  \, .
\label{eq:matchedscalevariation}
\end{equation}
In contrast to the case of NLO calculations, in Eq.~(\ref{eq:matchedscalevariation}) no distinction is made between the renormalization and factorization scales. One can then define an upper and lower scale uncertainty for the variation of each scale in  Eq.~(\ref{eq:matchedscalevariation}) as follows
\begin{align}
\Delta \Sigma^+_{{\rm NLO+NNLL}, i}  = & \max_{\kappa_i \in \{1/2,1,2\}} \left[\Sigma_{\rm NLO+NNLL}\left(\kappa_i \right)\right] - \Sigma_{\rm NLO+NNLL}\left(\kappa_i = 1\right)\, ,\nonumber \\
\Delta \Sigma^-_{{\rm NLO+NNLL}, i}  = & \min_{\kappa_i \in \{1/2,1,2\}} \left[\Sigma_{\rm NLO+NNLL}\left(\kappa_i \right)\right] - \Sigma_{\rm NLO+NNLL}\left(\kappa_i = 1\right)\, ,
\label{eq:scalevar}
\end{align}
for $i \in \{f,h,s\}$. In Eqs.~(\ref{eq:scalevar}) the two scales that are not varied are kept fixed to their central values: $\kappa_j = 1$ if $j\neq i$. 
The  residual theoretical uncertainty affecting a given resummed observable  is then obtained by
combining in quadrature the uncertainties associated to each of the  three scale variations in each of the histogram bins 
as done in Refs.~\cite{Broggio:2016zgg,Broggio:2016lfj,Broggio:2017kzi}. With reference to Eq.~(\ref{eq:scalevar}) one can then define the upper and lower scale uncertainty as
\begin{equation}
\Delta \Sigma^\pm_{{\rm NLO+NNLL}} = \pm \sqrt{\left( \Delta \Sigma^\pm_{{\rm NLO+NNLL}, f}\right)^2 + \left( \Delta \Sigma^\pm_{{\rm NLO+NNLL}, h}\right)^2 + \left( \Delta \Sigma^\pm_{{\rm NLO+NNLL}, s}\right)^2} \, .
\end{equation}

As discussed in Section~\ref{sec:results},
results with the two scale choices in Eqs.~\eqref{scale1} and \eqref{scale2} are
compatible with each other at the level of total cross sections, although
somewhat less so at the level of differential distributions.
Since there is no conclusive argument in favor of either 
scale choice,  we opt for taking the bin-by-bin average of the two results as the best
prediction for the central value of each given observable. Moreover, we use
the envelope of the uncertainty bands generated with the two scale choices
as an estimate of the missing higher-order corrections.
This combined-scale method  is particularly relevant in the case of $\ttw$ production,
where for the two choices in Eqs.~$\eqref{scale1}$ and $\eqref{scale2}$
individually one observes that the NNLL corrections lead to a reduction of scale uncertainty for the total cross section, but the difference 
between the central values obtained with the two scale
choices increases when NNLL corrections are accounted for. Hence, we think that considering
only a single central scale choice and its corresponding uncertainty band
underestimates the uncertainties due to missing higher orders.
On the other hand, we believe that our procedure of taking the envelope of the scale
uncertainty bands of the two calculations considered results in a
reliable and robust uncertainty estimate.

With the notation introduced above, one can rewrite Eq.~(\ref{eq:matchingnNLO}) as 
\begin{equation}
\Sigma_{\rm nNLO} (\mu_f) = \Sigma_{\rm NLO} (\mu_r = \mu_f) + \left[\Sigma_{\nNLOQCD} (\mu_f) -  \Sigma_{\nLOQCD} (\mu_f)\right] \, .
\label{eq:scalevnNLO}
\end{equation}
The scale uncertainty associated to this quantity is obtained varying $\mu_f$ in the range $\{\mu_f^0/2, 2 \mu_f^0\}$. Finally, in  calculations where the matching between NLO and NNLL corrections is carried out within the multiplicative approach, the scale uncertainty is obtained 
by applying the method described in Eqs.~\eqref{eq:matchedscalevariation} and \eqref{eq:scalevnNLO} to Eqs.~\eqref{eq:matchingmult} and \eqref{eq:matchingnNLOmult}.

\subsection{PDF uncertainties}

Results in Section~\ref{sec:results} are obtained by using the {\tt LUXqed17\_plus\_PDF4LHC15\_nnlo\_100}  PDF set  \cite{Manohar:2016nzj,Manohar:2017eqh}, which in turn was obtained starting from  the {\sc \small PDF4LHC} PDF set \cite{Butterworth:2015oua, Ball:2014uwa, Harland-Lang:2014zoa, Dulat:2015mca}. The PDFs in Refs.~\cite{Manohar:2016nzj,Manohar:2017eqh} include NLO QED effects in the DGLAP evolution \cite{deFlorian:2015ujt, deFlorian:2016gvk} and they provide the most precise determination of the photon PDF available to date.
In complete-NLO calculations, the PDF uncertainties are evaluated by means of {\sc\small MadGraph5\_aMC@NLO} thanks to the procedure introduced in~\cite{Frederix:2011ss}. In this way, one can calculate an observable for each PDF replica in a given PDF set.
The PDF uncertainties related to the NNLL resummation corrections, {\it i.e.}, the terms between square brackets in Eq.~\eqref{eq:matching}, are instead evaluated with an approximation. This is necessary because the evaluation of the NNLL resummation formulas for all of the PDFs in a given set would require an excessive amount of computer time. The approximation relies on the assumption that the relative PDF uncertainty associated to the part of the $\NLO_1$ corrections that does not depend on the $qg$ luminosity, denoted here by $\NLO_{1}^{{\rm no}-qg}$, is the same relative PDF uncertainty affecting $\Sigma_{\rm NLO+ NNLL} -\Sigma_{\rm NLO}$. In this approximation, $\NLO_1$ corrections arising from quark-gluon channel diagrams are excluded from the calculation of the relative error induced by PDFs in resummed calculations because this channel is subleading in the threshold limit.  Therefore, for each replica $i$ in the PDF set  we assume that 
\begin{eqnarray}
(\Sigma_{\rm NLO+ NNLL})_i &=&  (\Sigma_{\rm NLO})_i + (\Sigma_{\rm
NLO+ NNLL} -\Sigma_{\rm NLO})|_{\rm central} 
%\nonumber \\ &\times&
 \times \frac{\cS_{\NLO_{1}^{{\rm no}-qg}}|_i
}{\cS_{\NLO_{1}^{{\rm no}-qg}}|_{\rm central}
} \label{pdfrescale} \, ,
\end{eqnarray}
where the subscript ``central'' refers to the central PDF prediction. In conclusion, for each replica $i$ the value of $(\cS_{\NLO_{1}^{{\rm no}-qg}})|_i$ is evaluated via {\sc\small MadGraph5\_aMC@NLO}, rescaled as prescribed by Eq.~\eqref{pdfrescale} and added back to  $\Sigma_{\rm NLO}|_i$, to provide an estimate of the NLO+NNLL calculation carried out with the replica $i$ in the PDF set. Once an NLO+NNLL prediction is available for each replica $i$, PDF uncertainties are evaluated following the standard procedure for the PDF set considered. 

The same procedure can be employed  for the nNLO predictions via the substitution $\rm NLO+NNLL \rightarrow nNLO$ in Eq.~\eqref{pdfrescale}. In the case of the combination of NLO predictions and NNLL corrections in the multiplicative approach we do not evaluate PDF uncertainties, but it is reasonable to think that they would be similar in size to those calculated in the additive approach.

\section{Results}
\label{sec:results}
In this section we present predictions  for each of the four processes considered in this work, namely $\ttwp$,
$\ttwm$, $\tth$, and $\ttz$ production. We start by considering total cross sections and charge asymmetries, and then give results for differential distributions in Section~\ref{sec:diffdist}.

\subsection{Total cross sections and asymmetries}
The total cross sections for the  four processes, calculated within 
different perturbative approximations, can be found in the middle columns
of Tables~\ref{ttwp}-\ref{ttz}. 
Each table is subdivided in three sections:
\begin{itemize}
\item In the top section of each table the cross sections are evaluated with the $m(t\bar{t}V)$-based scale choices listed in Eq.~(\ref{scale1}). 

\item In the middle section of the tables the cross sections are evaluated with the $H_T$-based scale choices listed in Eq.~(\ref{scale2}). 

\item Finally,  the lower section of each table  shows the combination of the results for the two aforementioned scale choices. The results for the scale choices in 
Eqs.~\eqref{scale1} and \eqref{scale2} are combined as explained in Section~\ref{sec:scales}; the results for the total cross section listed in the lower portion of the tables represent one of the main results of this paper. 
\end{itemize}

In each of the three
parts of the tables, the predictions become more accurate
as one moves from the highest line to the lowest line. Each line starts with a label that indicates the accuracy of the calculations found on that line. For convenience, we summarize the notation introduced in Section~\ref{sec:calcframe} and used to label the various rows: 
\begin{itemize}
\item $\LOQCD$: rows labeled in this way are based on tree-level QCD  calculations, {\it i.e.}, they include only the
  $\LO_1$ contribution to the cross section.
\item $\NLOQCD$: lines labeled in this way include the $\LOQCD$ calculation added to the NLO QCD corrections. For example, for the total cross section
\begin{equation}
\sigma_{\rm NLO_{QCD}} \equiv \sigma_{\rm LO_1} + \check{\sigma}_{\rm NLO_1} \, ,
\end{equation}
where, consistent with the notation introduced in Section~\ref{sec:calcframe}, $\check{\sigma}_{\rm NLO_1}$ includes only terms of $\ord{\alpha_s^3 \alpha}$.
\item NLO: rows labeled in this way correspond to complete-NLO results, which include NLO QCD corrections ($\ord{\alpha_s^3 \alpha}$),  NLO EW  corrections ($\ord{\alpha_s^2 \alpha^2}$), and further
  subleading contributions. For example, for the case of the total cross section, one has 
  \begin{equation}
  \sigma_{\rm NLO} \equiv \sum_{i=1}^3 \sigma_{\rm LO_i} + \sum_{i=1}^4 \check{\sigma}_{\rm NLO_i} \, .
  \end{equation}
\item $\nNLOQCD$: indicates the approximate-NNLO predictions evaluated by adding to the $\NLOQCD$ results the contribution of the corrections of $\ord{\alpha_s^2}$ relative to the LO QCD cross section that are obtained from the NNLL resummation formula  for the QCD process.
\item nNLO: same as $\nNLOQCD$, but including in addition  the NLO EW corrections and further
  subleading contributions from the complete-NLO calculation.
\item $\NLOQCD$+NNLL: this label indicates the $\NLOQCD$ predictions improved by NNLL resummation.
\item NLO+NNLL: lines labeled this way include the complete-NLO predictions improved by NNLL resummation. These must be considered our most accurate predictions.
\end{itemize}

In the cases where beyond-NLO predictions are combined with NLO results,
{\it i.e.}, nNLO and NLO+NNLL calculations, the matching is carried out with the additive approach discussed in Section~\ref{sec:matching}. For total cross sections and charge asymmetries, which are the quantities considered in the  Tables \ref{ttwp}-\ref{ttz}, results based on the multiplicative approach
differ from the additive combination  by  less than 1\%.
For this reason, results obtained with the multiplicative approach are  not shown in the tables.
However, in the case of 
differential distributions discussed in Section~\ref{sec:diffdist}, the differences between additive approach and multiplicative approach are 
sometimes larger. Therefore, in that case, results obtained with the multiplicative approach are presented in a separate ratio inset in the figures. The labels employed to identify calculations in the multiplicative approach are the following:
\begin{itemize}
\item $\rm NLO \times NNLL$: this label indicates a calculations that includes the same corrections found in NLO+NNLL calculations,  but with $\NLOEW$ and purely QCD corrections combined in the multiplicative approach, according to Eq.~(\ref{eq:matchingmult}).
\item $\rm nNLO_{mult}$: this label indicates calculations analogous to nNLO but with $\NLOEW$ and  QCD corrections combined with the multiplicative approach, according to Eq.~(\ref{eq:matchingnNLOmult}).
\end{itemize}

\begin{table}[h]
\renewcommand{\arraystretch}{1.7}
\scriptsize
\begin{center}
\begin{tabular}{r r @{~} l @{} l r @{~} l @{} l}
\bottomrule
\multicolumn{7}{l}{$m(t \bar{t} W^+)$-based scales} \\
Order & \multicolumn{3}{c}{$\sigma$ [fb]} & \multicolumn{3}{c}{$A_C [\%]$}\\
\midrule
\LOQCDt & $225.45(1)$ & ${}_{ -39.41(-17.5 \%)}^{+ 51.61(+22.9 \%)}$ & ${}_{ -5.85(-2.6 \%)}^{+ 5.85(+2.6 \%)}$ & \multicolumn{3}{c}{$0$}\\
\NLOQCDt & $355.69(4)$ & ${}_{ -39.29(-11.0 \%)}^{+ 43.50(+12.2 \%)}$ & ${}_{ -8.12(-2.3 \%)}^{+ 8.12(+2.3 \%)}$ & $2.58(1)$ & ${}_{ -0.37(-14.3 \%)}^{+ 0.50(+19.4 \%)}$ & ${}_{ -0.08(-2.9 \%)}^{+ 0.08(+2.9 \%)}$ \\
NLO & $376.58(5)$ & ${}_{ -41.73(-11.1 \%)}^{+ 46.52(+12.4 \%)}$ & ${}_{ -8.02(-2.1 \%)}^{+ 8.02(+2.1 \%)}$ & $2.76(2)$ & ${}_{ -0.33(-12.0 \%)}^{+ 0.45(+16.1 \%)}$ & ${}_{ -0.09(-3.2 \%)}^{+ 0.09(+3.2 \%)}$ \\
\nNLOQCDt & $363.13(4)$ & ${}_{ -27.29(-7.5 \%)}^{+ 37.14(+10.2 \%)}$ & ${}_{ -8.3(-2.3 \%)}^{+ 8.3(+2.3 \%)}$ & $3.33(2)$ & ${}_{ -0.12(-3.6 \%)}^{+ 0.16(+4.7 \%)}$ & ${}_{
  -0.08(-2.4 \%)}^{+ 0.08(+2.4 \%)}$ \\
nNLO & $384.02(5)$ & ${}_{ -29.73(-7.7 \%)}^{+ 40.16(+10.5 \%)}$ & ${}_{ -8.20(-2.1 \%)}^{+ 8.20(+2.1 \%)}$ & $3.47(2)$ & ${}_{ -0.15(-4.3 \%)}^{+ 0.18(+5.1 \%)}$ & ${}_{ -0.09(-2.7 \%)}^{+ 0.09(+2.7 \%)}$ \\
\NLOQCDt+NNLL & $347.1(1)$ & ${}_{ -14.4(-4.2 \%)}^{+ 23.9(+6.9 \%)}$ & ${}_{ -7.9(-2.3
  \%)}^{+ 7.9(+2.3 \%)}$ & \multicolumn{3}{c}{--} \\
NLO+NNLL & $368.0(1)$ & ${}_{ -16.2(-4.4 \%)}^{+ 26.5(+7.2 \%)}$ & ${}_{ -7.8(-2.1 \%)}^{+ 7.8(+2.1 \%)}$  & \multicolumn{3}{c}{--}\\
\bottomrule
\multicolumn{7}{l}{$H_T$-based scales} \\
Order & \multicolumn{3}{c}{$\sigma$ [fb]} & \multicolumn{3}{c}{$A_C [\%]$}\\
\midrule
\LOQCDt & $241.146(9)$ & ${}_{ -43.182(-17.9 \%)}^{+ 57.030(+23.6 \%)}$ & ${}_{ -6.367(-2.6 \%)}^{+ 6.367(+2.6 \%)}$ & \multicolumn{3}{c}{$0$}\\
\NLOQCDt & $375.64(4)$ & ${}_{ -42.76(-11.4 \%)}^{+ 47.98(+12.8 \%)}$ & ${}_{ -8.43(-2.2 \%)}^{+ 8.4(+2.2 \%)}$ & $2.78(1)$ & ${}_{ -0.41(-14.9 \%)}^{+ 0.56(+20.3 \%)}$ & ${}_{ -0.08(-2.9 \%)}^{+ 0.08(+2.9 \%)}$  \\
NLO & $397.90(6)$ & ${}_{ -45.48(-11.4 \%)}^{+ 51.39(+12.9 \%)}$ & ${}_{ -8.32(-2.1 \%)}^{+ 8.3(+2.1 \%)}$ & $2.94(2)$ & ${}_{ -0.38(-13.0 \%)}^{+ 0.51(+17.7 \%)}$ & ${}_{ -0.10(-3.2 \%)}^{+ 0.10(+3.2 \%)}$ \\
\nNLOQCDt & $380.31(4)$ & ${}_{ -32.34(-8.5 \%)}^{+ 42.52(+11.2 \%)}$ & ${}_{ -8.55(-2.2 \%)}^{+ 8.55(+2.2 \%)}$ & $3.26(3)$ & ${}_{ -0.02(-0.7 \%)}^{+ 0.17(+5.3 \%)}$ & ${}_{ -0.09(-2.6 \%)}^{+ 0.09(+2.6 \%)}$ \\
nNLO & $402.57(6)$ & ${}_{ -35.06(-8.7 \%)}^{+ 45.94(+11.4 \%)}$ & ${}_{ -8.44(-2.1 \%)}^{+ 8.44(+2.1 \%)}$ & $3.39(3)$ & ${}_{ -0.06(-1.8 \%)}^{+ 0.19(+5.7 \%)}$ & ${}_{ -0.10(-2.9 \%)}^{+ 0.10(+2.9 \%)}$ \\
\NLOQCDt+NNLL & $378.1(1)$ & ${}_{ -21.7(-5.7 \%)}^{+ 32.4(+8.6 \%)}$ & ${}_{ -8.5(-2.2 \%)}^{+ 8.5(+2.2 \%)}$ & \multicolumn{3}{c}{--}\\
NLO+NNLL & $400.4(1)$ & ${}_{ -23.4(-5.9 \%)}^{+ 35.3(+8.8 \%)}$ & ${}_{
  -8.4(-2.1 \%)}^{+ 8.4(+2.1 \%)}$ &  \multicolumn{3}{c}{--}\\
\bottomrule
\multicolumn{7}{l}{Combined scales} \\
Order & \multicolumn{3}{c}{$\sigma$ [fb]} & \multicolumn{3}{c}{$A_C [\%]$}\\
\midrule
\LOQCDt & $233.297(8)$ & ${}_{ -47.26(-20.3 \%)}^{+ 64.88(+27.8 \%)}$ & ${}_{ -6.16(-2.6 \%)}^{+ 6.16(+2.6 \%)}$ &  \multicolumn{3}{c}{$0$}\\
\NLOQCDt & $365.66(3)$ & ${}_{ -49.27(-13.5 \%)}^{+ 57.95(+15.85 \%)}$ & ${}_{ -8.35(-2.3 \%)}^{+ 8.35(+2.3 \%)}$ & $2.68(1)$ & ${}_{ -0.47(-17.4 \%)}^{+ 0.66(+24.6 \%)}$ & ${}_{ -0.08(-2.9 \%)}^{+ 0.08(+2.9 \%)}$ \\
NLO & $387.24(4)$ & ${}_{ -52.39(-13.5 \%)}^{+ 62.05(+16.0 \%)}$ & ${}_{ -8.25(-2.1 \%)}^{+ 8.25(+2.1 \%)}$ & $2.85(1)$ & ${}_{ -0.42(-14.7 \%)}^{+ 0.60(+21.1 \%)}$ & ${}_{ -0.09(-3.2 \%)}^{+ 0.09(+3.2 \%)}$ \\
\nNLOQCDt & $371.72(3)$ & ${}_{ -35.88(-9.7 \%)}^{+ 51.11(+13.8 \%)}$ & ${}_{ -8.50(-2.3 \%)}^{+ 8.50(+2.3 \%)}$ & $3.30(2)$ & ${}_{ -0.08(-2.5 \%)}^{+ 0.19(+5.8 \%)}$ & ${}_{ -0.09(-2.6 \%)}^{+ 0.09(+2.6 \%)}$ \\
nNLO & $393.29(4)$ & ${}_{ -39.00(-9.9 \%)}^{+ 55.21(+14.0 \%)}$ & ${}_{ -8.40(-2.1 \%)}^{+ 8.40(+2.1 \%)}$ & $3.43(2)$ & ${}_{ -0.11(-3.3 \%)}^{+ 0.21(+6.2 \%)}$ & ${}_{ -0.10(-2.9 \%)}^{+ 0.10(+2.9 \%)}$ \\
\NLOQCDt+NNLL & $362.59(8)$ & ${}_{ -29.95(-8.3 \%)}^{+ 47.94(+13.2 \%)}$ & ${}_{ -8.26(-2.3 \%)}^{+ 8.26(+2.3 \%)}$
 &   \multicolumn{3}{c}{--}\\
NLO+NNLL & $384.17(9)$ & ${}_{ -32.36(-8.4 \%)}^{+ 51.52(+13.4 \%)}$ & ${}_{ -8.16(-2.1 \%)}^{+ 8.16(+2.1 \%)}$ &  \multicolumn{3}{c}{--} \\
\bottomrule
\end{tabular}
\end{center}
\caption{Cross section and charge asymmetry for $t \bar t W^+$ production for the
  13~TeV LHC at various accuracies. The top portion of the table corresponds to scales based on
  $m(t\bar{t}W^+)$, the middle portion on $H_T$. The lower part
  contains predictions based on the combination of the results for  the two scale choices. The first number in brackets corresponds
  to the statistical uncertainty in the  Monte Carlo integration. The first number in the subscript/superscript
  is the uncertainty due to  scale variations (the number in the bracket is the uncertainty expressed as a percentage of the central value). The last number in the subscript/superscript is  the PDF uncertainty.\label{ttwp}}
\end{table}

\begin{table}[!t]
\renewcommand{\arraystretch}{1.7}
\scriptsize
\begin{center}
\begin{tabular}{r r @{~} l @{} l r @{~} l @{} l}
\bottomrule
\multicolumn{7}{l}{$m(t \bar{t} W^-)$-based scales} \\
Order & \multicolumn{3}{c}{$\sigma$ [fb]} & \multicolumn{3}{c}{$A_C [\%]$}\\
\midrule
\LOQCDt & $114.305(6)$ & ${}_{ -20.056(-17.5 \%)}^{+ 26.261(+23.0 \%)}$ &
${}_{ -3.563(-3.1 \%)}^{+ 3.563(+3.1 \%)}$ &\multicolumn{3}{c}{$0$} \\
\NLOQCDt & $181.65(2)$ & ${}_{ -20.36(-11.2 \%)}^{+ 22.74(+12.5 \%)}$ & ${}_{ -5.20(-2.9 \%)}^{+ 5.20(+2.9 \%)}$ & $2.04(1)$ & ${}_{ -0.30(-14.7 \%)}^{+ 0.41(+19.9 \%)}$ & ${}_{ -0.05(-2.5 \%)}^{+ 0.05(+2.5 \%)}$ \\
NLO & $193.26(2)$ & ${}_{ -21.81(-11.3 \%)}^{+ 24.55(+12.7 \%)}$ & ${}_{ -5.29(-2.7 \%)}^{+ 5.29(+2.7 \%)}$ & $2.04(2)$ & ${}_{ -0.27(-13.2 \%)}^{+ 0.37(+18.1 \%)}$ & ${}_{ -0.05(-2.3 \%)}^{+ 0.05(+2.3 \%)}$ \\
\nNLOQCDt & $186.20(2)$ & ${}_{ -13.67(-7.34 \%)}^{+ 18.89(+10.14 \%)}$ & ${}_{ -5.33(-2.9 \%)}^{+ 5.33(+2.9 \%)}$ & $2.69(2)$ & ${}_{ -0.11(-4.0 \%)}^{+ 0.09(+3.5 \%)}$ & ${}_{ -0.06(-2.0 \%)}^{+ 0.06(+2.0 \%)}$ \\
nNLO & $197.80(3)$ & ${}_{ -15.11(-7.6 \%)}^{+ 20.68(+10.5 \%)}$ & ${}_{ -5.42(-2.7 \%)}^{+ 5.42(+2.7 \%)}$ & $2.64(2)$ & ${}_{ -0.12(-4.4 \%)}^{+ 0.10(+4.0 \%)}$ & ${}_{ -0.05(-1.8 \%)}^{+ 0.05(+1.8 \%)}$ \\
\NLOQCDt+NNLL & $178.16(4)$ & ${}_{ -7.13(-4.0 \%)}^{+ 12.29(+6.9 \%)}$ & ${}_{ -5.09(-2.9 \%)}^{+ 5.09(+2.9 \%)}$ & \multicolumn{3}{c}{--}\\
NLO+NNLL & $189.77(5)$ & ${}_{ -8.09(-4.3 \%)}^{+ 13.82(+7.3 \%)}$ & ${}_{ -5.19(-2.7 \%)}^{+ 5.19(+2.7 \%)}$ & \multicolumn{3}{c}{--}\\
\bottomrule
\multicolumn{7}{l}{$H_T$-based scales} \\
Order & \multicolumn{3}{c}{$\sigma$ [fb]} & \multicolumn{3}{c}{$A_C [\%]$}\\
\midrule
\LOQCDt & $121.754(5)$ & ${}_{ -21.843(-17.9 \%)}^{+ 28.824(+23.7 \%)}$ &
${}_{ -3.853(-3.2 \%)}^{+ 3.854(+3.2 \%)}$ & \multicolumn{3}{c}{$0$}\\
\NLOQCDt & $191.40(2)$ & ${}_{ -22.13(-11.6 \%)}^{+ 25.09(+13.1 \%)}$ & ${}_{ -5.44(-2.8 \%)}^{+ 5.44(+2.8 \%)}$ & $2.19(1)$ & ${}_{ -0.33(-15.1 \%)}^{+ 0.45(+20.5 \%)}$ & ${}_{ -0.06(-2.5 \%)}^{+ 0.06(+2.5 \%)}$  \\
NLO & $203.81(3)$ & ${}_{ -23.74(-11.6 \%)}^{+ 27.13(+13.3 \%)}$ & ${}_{ -5.53(-2.7 \%)}^{+ 5.53(+2.7 \%)}$ & $2.16(2)$ & ${}_{ -0.30(-13.7 \%)}^{+ 0.41(+18.9 \%)}$ & ${}_{ -0.05(-2.4 \%)}^{+ 0.05(+2.4 \%)}$ \\
\nNLOQCDt & $194.43(2)$ & ${}_{ -16.14(-8.3 \%)}^{+ 21.73(+11.2 \%)}$ & ${}_{ -5.52(-2.8 \%)}^{+ 5.52(+2.8 \%)}$ & $2.60(2)$ & ${}_{ -0.01(-0.4 \%)}^{+ 0.14(+5.5 \%)}$ & ${}_{ -0.06(-2.3 \%)}^{+ 0.06(+2.3 \%)}$  \\
nNLO & $206.83(3)$ & ${}_{ -17.76(-8.6 \%)}^{+ 23.78(+11.5 \%)}$ & ${}_{ -5.62(-2.7 \%)}^{+ 5.62(+2.7 \%)}$  & $2.54(2)$ & ${}_{ -0.03(-1.0 \%)}^{+ 0.15(+6.0 \%)}$ & ${}_{ -0.05(-2.2 \%)}^{+ 0.05(+2.2 \%)}$ \\
\NLOQCDt+NNLL & $193.33(5)$ & ${}_{ -10.65(-5.5 \%)}^{+ 16.66(+8.6 \%)}$ & ${}_{ -5.49(-2.8 \%)}^{+ 5.49(+2.8 \%)}$ & \multicolumn{3}{c}{--}\\
NLO+NNLL & $205.73(5)$ & ${}_{ -11.68(-5.7 \%)}^{+ 18.43(+9.0 \%)}$ & ${}_{ -5.59(-2.7 \%)}^{+ 5.59(+2.7 \%)}$  & \multicolumn{3}{c}{--}\\
\bottomrule
\multicolumn{7}{l}{Combined scales} \\
Order & \multicolumn{3}{c}{$\sigma$ [fb]} & \multicolumn{3}{c}{$A_C [\%]$}\\
\midrule
\LOQCDt & $118.030(4)$ & ${}_{ -23.781(-20.1 \%)}^{+ 32.548(+27.6 \%)}$ & ${}_{ -3.735(-3.2 \%)}^{+ 3.736(+3.2 \%)}$ & \multicolumn{3}{c}{$0$}\\
\NLOQCDt & $186.53(1)$ & ${}_{ -25.24(-13.5 \%)}^{+ 29.96(+16.1 \%)}$ & ${}_{ -5.34(-2.9 \%)}^{+ 5.34(+2.9 \%)}$ & $2.12(1)$ & ${}_{ -0.38(-17.7 \%)}^{+ 0.52(+24.8 \%)}$ & ${}_{ -0.05(-2.5 \%)}^{+ 0.05(+2.5 \%)}$ \\
NLO & $198.53(2)$ & ${}_{ -27.08(-13.6 \%)}^{+ 32.41(+16.3 \%)}$ & ${}_{ -5.43(-2.7 \%)}^{+ 5.43(+2.7 \%)}$ & $2.10(1)$ & ${}_{ -0.33(-15.63 \%)}^{+ 0.47(+22.3 \%)}$ & ${}_{ -0.05(-2.4 \%)}^{+ 0.05(+2.4 \%)}$ \\
\nNLOQCDt & $190.31(1)$ & ${}_{ -17.78(-9.3 \%)}^{+ 25.85(+13.6 \%)}$ & ${}_{ -5.45(-2.9 \%)}^{+ 5.45(+2.9 \%)}$ & $2.64(1)$ & ${}_{ -0.06(-2.4 \%)}^{+ 0.14(+5.2 \%)}$ & ${}_{ -0.06(-2.3 \%)}^{+ 0.06+2.3 \%)}$ \\
nNLO & $202.32(2)$ & ${}_{ -19.63(-9.7 \%)}^{+ 28.29(+14.0 \%)}$ & ${}_{ -5.54(-2.7 \%)}^{+ 5.54(+2.7 \%)}$ & $2.59(1)$ & ${}_{ -0.08(-3.0 \%)}^{+ 0.16(+6.0 \%)}$ & ${}_{ -0.06(-2.2 \%)}^{+ 0.06(+2.2 \%)}$ \\
\NLOQCDt+NNLL & $185.75(3)$ & ${}_{ -14.71(-7.9 \%)}^{+ 24.25(+13.1 \%)}$ & ${}_{ -5.31(-2.9 \%)}^{+ 5.31(+2.9 \%)}$ &  \multicolumn{3}{c}{--}\\
NLO+NNLL & $197.75(4)$ & ${}_{ -16.07(-8.1 \%)}^{+ 26.41(+13.4 \%)}$ & ${}_{ -5.41(-2.7 \%)}^{+ 5.41(+2.7 \%)}$ &  \multicolumn{3}{c}{--}\\
\bottomrule
\end{tabular}
\end{center}
\caption{Total cross section and charge asymmetry for $\ttwm$ production. Same structure as in Table~\ref{ttwp}.\label{ttwm}}
\end{table}

\boldmath
\subsubsection{$\ttwp$ and $\ttwm$}
\unboldmath

Values for the total cross sections for the $\ttwp$ and $\ttwm$
processes are shown in Tables~\ref{ttwp} and~\ref{ttwm}, respectively. The results for
$\ttwm$ are quantitatively rather similar to the results for $\ttwp$; consequently,
 we comment almost exclusively on the latter.

The central value of the $\LOQCD$ cross section for $\ttwp$ production is about
$225$~fb or $241$~fb for the $\MttW$-based or $H_T$-based scale choices,
respectively. This difference is well captured by the uncertainty due to scale variation, which is roughly ${}_{-18\%}^{+23\%}$ in both cases. Since
the central values for the two scale choices differ by about $7\%$, when
combining the two scale choices  the cross section  gets a value of 
$233$~fb, with a slightly increased scale dependence of ${}_{-20\%}^{+27\%}$. As discussed in Section~\ref{sec:phenointro}, the large NLO QCD corrections to this process are  due
to the opening of the $qg$-induced real-emission channel. This contribution is
particularly sizable since the gluon luminosity is rather large at the
LHC.  Indeed, the $\NLOQCD$
predictions are more than 50\% larger than the $\LOQCD$ cross section. The corrections
are only slightly larger for the $H_T$-based scale choice than for the $\MttW$-based choice. This fact brings the two $\NLOQCD$ cross section results closer to each other (in relative
terms) as compared to the $\LOQCD$ cross sections, as expected from perturbation theory. At
$\NLOQCD$ the relative uncertainty from scale variation is
${}_{-13.5\%}^{+15.9\%}$ for the combined-scales result, which is
significantly smaller than the  corresponding $\LOQCD$ uncertainty. The inclusion of the EW corrections further increases the cross section  by about
$6\%$, mainly due to $\NLO_3$ corrections. Even though
this contribution is suppressed by a factor $(\alpha/\alpha_s)^{2}$ w.r.t.~the
$\NLO_1$ corrections,  there is a large enhancement due to the opening of the
$t$-channel-enhanced $t W \to t W$ scattering contribution ~\cite{Dror:2015nkp, Frederix:2017wme}.

 For both the $\MttW$ and $H_T$-based
scale choices, approximate NNLO   corrections increase the cross section.  These corrections for the $\MttW$-based scale choice are
slightly larger than for the $H_T$-based scale choice. Therefore, the central values
of the nNLO${}_\textrm{QCD}$ cross sections for the two scale choices are closer to each other than the values of the 
NLO${}_{\textrm{QCD}}$ cross sections. 
By including the approximate NNLO QCD corrections  the
cross-section scale dependence is reduced  to  $\sim {}_{-10\%}^{+14\%}$
for the combined-scale calculation. Hence, perturbation theory seems to converge well. On the other hand, if one considers 
 NNLL resummed results, either matched to $\NLOQCD$ or to complete-NLO
predictions,  a slightly different picture emerges. For both scale choices, the
corrections due to resummation are small and well-behaved. Moreover, they 
reduce significantly the scale dependence, to about ${}_{-4\%}^{+7\%}$ and ${}_{-6\%}^{+8\%}$
for the total cross section obtained with the $\MttW$ and $H_T$-based scales,
respectively. However, the corrections move the cross-section central values further apart from each other
than in the case of NLO${}_\textrm{QCD}$ calculations.  The
corrections for the $\MttW$-based scales are negative, while for the
$H_T$-based scales the corrections are positive. The net effect is that the central value
for the combined-scale result is very much compatible with the corresponding
NLO${}_\textrm{QCD}$ predictions, with a scale uncertainty equal to
${}_{-8\%}^{+13\%}$, which is  larger than the
NLO${}_\textrm{QCD}$+NNLL scale uncertainties for the two separate scale
choices. 
The reason behind this feature resides in  the origin of the NLO QCD corrections. The $\NLO_1$ term is  dominated by hard radiation and especially by the $qg$ initial-state contribution. 
Therefore, in this case,  the reduction of the scale dependence due to the resummation of soft emission does not reflect the theoretical uncertainty associated to missing higher-order corrections. However, by combining the results obtained with the two different scale choices, one obtains a more reliable estimate of the uncertainty due to missing higher order corrections. Nevertheless, by following this approach an improvement in the scale uncertainty in the combined-scales results is observed; the scale uncertainty
affecting  the combined-scales NLO${}_\textrm{QCD}$ cross section is larger than the scale uncertainty for the NLO${}_\textrm{QCD}$+NNLL  cross section. Since scale uncertainties primarily affect QCD corrections, this argument still holds when EW effects are included, as can be seen by comparing the scale uncertainties of $\NLO$, $\rm nNLO$, and  $\rm NLO+NNLL$ calculations of the cross section. 
 
The PDF uncertainties on the total $\ttw$ cross section are significantly smaller than the corresponding scale uncertainties,
and are at the level of $\pm 2\%$ for $\ttwp$ production and $\pm 3\%$ for $\ttwm$ production.

\medskip

The 
third column in Tables.~\ref{ttwp} and~\ref{ttwm} shows predictions for the charge asymmetry
$A_C$.  At the LHC, the charge asymmetry  is defined as 
\begin{equation}
  A_C=\frac{\sigma(\Delta>0)-\sigma(\Delta<0)}{\sigma(\Delta>0)+\sigma(\Delta<0)}\, , \label{eq:asym}
\end{equation}
 where $\Delta \equiv |y(t)|-|y(\bar{t})|$, and $y(t)$ ($y(\bar{t})$) indicates the top-quark (antiquark) rapidity in the laboratory frame. Consequently, $\Delta$ is positive when the top is emitted less centrally than the antitop. 
An analogous asymmetry was measured at the Tevatron for top-pair
production\footnote{At $p \bar p$ colliders such as the Tevatron  the relevant observable was the  forward-backward asymmetry, defined as in Eq.~(\ref{eq:asym}) but with  $\Delta \equiv y(t)-y(\bar{t})$.}. The Tevatron asymmetry received  considerable attention due  to a tension between the measured asymmetry and the SM predictions~\cite{Aaltonen:2011kc}, which were initially 
known only at the lowest non-vanishing order (NLO QCD)  \cite{Kuhn:1998kw}. The tension could be interpreted as a BSM effect. With improved measurements and especially with the calculation of NLO EW \cite{Hollik:2011ps} and then NNLO QCD corrections \cite{Czakon:2014xsa, Czakon:2017lgo}, the tension between  theory
  predictions and experimental data decreased considerably.  It is therefore essential to have precise predictions for such observables. 
  
  At the LHC, the $A_C$ asymmetry for top-pair production is rather small (see {\it e.g.}~\cite{Czakon:2017lgo}) due to the fact that the cross section is dominated by the gluon-fusion channel, which is charge symmetric. It is therefore interesting to consider top-pair production in association with a $W$ boson~\cite{Maltoni:2014zpa, Maltoni:2015ena}. Since at lowest order the $W$ boson only couples to initial state
quarks, the contribution of the $gg$ channel to the asymmetry is suppressed  and only enters at NNLO and beyond. As a result of this situation,  the asymmetry in $\ttw$ production is 
significantly larger than in the case of top-pair production. For $\ttwp$ production at $\NLOQCD$, which is the lowest  perturbative order for which the
asymmetry is non-zero, one finds that $A_C$ is equal to about $2.6\%$ and $2.8\%$
for the $\MttW$ and $H_T$-based scale choices, respectively, with scale uncertainties of  ${}_{-15\%}^{+20\%}$~\footnote{Both for scale and PDF uncertainties a full correlation is assumed for the numerator and the denominator of Eq.~\eqref{eq:asym}. In the case of combined-scale predictions we proceed similarly to the case of total cross sections, by looking at the envelope of the $\MttW$ and $H_T$-based scale choices directly for $A_C$.}. The inclusion of 
the EW corrections increases the asymmetry by a
small amount, about $0.16$ percent.

In our framework, it is not possible to evaluate the charge asymmetry to $\rm NLO+NNLL$ or $\rm NLO_{QCD}+NNLL$
accuracy, since the resummation is carried out inclusively w.r.t.~the
rapidities.
Hence, the nNLO calculations are the most accurate predictions for the charge asymmetry that we present in this paper. On the other hand, the NNLO QCD corrections to $\ttwp$ (and $\ttwm$) production involve $gg \to t \bar t W q \bar q'$ processes, which are expected to be large due to the $gg$ luminosity and are completely symmetric, so that they contribute only to the denominator of Eq.~\eqref{eq:asym}. These effects cannot be estimated via scale variations and may substantially alter the $A_C$ prediction for $\ttwp$ (and $\ttwm$).
However, when the approximate NNLO corrections are included, the $\ttwp$ charge asymmetry increases by $0.6$ and $0.8$ percent for the $\MttW$ and $H_T$-based scale choices, bringing the central values for the two scale choices rather close to each other. More
significantly, by including these terms, which constitute the first order
correction to the asymmetry, the scale dependence  is reduced to
${}_{-3\%}^{+6\%}$ for the combined-scales prediction. In this case, the uncertainties
coming from the PDFs can no longer be neglected, since they are  similar in size the scale uncertainties ($\pm 2.9\%$ for $\ttwp$
production). 
Similar remarks apply to the charge asymmetry calculation in $\ttwm$ production.

\begin{table}[!t]
\renewcommand{\arraystretch}{1.7}
\scriptsize
\begin{center}
\begin{tabular}{r r @{~} l @{} l r @{~} l @{} l}
\bottomrule
\multicolumn{7}{l}{$m(t \bar{t} H)$-based scales} \\
Order & \multicolumn{3}{c}{$\sigma$ [fb]} & \multicolumn{3}{c}{$A_C [\%]$}\\
\midrule
\LOQCDt & $327.65(4)$ & ${}_{ -68.46(-20.9 \%)}^{+ 94.18(+28.7 \%)}$ & ${}_{ -7.11(-2.2
  \%)}^{+ 7.11(+2.2 \%)}$ & \multicolumn{3}{c}{0}\\
\NLOQCDt & $463.70(8)$ & ${}_{ -49.72(-10.7 \%)}^{+ 45.1(+9.7 \%)}$ & ${}_{ -11.08(-2.4 \%)}^{+ 11.08(+2.4 \%)}$ & $0.84(2)$ & ${}_{ -0.13(-15.8 \%)}^{+ 0.19(+22.2 \%)}$ & ${}_{ -0.04(-4.2 \%)}^{+ 0.04(+4.2 \%)}$ \\
NLO & $475.68(8)$ & ${}_{ -51.11(-10.7 \%)}^{+ 46.94(+9.9 \%)}$ & ${}_{ -11.21(-2.4 \%)}^{+ 11.21(+2.4 \%)}$ & $1.01(2)$ & ${}_{ -0.14(-13.6 \%)}^{+ 0.19(+19.0 \%)}$ & ${}_{ -0.04(-4.0 \%)}^{+ 0.04(+4.0 \%)}$ \\
\nNLOQCDt & $490.38(8)$ & ${}_{ -9.61(-2.0 \%)}^{+ 18.46(+3.8 \%)}$ & ${}_{ -11.82(-2.4 \%)}^{+ 11.82(+2.4 \%)}$ & $0.79(5)$ & ${}_{ - 0.00( - 0.0 \%)}^{+ 0.30(+38.5 \%)}$ & ${}_{ -0.04(-5.1 \%)}^{+ 0.04(+5.1 \%)}$ \\
nNLO & $502.36(8)$ & ${}_{ -10.99(-2.2 \%)}^{+ 20.27(+4.0 \%)}$ & ${}_{ -11.95(-2.4 \%)}^{+ 11.95(+2.4 \%)}$ & $0.95(5)$ & ${}_{ - 0.00(- 0.0 \%)}^{+ 0.28(+29.5 \%)}$ & ${}_{ -0.05(-4.7 \%)}^{+ 0.05(+4.7 \%)}$ \\
\NLOQCDt+NNLL & $479.1(1)$ & ${}_{ -24.2(-5.0 \%)}^{+ 29.0(+6.1 \%)}$ & ${}_{ -11.5(-2.4
  \%)}^{+ 11.5(+2.4 \%)}$ & \multicolumn{3}{c}{--}\\
NLO+NNLL & $491.1(1)$ & ${}_{ -24.0(-4.9 \%)}^{+ 27.8(+5.7 \%)}$ & ${}_{ -11.6(-2.4 \%)}^{+
  11.6(+2.4 \%)}$ & \multicolumn{3}{c}{--}\\
\bottomrule
\multicolumn{7}{l}{$H_T$-based scales} \\
Order & \multicolumn{3}{c}{$\sigma$ [fb]} & \multicolumn{3}{c}{$A_C [\%]$}\\
\midrule
\LOQCDt & $344.86(4)$ & ${}_{ -73.22(-21.2 \%)}^{+ 101.38(+29.4 \%)}$ & ${}_{ -7.61(-2.2
  \%)}^{+ 7.61(+2.2 \%)}$ & \multicolumn{3}{c}{0}\\
\NLOQCDt & $472.22(7)$ & ${}_{ -48.83(-10.3 \%)}^{+ 41.31(+8.7 \%)}$ & ${}_{ -11.41(-2.4 \%)}^{+ 11.41(+2.4 \%)}$ & $0.92(2)$ & ${}_{ -0.16(-17.1 \%)}^{+ 0.22(+23.9 \%)}$ & ${}_{ -0.04(-4.2 \%)}^{+ 0.04(+4.2 \%)}$ \\
NLO & $484.31(7)$ & ${}_{ -50.24(-10.4 \%)}^{+ 43.15(+8.9 \%)}$ & ${}_{ -11.55(-2.4 \%)}^{+ 11.55(+2.4 \%)}$ & $1.09(2)$ & ${}_{ -0.16(-14.7 \%)}^{+ 0.23(+20.9 \%)}$ & ${}_{ -0.04(-4.0 \%)}^{+ 0.04(+4.0 \%)}$  \\
\nNLOQCDt & $490.17(8)$ & ${}_{ -8.95(-1.8 \%)}^{+ 15.35(+3.1 \%)}$ & ${}_{ -11.92(-2.4 \%)}^{+ 11.92(+2.4 \%)}$ & $0.94(5)$ & ${}_{ -0.09(-9.4 \%)}^{+ 0.003(+0.3 \%)}$ & ${}_{ -0.04(-4.6 \%)}^{+ 0.04(+4.6 \%)}$ \\
nNLO & $502.26(7)$ & ${}_{ -10.37(-2.1 \%)}^{+ 17.19(+3.4 \%)}$ & ${}_{ -12.06(-2.4 \%)}^{+ 12.06(+2.4 \%)}$ & $1.11(5)$ & ${}_{ -0.11(-9.6 \%)}^{+ 0.03(+2.5 \%)}$ & ${}_{ -0.05(-4.3 \%)}^{+ 0.05(+4.3 \%)}$ \\
\NLOQCDt+NNLL & $489.58(9)$ & ${}_{ -22.54(-4.6 \%)}^{+ 34.35(+7.0 \%)}$ & ${}_{
  -11.91(-2.4 \%)}^{+ 11.91(+2.4 \%)}$ & \multicolumn{3}{c}{--}\\
NLO+NNLL & $501.67(9)$ & ${}_{ -22.54(-4.5 \%)}^{+ 33.34(+6.6 \%)}$ & ${}_{ -12.05(-2.4
  \%)}^{+ 12.05(+2.4 \%)}$ & \multicolumn{3}{c}{--}\\
\bottomrule
\multicolumn{7}{l}{Combined scales} \\
Order & \multicolumn{3}{c}{$\sigma$ [fb]} & \multicolumn{3}{c}{$A_C [\%]$}\\
\midrule
\LOQCDt & $336.25(3)$ & ${}_{ -77.07(-22.9 \%)}^{+ 109.98(+32.7 \%)}$ & ${}_{ -7.42(-2.2 \%)}^{+ 7.42(+2.2 \%)}$ & \multicolumn{3}{c}{0}\\
\NLOQCDt & $467.96(5)$ & ${}_{ -53.98(-11.5 \%)}^{+ 45.57(+9.7 \%)}$ & ${}_{ -11.31(-2.4 \%)}^{+ 11.31(+2.4 \%)}$ & $0.88(1)$ & ${}_{ -0.17(-19.2 \%)}^{+ 0.25(+28.9 \%)}$ & ${}_{ -0.04(-4.2 \%)}^{+ 0.04(+4.2 \%)}$ \\
NLO & $479.99(5)$ & ${}_{ -55.42(-11.5 \%)}^{+ 47.46(+9.9 \%)}$ & ${}_{ -11.45(-2.4 \%)}^{+ 11.45(+2.4 \%)}$ & $1.05(1)$ & ${}_{ -0.18(-16.8 \%)}^{+ 0.27(+25.5 \%)}$ & ${}_{ -0.04(-4.0 \%)}^{+ 0.04(+4.0 \%)}$ \\
\nNLOQCDt & $490.27(6)$ & ${}_{ -9.50(-1.9 \%)}^{+ 18.56(+3.8 \%)}$ & ${}_{ -11.93(-2.4 \%)}^{+ 11.93(+2.4 \%)}$ & $0.87(4)$ & ${}_{ -0.01(-1.5 \%)}^{+ 0.23(+26.4 \%)}$ & ${}_{ -0.04(-5.1 \%)}^{+ 0.04(+5.1 \%)}$ \\
nNLO & $502.31(6)$ & ${}_{ -10.95(-2.2 \%)}^{+ 20.32(+4.0 \%)}$ & ${}_{ -12.06(-2.4 \%)}^{+ 12.06(+2.4 \%)}$ & $1.03(4)$ & ${}_{ -0.03(-2.6 \%)}^{+ 0.20(+19.5 \%)}$ & ${}_{ -0.05(-4.7 \%)}^{+ 0.05(+4.7 \%)}$ \\
\NLOQCDt+NNLL & $484.33(7)$ & ${}_{ -29.43(-6.1 \%)}^{+ 39.60(+8.2 \%)}$ & ${}_{ -11.78(-2.4 \%)}^{+ 11.78(+2.4 \%)}$ & \multicolumn{3}{c}{--}\\
NLO+NNLL & $496.36(7)$ & ${}_{ -29.35(-5.9 \%)}^{+ 38.64(+7.8 \%)}$ & ${}_{ -11.92(-2.4 \%)}^{+ 11.92(+2.4 \%)}$ & \multicolumn{3}{c}{--}\\
\bottomrule
\end{tabular}
\end{center}
\caption{Total cross section and charge asymmetry for $\tth$ production. Same structure as in Table~\ref{ttwp}. \label{tth}}  
\end{table}

\boldmath
\subsubsection{$\tth$}
\unboldmath

The total cross section for $\tth$ production is shown in the second column of  Table~\ref{tth}.   
When NLO QCD corrections are included, the cross-section central values obtained with the two scale choices differ by $8.5~$fb, roughly half of the difference between the central values of the cross section calculated with the two scale choices at $\LOQCD$.
For both scale choices,  EW corrections increase the cross section
by $2.5\%$ w.r.t.~the $\NLOQCD$ result; this is a small correction when compared to the scale uncertainty. Indeed, 
although the scale uncertainty at $\NLOQCD$ is more than a factor two smaller than at $\LOQCD$, it remains of the order of $\pm 10\%$.

When QCD corrections beyond NLO are included, the agreement between the  predictions obtained with the two scale choices is very good:
for $\rm nNLO_{QCD}$ and $\rm nNLO$ calculations the difference between the two values of the cross section is below the permille. 
The scale uncertainties, which are of the order of ${}_{-2\%}^{+4\%}$ for $\rm nNLO$ calculations, are significantly reduced w.r.t.~NLO calculations. Compared to these small uncertainties, 
EW corrections can no longer be neglected. 

In contrast to the $\ttw$ processes, the 
NLO+NNLL cross sections come with a larger scale uncertainty
than the nNLO ones and the central values with
$\MttH$ and $H_T$-based scale choices are further apart than at nNLO.
 By combining results for the two scale choices one obtains a total cross section of approximately $500~$fb with a scale uncertainty just below the ${}_{-6\%}^{+8\%}$ level.
The scale uncertainty in the resummed calculations is obtained by separately varying three different non-physical scales, while the scale uncertainty associated to approximate NNLO results is obtained by varying only one scale. For this and other reasons, as discussed in Refs.~\cite{Broggio:2016zgg, Broggio:2016lfj,Broggio:2017kzi}, we consider  NLO+NNLL predictions to be more complete and reliable than the approximate NNLO predictions. Hence, the cross section in the last line of Table~\ref{tth} should be considered the most accurate prediction for the $\tth$ total cross section presented in this paper. 

\medskip

The charge asymmetry for the top and the antitop quarks in $\tth$ production is given in the third column of Table~\ref{tth}. As expected, the asymmetry for $\tth$ production is smaller than for $\ttw$  production, since
the latter does not contain (up to NLO) the large and symmetric $gg$-induced contributions, which enter only in the denominator of Eq.~\eqref{eq:asym}. The difference between the central values of the asymmetry calculated with the $\MttH$ and $H_T$-based scale choices is small. By comparing NLO and $\NLOQCD$ predictions it is possible to see  that the contribution of the EW corrections to the asymmetry is sizable. 
Going beyond NLO, the $\rm nNLO_{QCD}$ and $\rm nNLO$ 
predictions show very asymmetric uncertainty bands. For the 
$\MttH$-based scale choice, the central value of the asymmetry lies at the lower edge of the uncertainty band and the overall size of the band does not decrease significantly compared to
NLO. For the $H_T$-based scale choice, the band in the nNLO and $\nNLOQCD$ calculations does decrease in size w.r.t.~the NLO calculation, but its central value lies near the upper edge of the uncertainty band. The combination of the $\MttH$ and $H_T$-based calculations leads to a nNLO result that has a central value close to the NLO calculation, but with a significantly smaller scale dependence.  However, overall, the charge asymmetry is rather small for $\tth$ production and can be challenging to measure.

The PDF uncertainties are small for $\tth$ production, of the
order of $\pm 2.4\%$ for the total cross sections and slightly larger for the charge asymmetry.

\begin{table}[!t]
\renewcommand{\arraystretch}{1.7}
\scriptsize
\begin{center}
\begin{tabular}{r r @{~} l @{} l r @{~} l @{} l}
\bottomrule
\multicolumn{7}{l}{$m(t \bar{t} Z)$-based scales} \\
Order & \multicolumn{3}{c}{$\sigma$ [fb]} & \multicolumn{3}{c}{$A_C [\%]$}\\
\midrule
\LOQCDt & $463.90(4)$ & ${}_{ -96.96(-20.9 \%)}^{+ 133.53(+28.8 \%)}$ & ${}_{ -10.30(-2.2 \%)}^{+ 10.30(+2.2 \%)}$ & $-0.10(1)$&${}_{ -0.004(+4.4 \%)}^{+ 0.005(-4.7 \%)}$&${}_{ -0.02(16.3 \%)}^{+ 0.02(-16.3 \%)}$\\
\NLOQCDt & $732.9(1)$ & ${}_{ -90.1(-12.3 \%)}^{+ 92.7(+12.6 \%)}$ & ${}_{ -17.0(-2.3 \%)}^{+ 17.0(+2.3 \%)}$ & $0.76(2)$ & ${}_{ -0.12(-16.1 \%)}^{+ 0.16(+21.6 \%)}$ & ${}_{ -0.05(-6.3 \%)}^{+ 0.05(+6.3 \%)}$ \\
NLO & $741.5(1)$ & ${}_{ -89.9(-12.1 \%)}^{+ 92.3(+12.4 \%)}$ & ${}_{ -17.2(-2.3 \%)}^{+ 17.2(+2.3 \%)}$ & $0.85(2)$ & ${}_{ -0.12(-13.9 \%)}^{+ 0.16(+18.80 \%)}$ & ${}_{ -0.05(-5.3 \%)}^{+ 0.05(+5.3 \%)}$ \\
\nNLOQCDt & $811.9(1)$ & ${}_{ -24.7(-3.0 \%)}^{+ 36.7(+4.5 \%)}$ & ${}_{ -18.9(-2.3 \%)}^{+ 18.9(+2.3 \%)}$ & $0.91(6)$ & ${}_{ -0.03(-2.9 \%)}^{+ 0.06(+6.8 \%)}$ & ${}_{ -0.05(-5.9 \%)}^{+ 0.05(+5.9 \%)}$ \\
nNLO & $820.5(1)$ & ${}_{ -24.4(-3.0 \%)}^{+ 36.4(+4.4 \%)}$ & ${}_{ -19.1(-2.3 \%)}^{+ 19.1(+2.3 \%)}$ & $0.99(6)$ & ${}_{ -0.02(-2.3 \%)}^{+ 0.06(+5.8 \%)}$ & ${}_{ -0.05(-5.2 \%)}^{+ 0.05(+5.2 \%)}$ \\
\NLOQCDt+NNLL & $790.7(2)$ & ${}_{ -66.2(-8.4 \%)}^{+ 61.5(+7.8 \%)}$ & ${}_{ -18.4(-2.3 \%)}^{+ 18.4(+2.3 \%)}$ & \multicolumn{3}{c}{--}\\
NLO+NNLL & $799.3(2)$ & ${}_{ -66.3(-8.3 \%)}^{+ 61.7(+7.7 \%)}$ & ${}_{ -18.6(-2.3 \%)}^{+ 18.6(+2.3 \%)}$ & \multicolumn{3}{c}{--}\\
\bottomrule
\multicolumn{7}{l}{$H_T$-based scales} \\
Order & \multicolumn{3}{c}{$\sigma$ [fb]} & \multicolumn{3}{c}{$A_C [\%]$}\\
\midrule
\LOQCDt & $504.63(8)$ & ${}_{ -108.36(-21.5 \%)}^{+ 150.89(+29.9 \%)}$ & ${}_{ -11.52(-2.3 \%)}^{+ 11.2(+2.3 \%)}$  &$-0.09(2)$&${}_{ -0.005(+5.8 \%)}^{+ 0.005(-6.2 \%)}$&${}_{ -0.02(+17.7 \%)}^{+ 0.02(-17.7 \%)}$ \\
\NLOQCDt & $769.5(3)$ & ${}_{ -93.6(-12.2 \%)}^{+ 92.7(+12.1 \%)}$ & ${}_{ -18.2(-2.4 \%)}^{+ 18.2(+2.4 \%)}$ & $0.82(4)$ & ${}_{ -0.13(-16.6 \%)}^{+ 0.20(+24.5 \%)}$ & ${}_{ -0.05(-5.9 \%)}^{+ 0.05(+5.9 \%)}$ \\
NLO & $777.4(3)$ & ${}_{ -93.2(-12.0 \%)}^{+ 92.1(+11.8 \%)}$ & ${}_{ -18.3(-2.4 \%)}^{+ 18.3(+2.4 \%)}$ & $0.90(4)$ & ${}_{ -0.13(-14.1 \%)}^{+ 0.19(+21.7 \%)}$ & ${}_{ -0.05(-5.1 \%)}^{+ 0.05(+5.1 \%)}$ \\
\nNLOQCDt & $822.3(3)$ & ${}_{ -25.2(-3.1 \%)}^{+ 37.1(+4.5 \%)}$ & ${}_{ -19.5(-2.4 \%)}^{+ 19.5(+2.4 \%)}$ & $1.00(5)$ & ${}_{ -0.05(-4.7 \%)}^{+0.00(+0.0 \%)}$ & ${}_{ -0.05(-5.3 \%)}^{+ 0.05(+5.3 \%)}$ \\
nNLO & $830.2(3)$ & ${}_{ -24.7(-3.0 \%)}^{+ 36.5(+4.4 \%)}$ & ${}_{ -19.6(-2.4 \%)}^{+ 19.6(+2.4 \%)}$ & $1.08(5)$ & ${}_{ -0.05(-4.5 \%)}^{+0.00(+0.0 \%)}$ & ${}_{ -0.05(-4.7 \%)}^{+ 0.05(+4.7 \%)}$ \\
\NLOQCDt+NNLL & $814.5(3)$ & ${}_{ -51.8(-6.4 \%)}^{+ 77.4(+9.5 \%)}$ & ${}_{ -19.3(-2.4 \%)}^{+ 19.3(+2.4 \%)}$ & \multicolumn{3}{c}{--}\\
NLO+NNLL & $822.5(3)$ & ${}_{ -51.9(-6.3 \%)}^{+ 77.7(+9.4 \%)}$ & ${}_{
  -19.4(-2.4 \%)}^{+ 19.4(+2.4 \%)}$ & \multicolumn{3}{c}{--} \\
\bottomrule
\multicolumn{7}{l}{Combined scales} \\
Order & \multicolumn{3}{c}{$\sigma$ [fb]} & \multicolumn{3}{c}{$A_C [\%]$}\\
\midrule
\LOQCDt & $484.26(4)$ & ${}_{ -117.32(-24.2 \%)}^{+ 171.26(+35.4 \%)}$ & ${}_{ -11.05(-2.3 \%)}^{+ 11.05(+2.3 \%)}$ &
$-0.09(1)$&${}_{-0.009( +9.9 \%)}^{+0.01(-11.1 \%)}$&${}_{ -0.02( +16.3 \%)}^{+ 0.02(-16.3 \%)}$\\
\NLOQCDt & $751.2(1)$ & ${}_{ -108.5(-14.4 \%)}^{+ 111.1(+14.8 \%)}$ & ${}_{ -17.7(-2.4 \%)}^{+ 17.7(+2.4 \%)}$ & $0.79(2)$ & ${}_{ -0.15(-19.1 \%)}^{+ 0.23(+29.0 \%)}$ & ${}_{ -0.05(-6.3 \%)}^{+ 0.05(+6.3 \%)}$ \\
NLO & $759.5(1)$ & ${}_{ -107.8(-14.2 \%)}^{+ 110.1(+14.5 \%)}$ & ${}_{ -17.9(-2.4 \%)}^{+ 17.9(+2.4 \%)}$ & $0.87(2)$ & ${}_{ -0.14(-16.2 \%)}^{+ 0.22(+25.0 \%)}$ & ${}_{ -0.05(-5.3 \%)}^{+ 0.05(+5.3 \%)}$ \\
\nNLOQCDt & $817.1(1)$ & ${}_{ -29.9(-3.7 \%)}^{+ 42.3(+5.2 \%)}$ & ${}_{ -19.3(-2.4 \%)}^{+ 19.3(+2.4 \%)}$ & $0.96(4)$ & ${}_{ -0.07(-7.5 \%)}^{+ 0.02(+1.7 \%)}$ & ${}_{ -0.06(-5.8 \%)}^{+ 0.06(+5.8 \%)}$ \\
nNLO & $825.4(1)$ & ${}_{ -29.3(-3.5 \%)}^{+ 41.3(+5.0 \%)}$ & ${}_{ -19.5(-2.4 \%)}^{+ 19.5(+2.4 \%)}$ & $1.03(4)$ & ${}_{ -0.07(-6.3 \%)}^{+ 0.01(+1.4 \%)}$ & ${}_{ -0.05(-5.2 \%)}^{+ 0.05(+5.2 \%)}$ \\
\NLOQCDt+NNLL & $802.6(2)$ & ${}_{ -78.1(-9.7 \%)}^{+ 89.4(+11.1 \%)}$ & ${}_{ -19.0(-2.4 \%)}^{+ 19.0(+2.4 \%)}$ & \multicolumn{3}{c}{--} \\
NLO+NNLL & $810.9(2)$ & ${}_{ -77.8(-9.6 \%)}^{+ 89.2(+11.0\%)}$ & ${}_{ -19.1(-2.4 \%)}^{+ 19.1(+2.4 \%)}$ & \multicolumn{3}{c}{--} \\
\bottomrule
\end{tabular}
\end{center}
\caption{Total cross section and charge symmetry for $\ttz$ production. Same structure as in Table~\ref{ttwp}.\label{ttz}}  
\end{table}

\pagebreak

\boldmath
\subsubsection{$\ttz$}
\unboldmath

Results for the $\ttz$ production total cross section are listed in the second column of Table~\ref{ttz}. Similarly to the processes considered so far,
 the predictions show good perturbative convergence for both the
$\MttZ$ and $H_T$-based scale choices. The difference between the values of the cross section obtained with the two scale choices is large at $\LOQCD$: at that order, the $H_T$-based scale choice leads to a cross section that is about 9\% larger than the one found with the 
$\MttZ$-based scales. The difference between the two scale choices at NLO is reduced to less than 5\%; nNLO calculations further reduce it
to just over 1\%. For all of the perturbative orders considered, the uncertainty bands obtained through scale variations are compatible with
these differences. In the results that combine the two scale choices (lower part of Table~\ref{ttz})  the uncertainty bands at $\LOQCD$ are ${}_{-24\%}^{+35\%}$, and they reduce to ${}_{-14\%}^{+15\%}$ at NLO and to ${}_{-4\%}^{+5\%}$ at nNLO.
Similar to $\tth$ production, the resummed calculation produces a
larger difference in central values for
the total cross section calculated with $\MttZ$ and $H_T$-based scales  than the nNLO one, and also has larger uncertainty bands. Again, this fact indicates that scale variation in nNLO calculations does not lead to a reliable estimate of the uncertainty associated to missing higher-order corrections.   For this reason, the most accurate and reliable  prediction for the total cross section in $\ttz$ production is given by the combined-scales calculation at NLO+NNLL accuracy, yielding
a total cross section of about $811$~fb, with an uncertainty from missing
higher orders corrections of ${}_{-10\%}^{+11\%}$. This prediction
includes the contributions of the EW corrections, which are rather small for the total cross sections. In fact, NLO calculations increase
the total cross section by about 1\% w.r.t.~$\NLOQCD$ calculations; this difference falls well within the theory uncertainty band.

\medskip

Similarly to the case of $\tth$ production, the charge asymmetry is small for $\ttz$
production. This is expected since the cross section is dominated by the gluon-fusion channel.\footnote{However, observe  that a small and negative charge asymmetry is present in  $t \bar t Z$ production already at $\LOQCD$ \cite{Maltoni:2015ena}.}
The contribution of EW corrections to the asymmetry is 
not negligible. However, contrary to $\tth$ production, the approximate nNLO QCD corrections
 do play a prominent role: indeed they increase the asymmetry by about
20\%. Furthermore, similarly to $\ttw$
production, the asymmetry calculated to nNLO  has smaller scale-uncertainty bands than the asymmetry calculated to NLO. The nNLO asymmetry in $\ttz$ production is about 1\%
with a scale uncertainty of ${}_{-0.07}^{+0.01}$.

For the total cross section, the uncertainties due the PDFs are of the order of $2.3-2.4\%$, independently from the level of accuracy in the theory predictions. They are therefore smaller than the residual scale uncertainty in all cases.  However, as for $\tth$ production, they are slightly 
larger for the charge asymmetry. Given the fact that, for the charge asymmetry, the scale uncertainties  are
much smaller for $\ttz$ production than for the
$pp\to\tth$ process, the PDF uncertainty  cannot be
neglected for the former process.

\subsection{Differential distributions \label{sec:diffdist}}

In this section, we present predictions for  binned differential distributions for invariant mass, transverse momentum, and rapidity observables. Transverse momentum and invariant mass distributions are evaluated up to NLO+NNLL accuracy
(NNLL-resummed results, matched to the complete-NLO predictions). Rapidity distributions are evaluated up to nNLO (approximate NNLO QCD results, matched to complete-NLO calculations),
since the NNLL resummed results  in our framework are integrated over rapidities.  Only predictions that combine  calculations carried out with $\MttV$-based  and $H_T$-based scale choices are shown in Figures~\ref{fig:Mttxv}-\ref{fig:ytx}.

In each of the figures there are four
plots. Each of them corresponds to one of the four processes considered in this work: the top-left  plot of each figure refers to
 $\ttwp$ production, the top-right plot shows  the $\ttwm$ process, the
bottom-left plot  refers to $\tth$ production, and the bottom-right plot shows the $\ttz$ process.  

Each of these plots has the same layout, consisting of a  main (top) panel and three ratio
insets below it. The four components of each plot show the following information:
\begin{itemize}
\item The top panel shows the absolute predictions for the differential distribution. The central value of the distribution in each bin (calculated to NLO+NNLL accuracy for the invariant mass and
transverse momentum distributions, and to nNLO for the rapidity distributions), is plotted in blue.  The differential distribution calculated to 
complete-NLO (denoted simply as NLO) is plotted in red. In this  panel
the vertical axis indicates the \emph{cross section per bin}. Consequently, the total cross section is simply the sum of the heights of the distribution in each bin (including the bins that fall outside the range shown). For the invariant mass and
transverse momentum distributions the horizontal axis is logarithmic.

\item In the  ratio inset just below the main panel, the NLO+NNLL (or nNLO in the case of rapidity distributions) and NLO predictions are
shown as a ratio w.r.t.~the central value of the NLO calculations. Here, also
the uncertainties from scale variations (dark shaded band) and PDFs (light
shaded band) are shown, with the latter added linearly\footnote{By linearly adding scale and PDF uncertainties we adopt a conservative approach, assuming full correlation among the two classes of effects.} to the former. The purpose of this inset is to show the impact of the soft emission corrections on the shape of the distribution.

\item The
middle ratio inset shows the difference between the additive and the multiplicative
combination for the $\rm NLO_{QCD}+NNLL$ and the EW
corrections, together with the
corresponding scale uncertainties. PDF uncertainties are not shown in this panel. In particular, the inset shows the ratio between the $\rm NLO+NNLL$ distribution ($\rm nNLO$ for rapidities) and the central value of the $\rm NLO+NNLL$ ($\rm nNLO$) calculation in each bin, as well as the ratio between the $\rm NLO\times NNLL$ distribution ($\rm nNLO_{mult}$ for rapidities) and the central value of the $\rm NLO+NNLL$ ($\rm nNLO$) calculation in each bin. Multiplicative results are shown as a
dark yellow band, while the additive-approach results are shown, as before, by a blue band. 

\item In the lower inset the effect of  EW corrections is shown by plotting 
the  $\rm NLO+NNLL$ ($\rm nNLO$ for rapidity distributions) calculations and the $\NLOQCD$+NNLL (or, in the case of rapidity distributions, to $\nNLOQCD$) calculations, both divided by the central value of the $\NLOQCD$+NNLL  ($\nNLOQCD$) calculation in each bin. Also in the lower inset, NLO+NNLL (nNLO) calculations are indicated by the
blue band, while results at $\NLOQCD$+NNLL ($\nNLOQCD$)  accuracy are shown by the brown band. 
\end{itemize}

\subsubsection{Invariant masses}

\begin{figure}[t]
\centering
\includegraphics[width=0.49\textwidth]{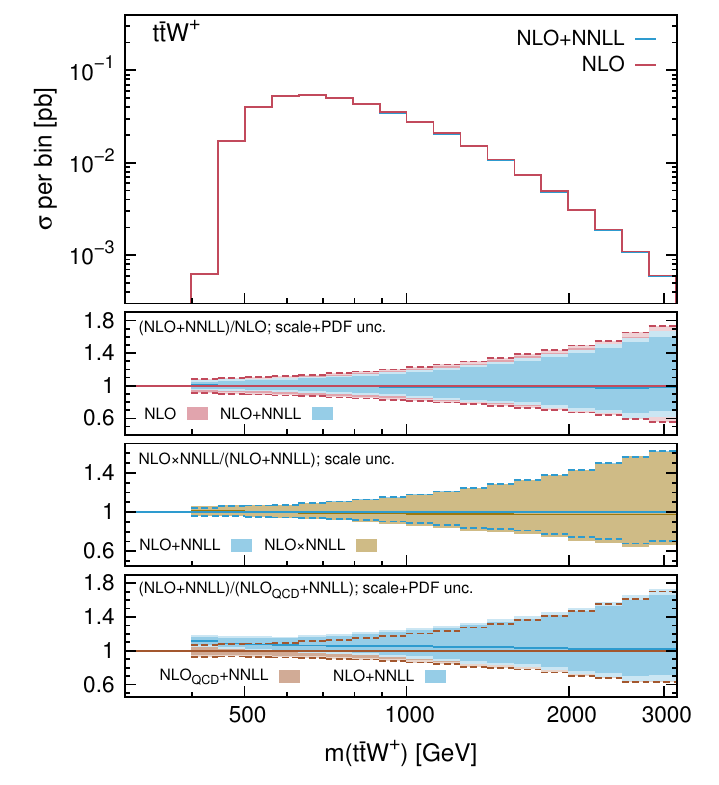}
\includegraphics[width=0.49\textwidth]{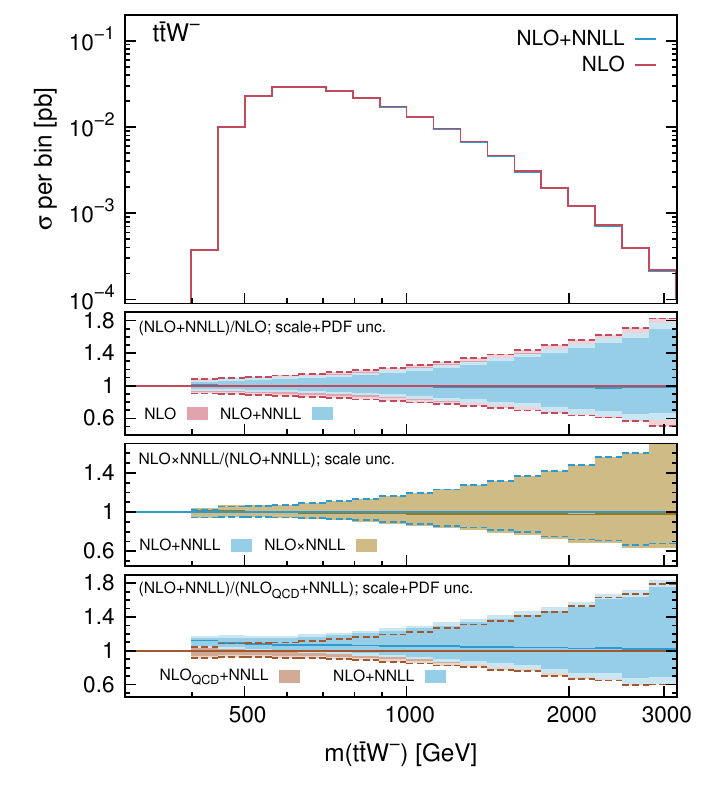}
\includegraphics[width=0.49\textwidth]{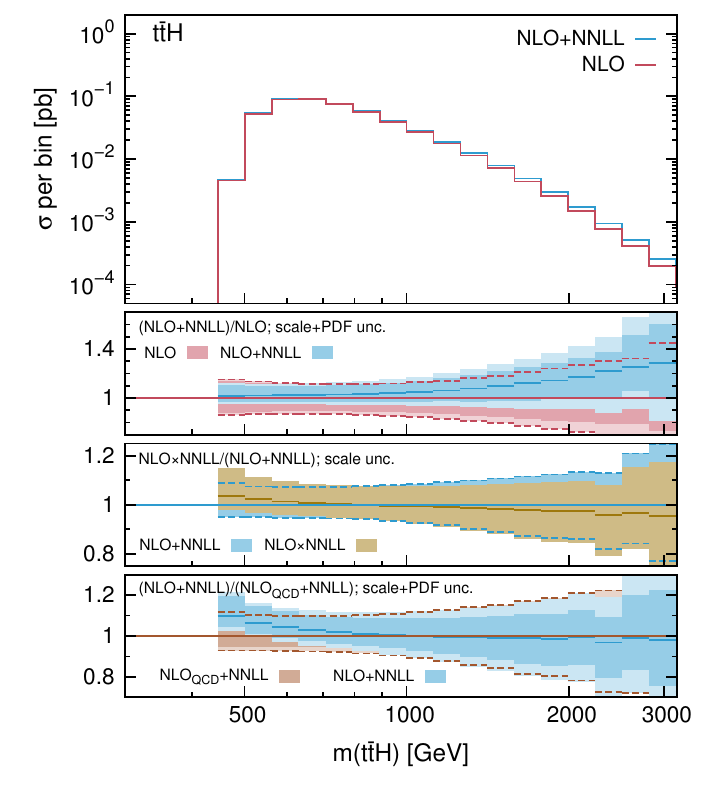}
\includegraphics[width=0.49\textwidth]{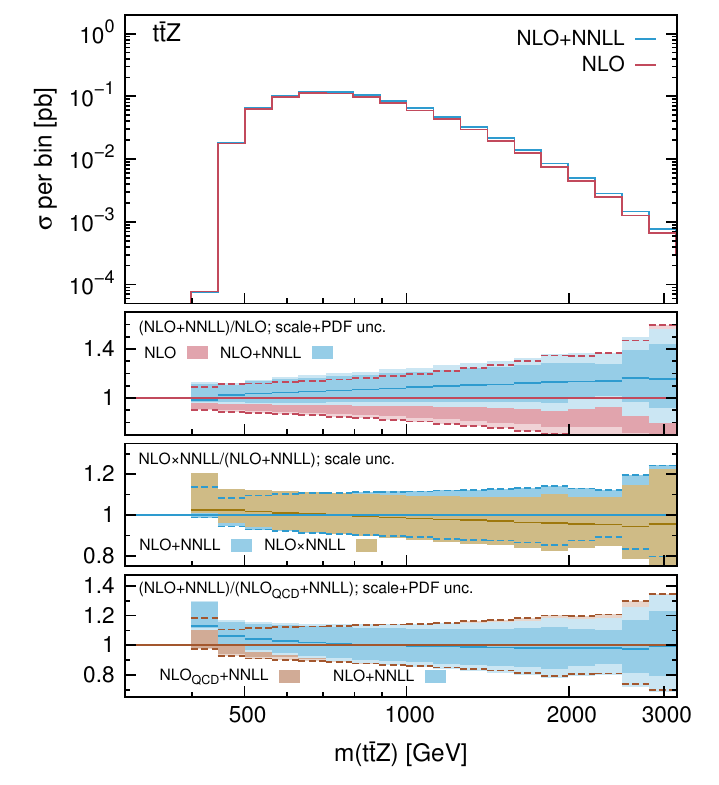}
\caption{Distribution of the invariant mass of the $t \bar t V$ system at 13 TeV. The upper plots  refer to $\ttwp$ (left) and $\ttwm$ (right), while the lower ones to $\tth$ (left) and $\ttz$ (right). In the first inset we focus on the resummation effects (NLO vs. NLO+NNLL), in the second one on the difference between the additive and multiplicative approach ($\rm NLO + NNLL$ vs. $\rm NLO \times NNLL$) including only scale uncertainties, and in the third on the impact of EW corrections ($\rm NLO_{QCD} + NNLL$ vs $\rm NLO + NNLL$). More details can be found in the main text.  }
\label{fig:Mttxv}
\end{figure}

The invariant mass distribution of the $\ttv$ system is shown in
Figure~\ref{fig:Mttxv}. The upper-left (upper-right) plot shows the  invariant mass of the $\ttwp$ ($\ttwm$) system. 
From the first ratio-inset it can be
seen that the resummation  has a relatively small impact  on the shape of the distribution  w.r.t.~the NLO calculation. Even though the uncertainties affecting the NLO+NNLL calculation are slightly smaller than the NLO uncertainties, they
remain large, in particular for large values of the invariant mass.
 The reason for the
rather large scale dependence at high invariant mass is that the predictions for the $\MttW$-based and
$H_T$-based scales differ significantly in this region of phase space. This observation applies to
NLO calculations as well as to NLO+NNLL accuracy calculations. 
The tail of the invariant mass distribution  is dominated by real radiation from quark emissions in the $qg$-initiated channel. Consequently, soft-gluon resummation cannot improve its description, since the $qg$ channel is subleading  in the threshold limit.

The difference
between the additive and multiplicative methods of combining EW corrections and QCD resummed calculations  is small, as  it can be seen by 
examining the middle inset in the two plots at the top of Figure~\ref{fig:Mttxv}; indeed, the blue and dark-yellow bands overlap almost entirely. Moreover, from the lower inset in the same plots it can be seen that   EW corrections have a significant impact on the distribution only for small invariant masses. Indeed,
the expected EW Sudakov suppression at large invariant masses is not observed since the NLO QCD corrections are rather large and dominated
by hard real-emission corrections. 
As shown in~\cite{Frederix:2017wme}, a jet-veto can
suppress the large QCD corrections, which results in an  enhancement of
the relative impact of EW corrections.

\medskip

The situation is somewhat different for the $\tth$ and $\ttz$ invariant mass
distributions, shown in the lower-left and lower-right plots of
Figure~\ref{fig:Mttxv}, respectively. Since the NLO corrections for these
processes are not dominated by the opening of new channels, the
resummation of soft radiation reduces
the scale dependence significantly. In addition, NLO+NNLL 
calculations lead to an increase of the  cross-section central value in each bin, ranging from a few percents for small invariant masses, to about $30\%$ ($20\%$) for $\tth$ ($\ttz$) production at
3~TeV.
By looking at the first inset in the lower line of Figure~\ref{fig:Mttxv}  one sees that for $\ttz$ production the entire uncertainty band at NLO+NNLL is
contained in the NLO uncertainty band over the whole mass range shown in the figure. For $\tth$ production,  the central
value of the NLO+NNLL distributions remains well within the NLO uncertainty, as one would expect
from a well-behaved perturbative expansion. However, the NLO+NNLL uncertainty band only has a partial overlap with the NLO uncertainty band in the far tail of the invariant mass distribution.

Given that resummation changes the central value of the NLO predictions, there is a slight dependence on how these effects are combined with the NLO corrections. As shown in the middle inset, the additive
and multiplicative approaches lead to  slightly different shapes for the $m(t\bar t H/Z)$ invariant mass distribution. This difference is marginal,
though, and remains well within the uncertainty band. Nevertheless, for large values of $m(t\bar t H/Z)$, this difference in shape amounts to a few percent. In this phase-space region, predictions in the multiplicative approach can be preferred, as discussed in Section~\ref{sec:matching}.  

EW corrections in the $\ttz$ and $\tth$ invariant mass distributions are more relevant near the production threshold, as was
already observed in~\cite{Frixione:2015zaa}. This effect  is due to a Sommerfeld enhancement arising, {\it e.g.}, from one-loop diagrams with Higgs propagators connecting two of the final-state particles \cite{Degrassi:2016wml, Maltoni:2017ims}, which contribute to the $\rm NLO_2$ corrections. In contrast with the analysis in~\cite{Frixione:2015zaa}, here also the
subleading EW contributions are included: the $\LO_3$ and $\NLO_3$
contributions are positive and almost completely cancel the negative $\NLO_2$
corrections at large invariant masses, resulting in a negligible difference
between the NLO+NNLL and $\NLOQCD$+NNLL predictions. Note that the multiplicative approach spoils this cancellation, since it rescales only the $\NLO_2$ corrections according to Eq.~(\ref{eq:matchingmult}).

\begin{figure}[t]
\centering
\includegraphics[width=0.49\textwidth]{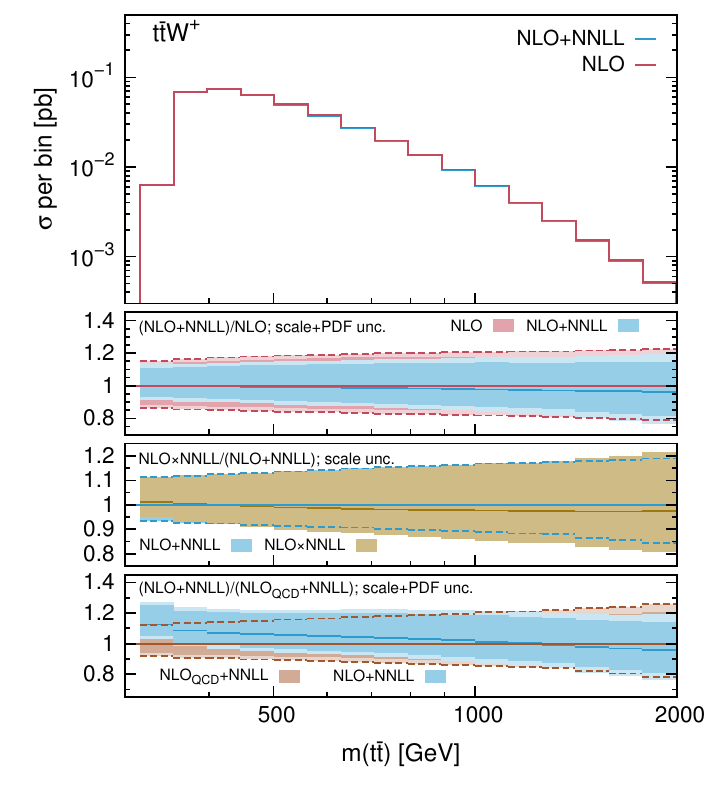}
\includegraphics[width=0.49\textwidth]{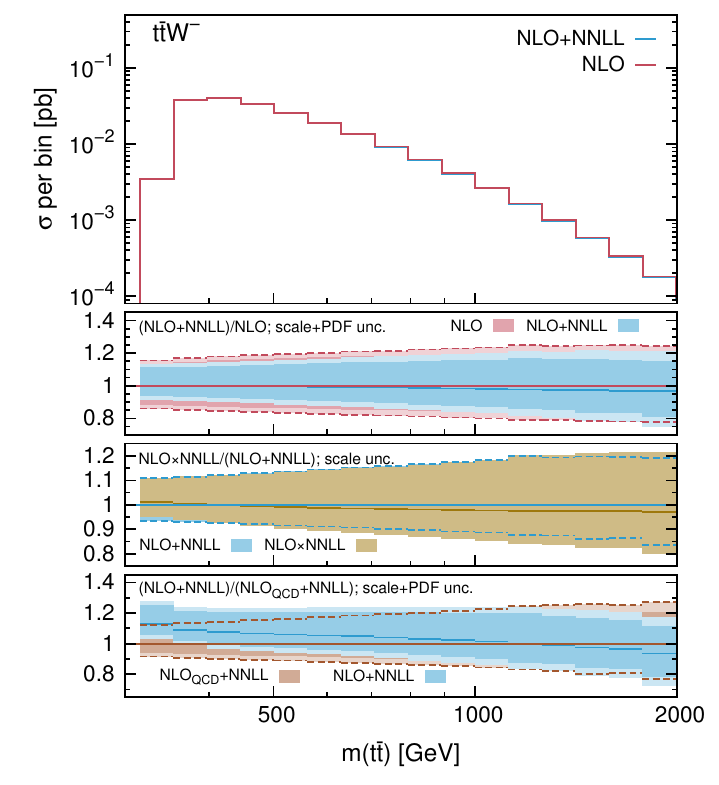}
\includegraphics[width=0.49\textwidth]{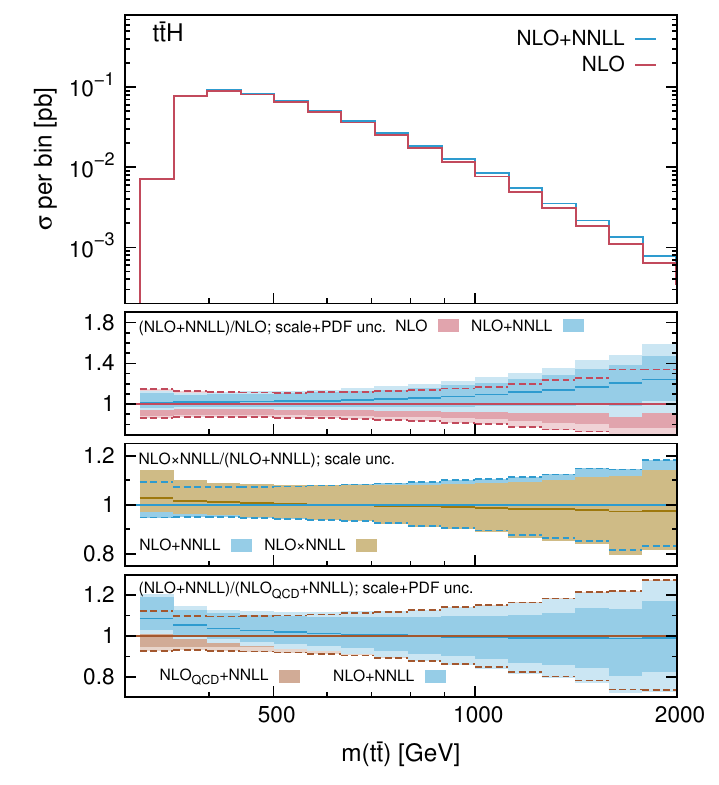}
\includegraphics[width=0.49\textwidth]{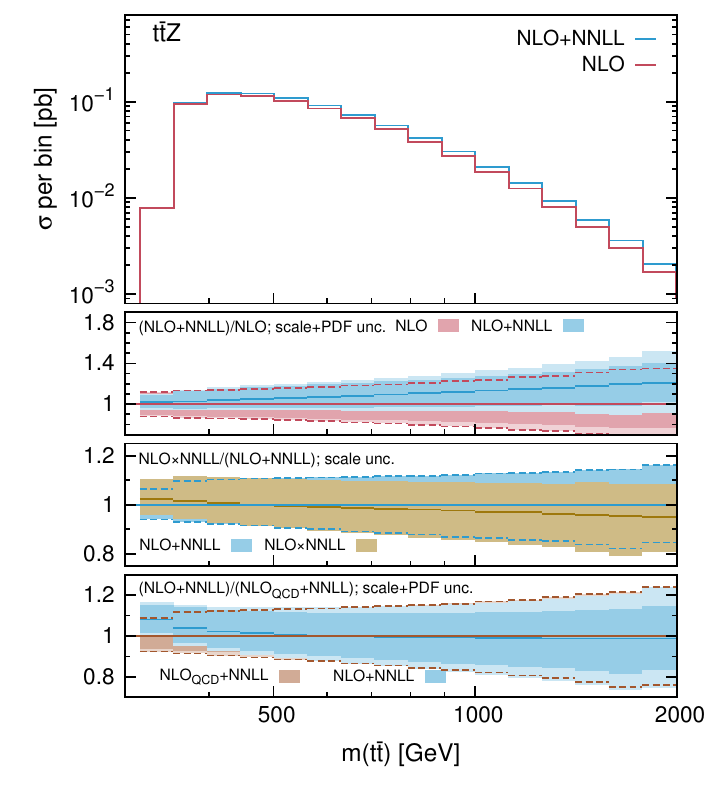}
\caption{Distributions differential w.r.t.~the invariant mass of the top-quark pair.
	Same structure as in Figure~\ref{fig:Mttxv}.}
\label{fig:Mttx}
\end{figure}

\medskip 

The distributions differential w.r.t.~the $t\bar{t}$ pair invariant mass are shown in Figure~\ref{fig:Mttx}. For the $\ttwp$ and $\ttwm$
processes (upper-left and upper-right plots, respectively) the corrections from threshold resummation to NNLL matched to the NLO corrections reduce the uncertainty band over the whole invariant mass range considered. The effect of the resummation on the central value of each bin is negligible. 
In comparison to the  $m(t\bar{t}W^\pm)$ invariant mass distributions (Figure~\ref{fig:Mttxv}), the scale-uncertainty
 band at large  top-pair invariant masses is narrower. The reason
is that the two scale choices ($\MttW$-based and $H_T$-based) give predictions that
lie much closer to each other, in comparison to the case of $m(t\bar{t}W^\pm)$ invariant mass distributions.

%\medskip
 On the other hand, for $\tth$ and $\ttz$ production, the
corrections to the $m(t\bar{t})$ invariant mass distributions due to resummation  are  similar to the case of the $m(t\bar{t}H/Z)$ invariant mass distribution,
 {\it i.e.}, they enhance the cross section over the full invariant mass range shown in the figure, starting with small effects at threshold and reaching up to about $20-30\%$ at
 $m(t\bar{t})=2~{\rm TeV}$.
The theory uncertainties at these large invariant masses are still
larger than the difference in central values between NLO and NLO+NNLL calculations, resulting in a stable
perturbative expansion. For all four processes, the difference between the
additive and multiplicative combination is small, even though there is a
trend: the multiplicative approach gives a slightly softer invariant mass
spectrum in the tail of the distribution. 

The lowest inset in each of the four plots in Figure~\ref{fig:Mttxv}   shows that  the EW corrections, included  in the NLO+NNLL predictions, distort the shape of the distributions calculated to $\NLOQCD$+NNLL accuracy. For all four processes, the EW corrections result in a
positive contribution to the cross section at small invariant masses, where they enhance  the distribution in each bin by $5-10\%$ for
$m(t\bar{t})<400$~GeV. At larger invariant masses the EW corrections are negligible in $\tth$ and $\ttz$ production. For $\ttwp$ and
$\ttwm$ production, the EW corrections remain positive up to  $\sim 1$~TeV, where they
turn negative, as expected from  EW Sudakov suppression. 

\subsubsection{Transverse momenta}

\begin{figure}[t]
\centering
\includegraphics[width=0.49\textwidth]{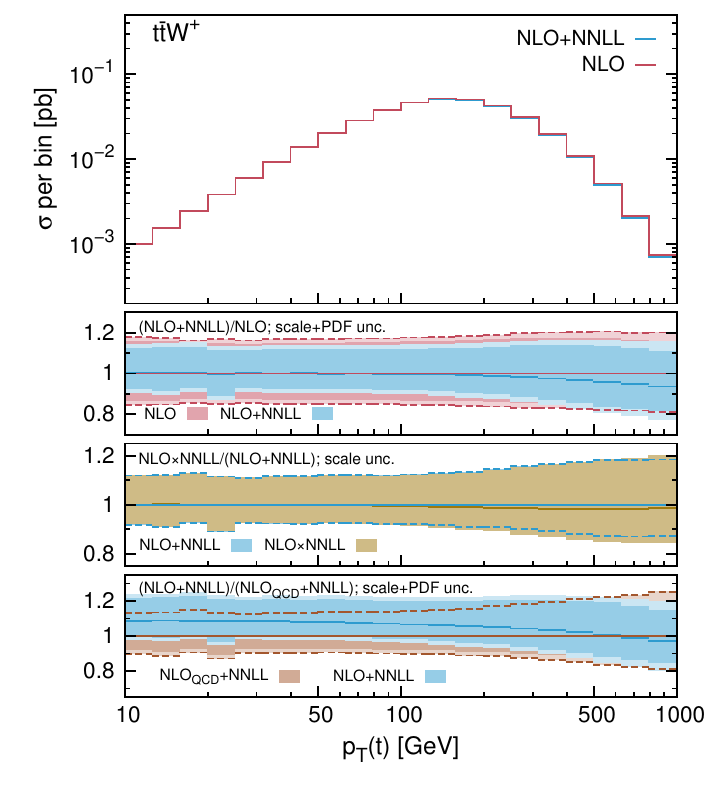}
\includegraphics[width=0.49\textwidth]{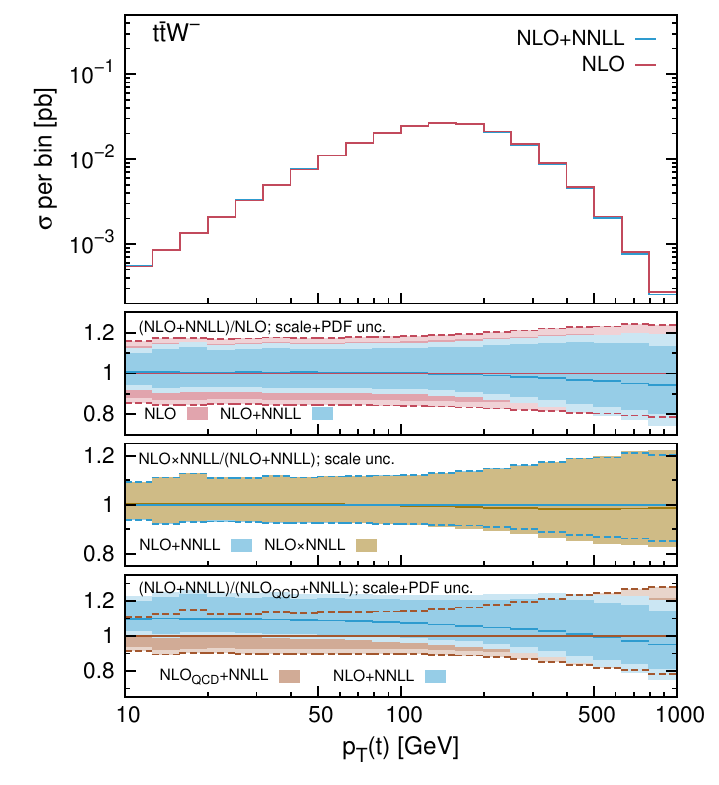}
\includegraphics[width=0.49\textwidth]{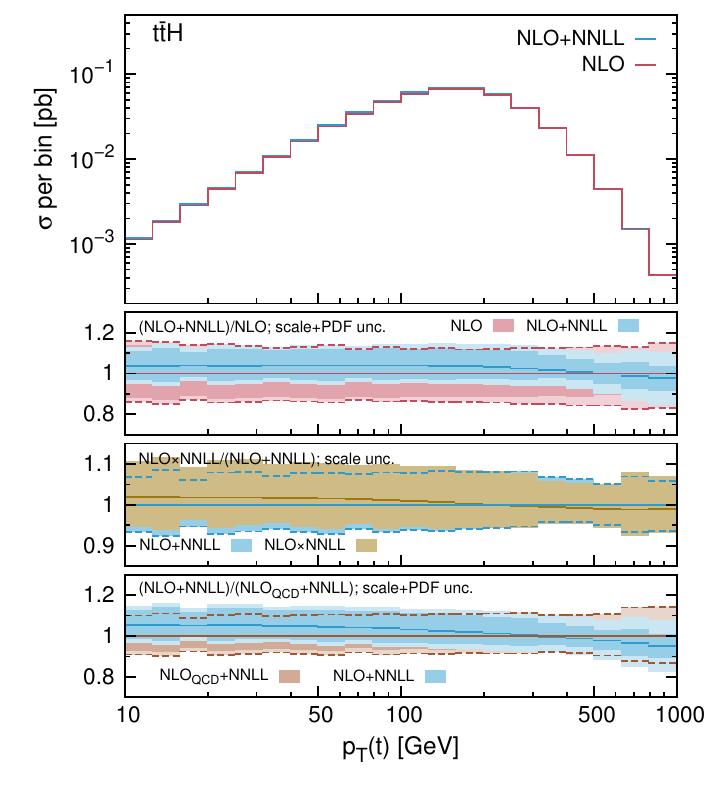}
\includegraphics[width=0.49\textwidth]{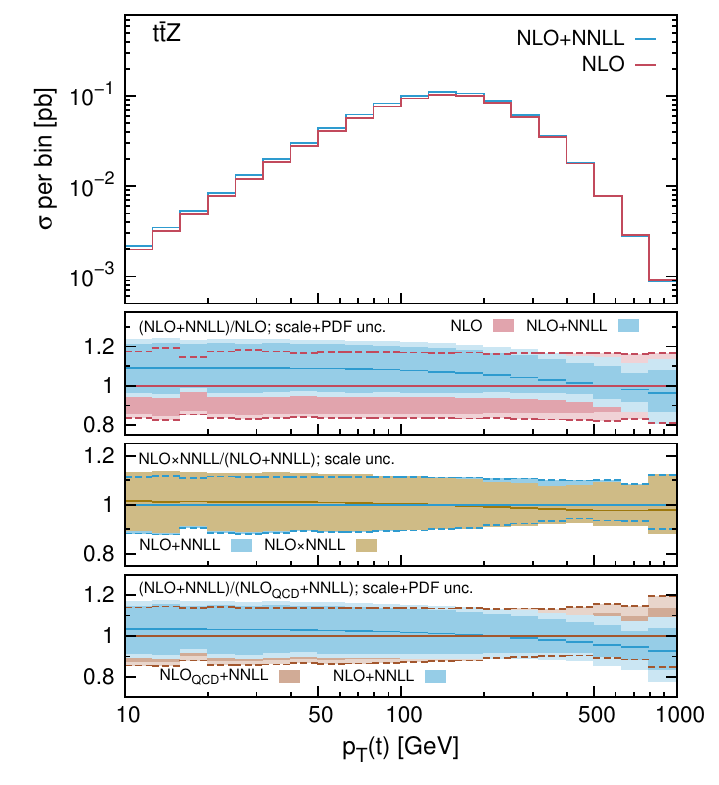}
\caption{Distributions differential w.r.t.~the transverse momentum of the top quark.
	Same structure as in Figure~\ref{fig:Mttxv}.}
\label{fig:pTt}
\end{figure}

\begin{figure}[t]
\centering
\includegraphics[width=0.49\textwidth]{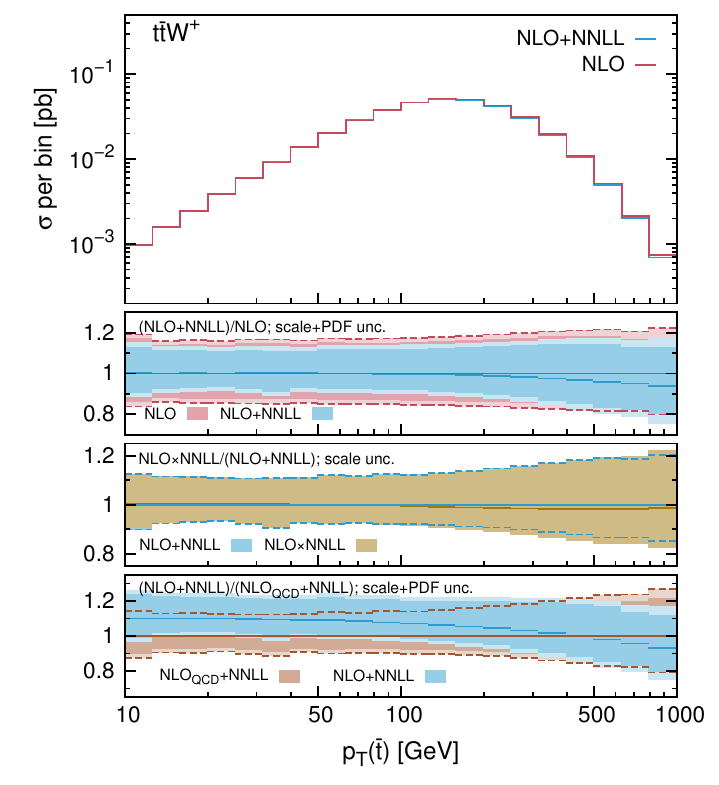}
\includegraphics[width=0.49\textwidth]{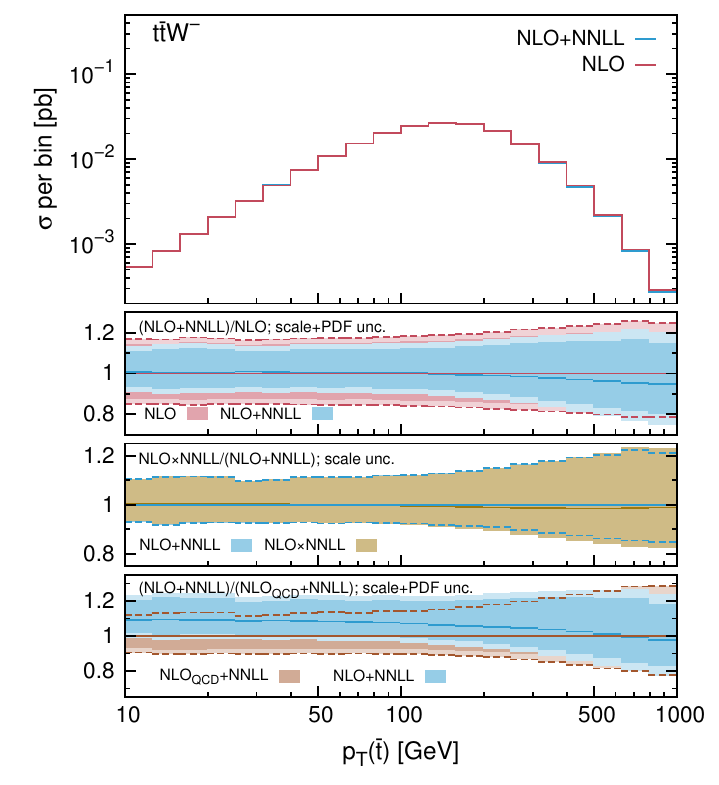}
\includegraphics[width=0.49\textwidth]{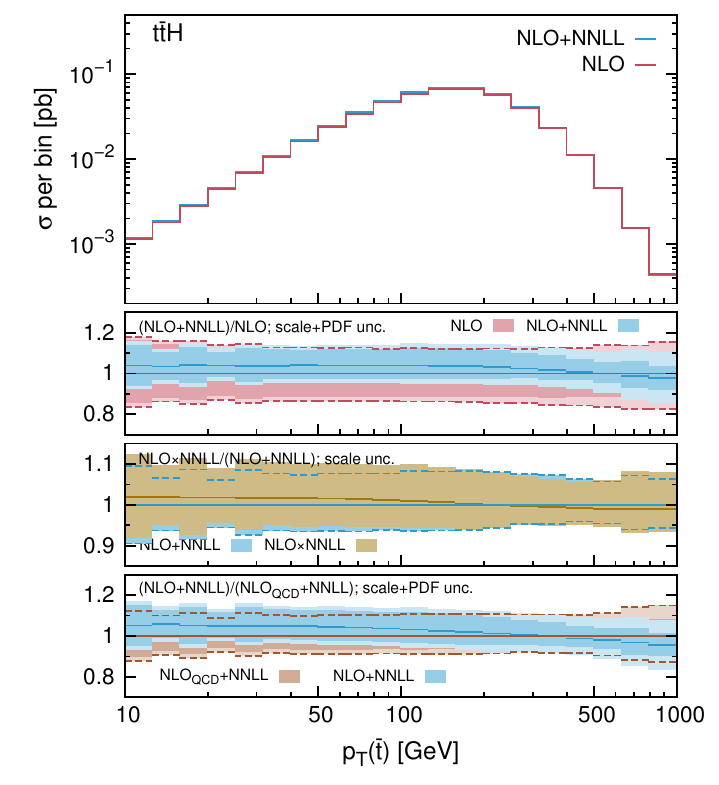}
\includegraphics[width=0.49\textwidth]{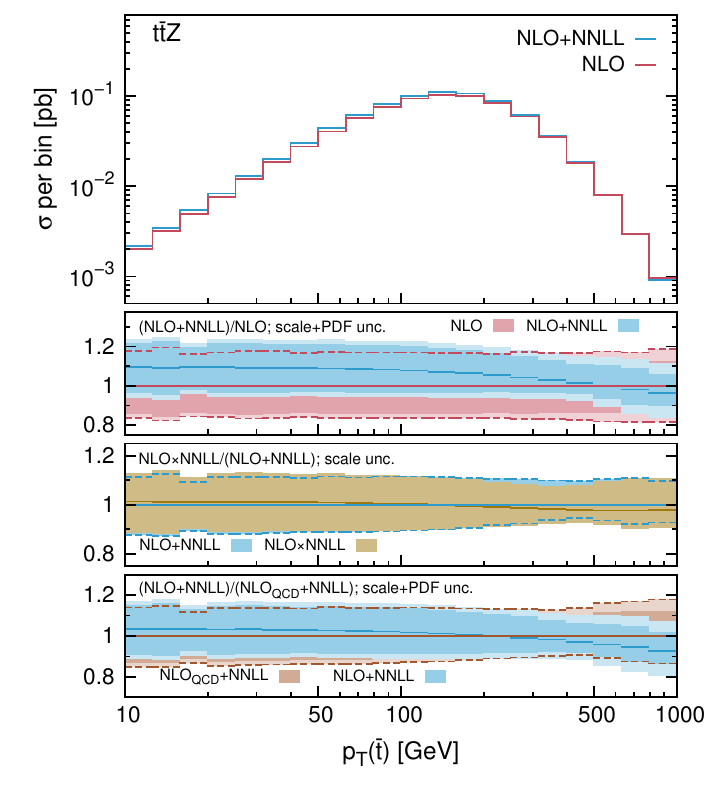}
\caption{Distributions differential w.r.t.~the transverse momentum of the antitop quark.
	Same structure as in Figure~\ref{fig:Mttxv}.}
\label{fig:pTtx}
\end{figure}

The distributions differential w.r.t.~the top quark (Figure~\ref{fig:pTt}) and antitop quark
(Figure~\ref{fig:pTtx}) transverse momentum are very similar. 
For these two observables, 
the NNLL resummation reduces the scale
uncertainty in comparison to NLO calculations. In addition, NNLL resummed results lead to slightly smaller cross sections at large transverse momenta
($p_T(t/\bar{t})>500$~GeV). For $\tth$ production and especially
for $\ttz$ production, NNLL resummation increases the cross section in comparison to NLO calculations for transverse momenta smaller than $\sim 500$~GeV. 
Hence, the corrections due to soft-gluon emission affect the shape of the distribution, even though the
NLO+NNLL  and NLO uncertainty bands always have a large overlap. Even more than in the case of invariant mass distributions, the differences between calculations carried out in  the additive and multiplicative approaches are marginal. 
The  EW corrections  show their typical behavior: for small transverse momenta, they induce a constant upward shift in the central value of the distribution in each bin, of the order of $5\%$ for $\tth$ and $\ttz$ production and somewhat
larger, $10\%$, for $\ttwp$ and $\ttwm$ production. These effects decrease rapidly as the transverse momentum increases, until
the EW corrections start lowering the QCD cross section in each bin for large transverse momenta. The large positive
correction at small transverse momenta in $\ttwp$ and $\ttwm$ production is
mainly due to the large $\NLO_3$ correction in these processes. In this region of phase space,
the EW corrections are of a size similar to the scale uncertainties of the resummed calculations; this results in a NLO+NNLL prediction for the distribution whose central value in each bin lies just within the uncertainty band of the $\NLOQCD$+NNLL calculation, and vice versa.  

\begin{figure}[t]
\centering
\includegraphics[width=0.49\textwidth]{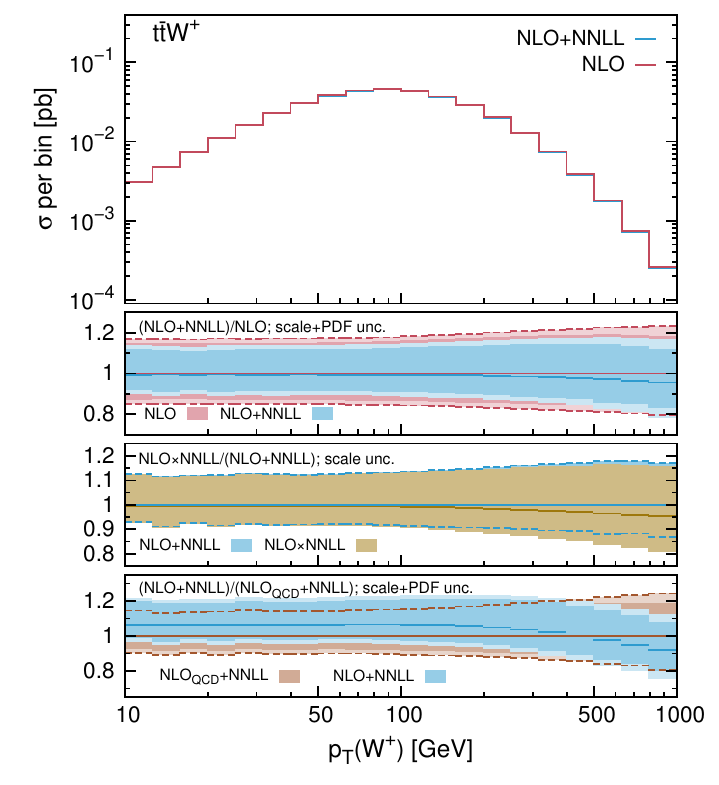}
\includegraphics[width=0.49\textwidth]{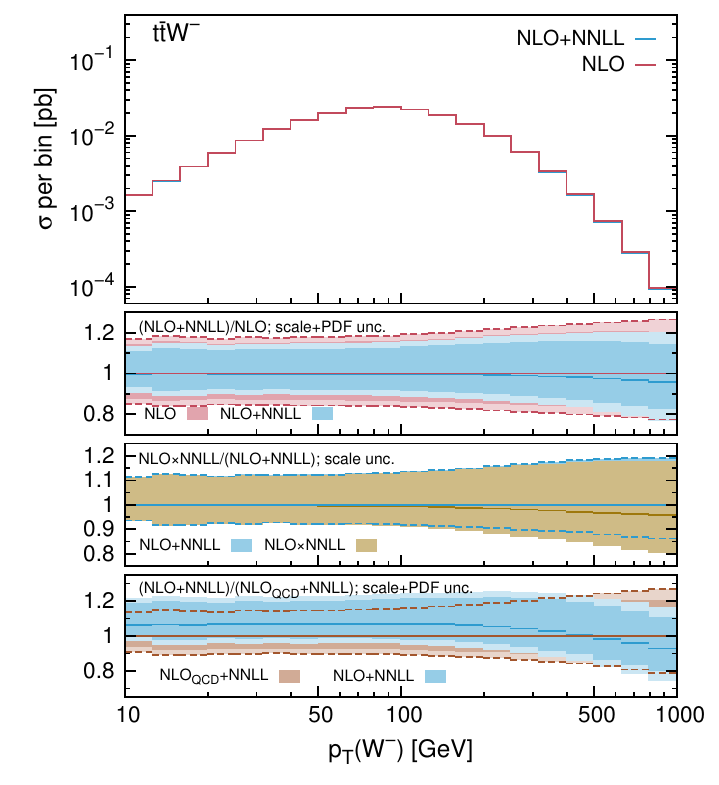}
\includegraphics[width=0.49\textwidth]{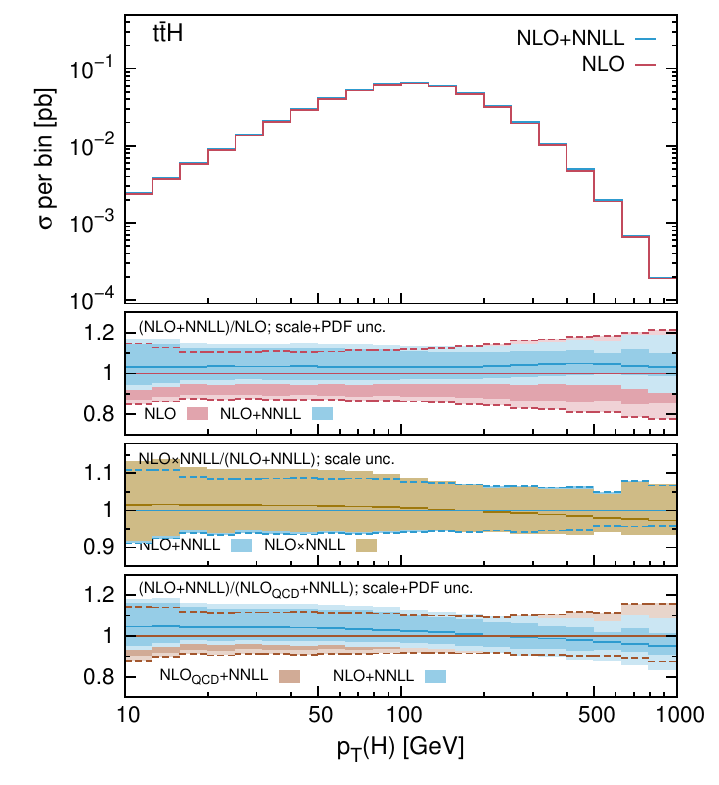}
\includegraphics[width=0.49\textwidth]{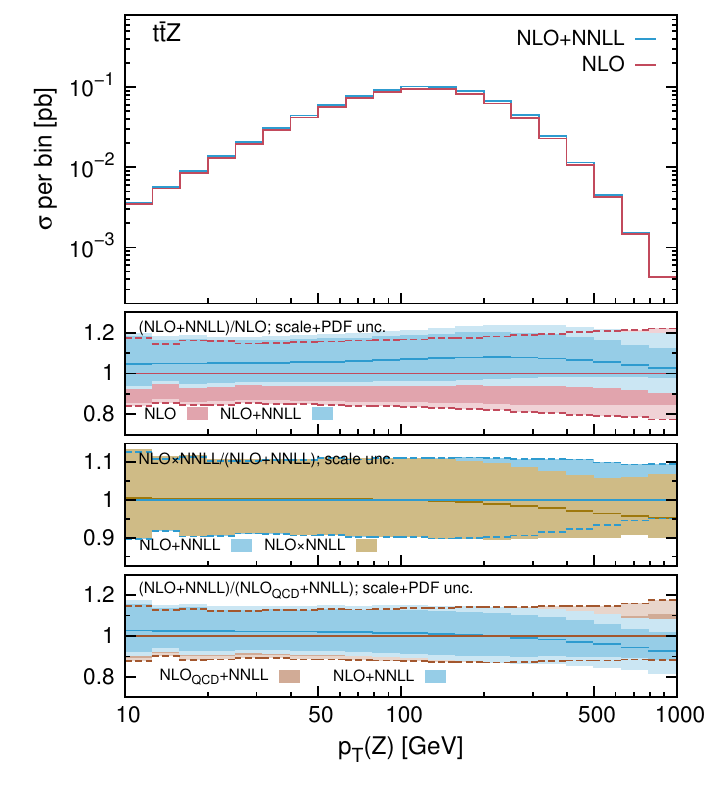}
\caption{Distributions differential w.r.t.~the transverse momentum of the EW boson.
	Same structure as in Figure~\ref{fig:Mttxv}.}
\label{fig:pTv}
\end{figure}

The distributions differential w.r.t.~the transverse momentum of the heavy EW boson are shown in Figure~\ref{fig:pTv}. Qualitatively, the corrections beyond $\NLOQCD$, either from
resummation of soft emission, or from EW corrections in fixed order perturbation theory, are very similar
to the corresponding corrections to the distributions for the transverse momenta of the top and the
antitop quarks. For this reason, many of  the remarks made in the discussion of those distributions apply
to Figure~\ref{fig:pTv} as well. 
There is, however, one important exception, {\it i.e.}, the non-negligible difference ($\sim 5 \%$ at $p_T(V)=1~{\rm TeV}$) between the additive and multiplicative matching of the resummed results with the NLO corrections at very large transverse momenta.
The reason is that in the tail of these distributions both the $\NLO_1$ and $\NLO_2$ corrections are large and therefore the difference between the two approaches, which is dominated by the product of these two corrections, is not negligible. 

\subsubsection{Rapidities}

\begin{figure}[t]
\centering
\includegraphics[width=0.49\textwidth]{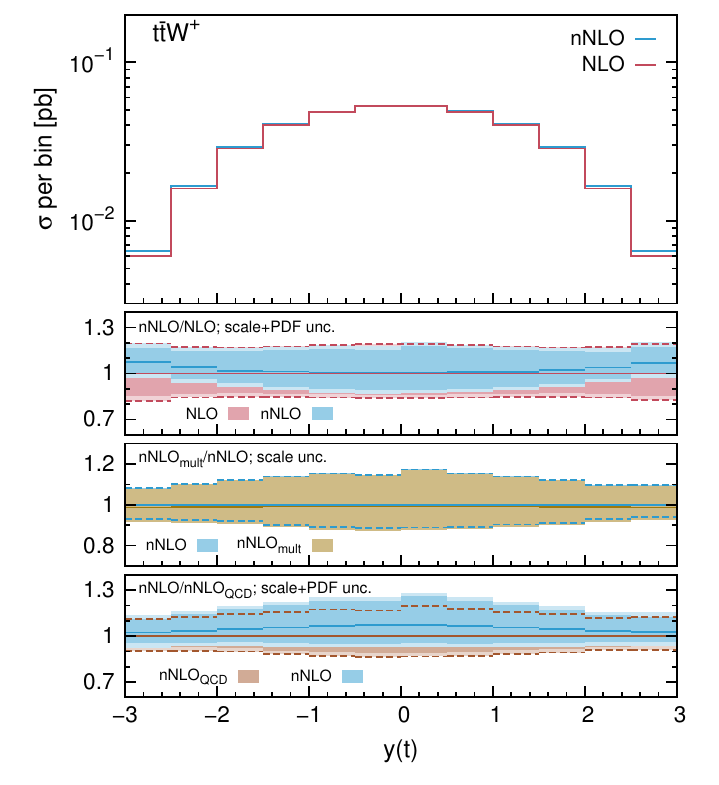}
\includegraphics[width=0.49\textwidth]{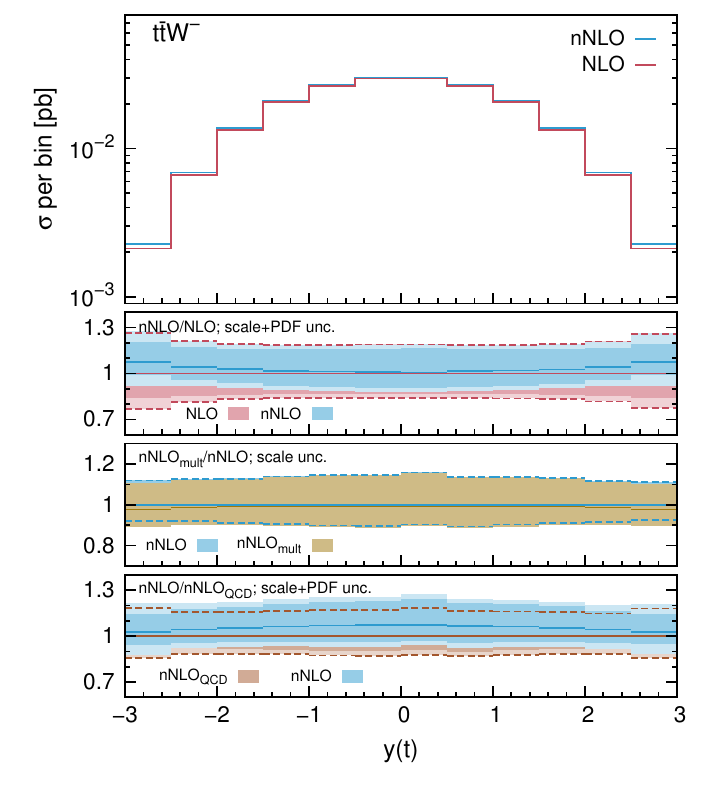}
\includegraphics[width=0.49\textwidth]{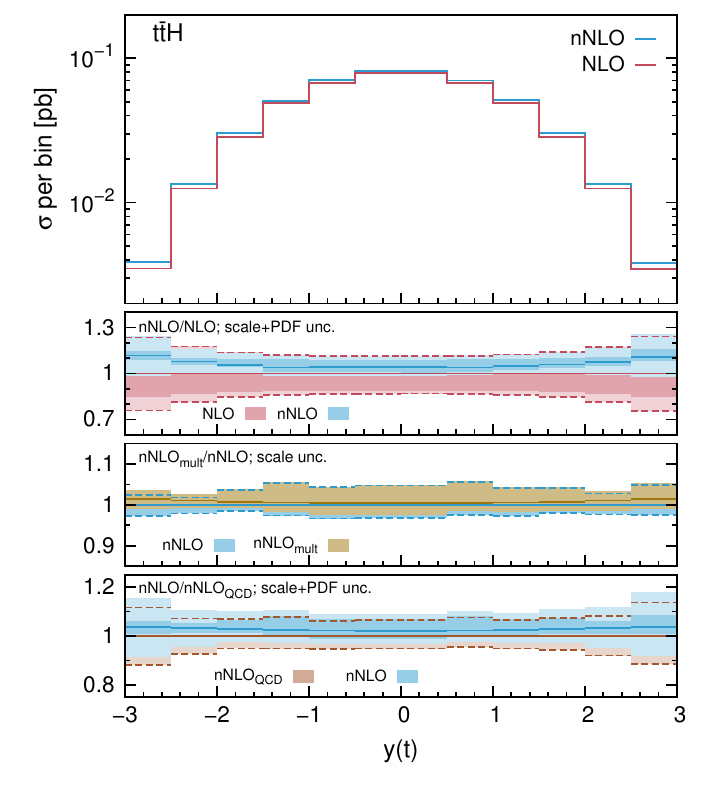}
\includegraphics[width=0.49\textwidth]{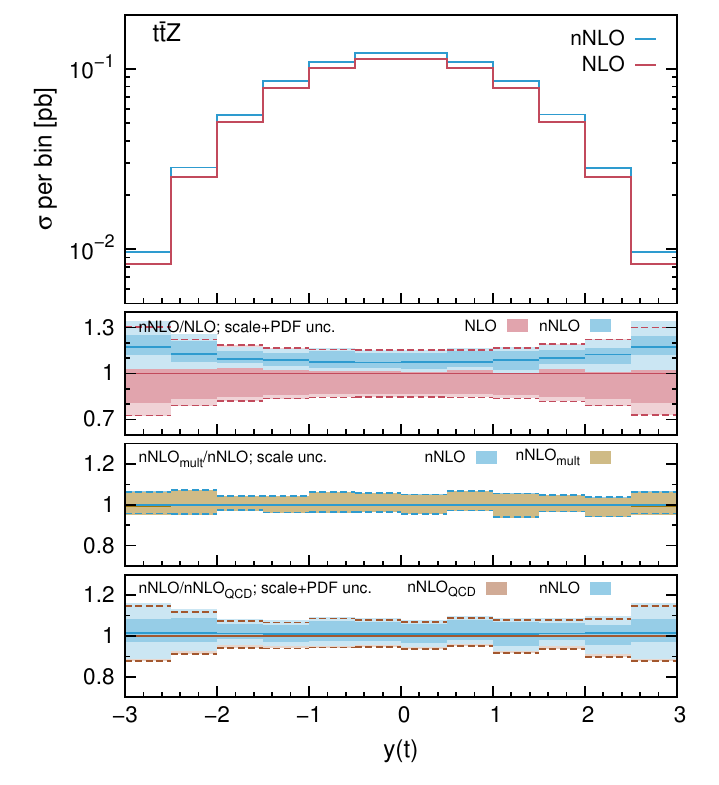}
\caption{Distributions differential w.r.t.~the top-quark rapidity. Same structure as in Figure~\ref{fig:Mttxv},  but with the substitution $\rm NLO_{QCD}+NNLL \rightarrow nNLO_{QCD}$, $\rm NLO+NNLL \rightarrow nNLO$, and $\rm NLO\times NNLL \rightarrow nNLO_{mult}$.}
\label{fig:yt}
\end{figure}

\begin{figure}[t]
\centering
\includegraphics[width=0.49\textwidth]{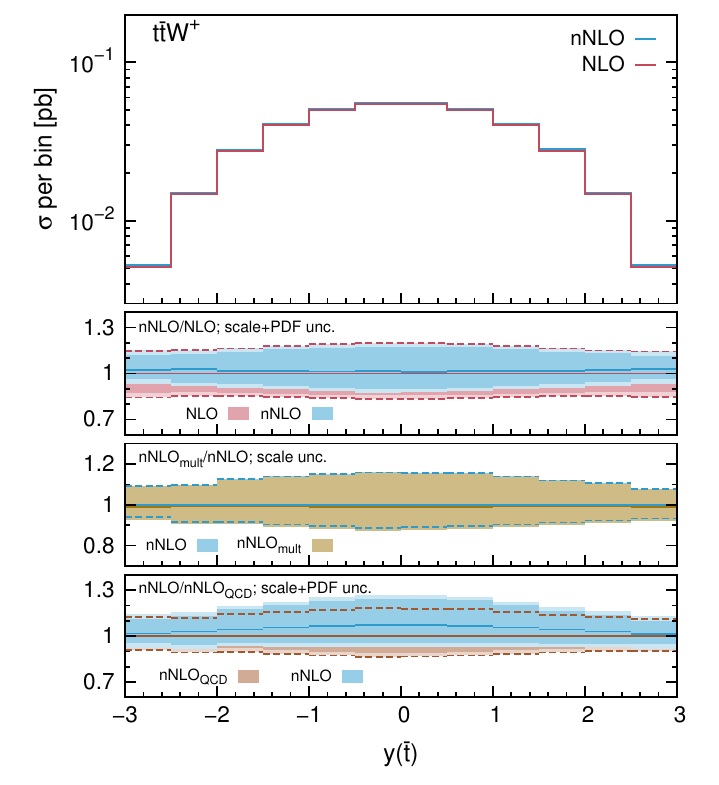}
\includegraphics[width=0.49\textwidth]{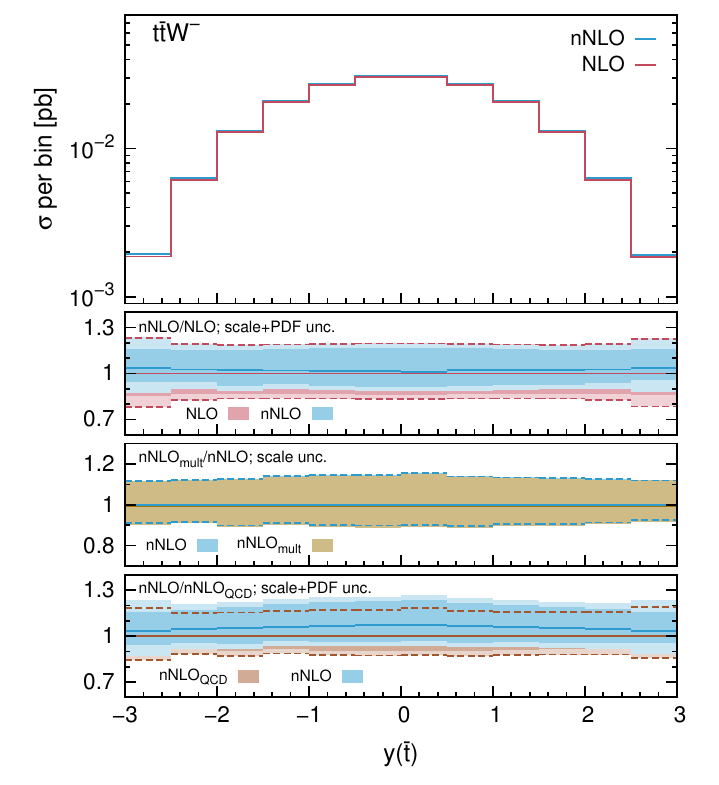}
\includegraphics[width=0.49\textwidth]{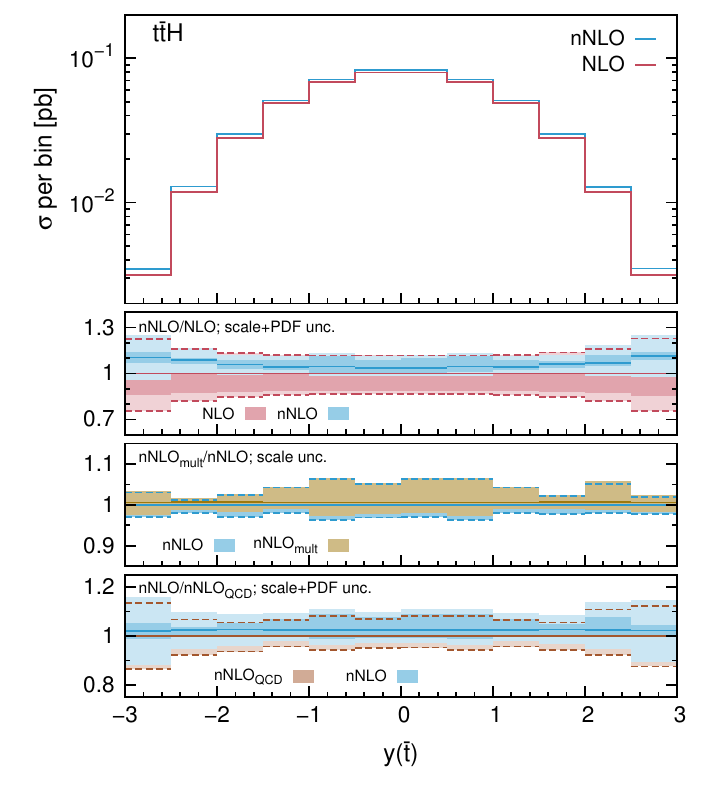}
\includegraphics[width=0.49\textwidth]{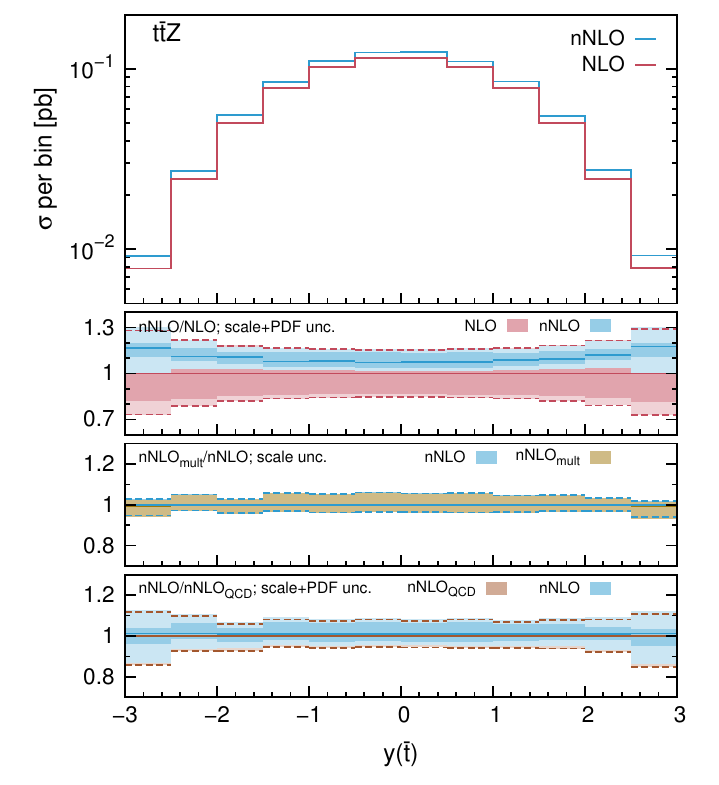}
\caption{Distributions differential w.r.t.~the antitop-quark rapidity Same structure as in Figure~\ref{fig:yt}.}
\label{fig:ytx}
\end{figure}

Figures~\ref{fig:yt} and~\ref{fig:ytx} show the predictions for the distributions differential w.r.t.~the top-quark and top-antiquark rapidities, respectively. 
In each individual process, antitop quarks are produced more centrally than top quarks. This fact is particularly  evident for the $\ttwp$ and $\ttwm$
processes. Indeed, this property is responsible for the large charge asymmetry
for these processes. 

In addition, by comparing the nNLO predictions to the NLO
predictions for the $y(t)$ differential distribution, (as it is done in first inset in each of the plots in Figure~\ref{fig:yt}), it can be seen that approximate NNLO corrections enhance the distributions at large forward and backward rapidities. On the contrary, the region of forward and backward rapidities in the  $y(\bar{t})$ differential distribution receives relatively large corrections only in $\ttz$ and $\tth$ production, but not in $\ttw$ production.
For $\ttw$ production, the different behavior of the nNLO corrections to the $y(t)$ and $y(\bar{t})$ differential distributions in the forward and backward rapidity regions is the cause of the relatively large  nNLO corrections to the charge asymmetry in these processes.

 For these distributions, the additive and multiplicative approaches to matching lead to almost identical results.
The EW corrections \ do not have a large impact on the shape of the top and antitop rapidity differential distributions, apart from the case of $t\bar{t}W^\pm$ production, where  the EW corrections enhance the small rapidity region of the distributions.
Finally we observe that, for the distributions shown in 
Figures~\ref{fig:yt} and~\ref{fig:ytx}, when one compares nNLO calculations to NLO calculations, distributions for $\tth$ and $\ttz$ production show a larger reduction of the relative size of the uncertainty bands  than  distributions for $\ttw$ production.

%\clearpage
\section{Conclusions \label{sec:conclusions}}

The purpose of this paper is to provide the most complete predictions to date for the total cross section and several differential distributions for the $\ttw$, $\ttz$, and $\tth$ production processes at the LHC operating at a center of mass energy of $13$~TeV.
In order to achieve this goal, we combined complete-NLO corrections, accounting for both QCD and electroweak effects, with the resummation of soft emission corrections to NNLL accuracy in QCD. The complete-NLO calculations were carried out with the most recent version of {\sc\small MadGraph5\_aMC@NLO}, while the NNLL resummation formulas were evaluated with an in-house parton level Monte Carlo code.

After considering theoretical uncertainties related to the choice of PDFs and to the residual dependence of the calculation on unphysical scales, we find the following predictions for the total cross section of the three processes
\begin{align}
\sigma_{\ttwp} &=  384.17(9)_{ -32.36(-8.4 \%) -8.16(-2.1 \%)}^{+
51.52(+13.4 \%)+ 8.16(+2.1 \%)} \, ,\nonumber \\
\sigma_{\ttwm} &=  197.75(4)_{ -16.07(-8.1 \%) -5.41(-2.7 \%)}^{+
26.41(+13.4 \%)+ 5.41(+2.7 \%)}\, , \nonumber \\
\sigma_{\tth} &=  496.36(7)_{ -29.35(-5.9 \%)-11.92(-2.4 \%)}^{+
38.64(+7.8 \%)+ 11.92(+2.4 \%)}\, ,\nonumber \\
\sigma_{\ttz} &= 810.9(2)_{ -77.8(-9.6 \%)-19.1(-2.4 \%)}^{+
89.2(+11.0\%)+ 19.1(+2.4 \%)} \, .
\end{align}
These predictions have NLO+NNLL accuracy. The number in parentheses next to the central value of the cross section indicates the statistical uncertainty due to the Monte Carlo integration. The first set of uncertainties is related to scale choices, and the second to PDFs.

Several differential distributions were analyzed in Section~\ref{sec:results}: the invariant masses of the $t \bar{t}$ and $\ttv$ systems, the transverse-mo\-men\-ta of the top quark, antitop quark and vector boson, all calculated to NLO+NNLL accuracy, and the rapidities of the top quark and antitop quark, evaluated to nNLO.
The behavior of the perturbative series and the relative size of the residual theoretical uncertainties indicate that the predictions for the observables considered here are stable and sufficiently accurate when compared to current and expected experimental errors.

\section*{Acknowledgments }

The authors are grateful to Marco Zaro for useful conversations.
The work of R.F., D.P.~and I.T.~is supported in part by the Alexander von Humboldt Foundation, in the framework of the Sofja Kovalevskaja Award Project ``Event Simulation for the Large Hadron Collider at High Precision''. 
D.P.~has been also supported by the Deutsche Forschungsgemeinschaft (DFG) through the Collaborative Research Centre SFB1258. The work of A.F. is supported in part by
the National Science Foundation under Grant No. PHY-1417354 and PSC CUNY Research
Award TRADA-61151-00 49.  A.B.~acknowledges the support by
the ERC Starting Grant REINVENT-714788. A.B.~would like to thank New York City College of Technology CUNY (grant PSC-CUNY 60185-00 48) for the kind hospitality in December 2018. R.F.~is supported in part by the Swedish Research Council under contract number 2016-05996. A.F.~would like to thank S.~Alioli and his group at Universit\`a degli Studi di Milano-Bicocca and INFN for their kind hospitality in the final stage of this work.

\bibliographystyle{JHEP}
\bibliography{article}

\providecommand{\href}[2]{#2}\begingroup\raggedright\begin{thebibliography}{100}

\bibitem{Abe:1995hr}
{\scshape CDF} collaboration, F.~Abe et~al., \emph{{Observation of top quark
  production in $\bar{p}p$ collisions}},
  \href{http://dx.doi.org/10.1103/PhysRevLett.74.2626}{\emph{Phys. Rev. Lett.}
  {\bf 74} (1995) 2626--2631}, [\href{http://arxiv.org/abs/hep-ex/9503002}{{\tt
  hep-ex/9503002}}].

\bibitem{D0:1995jca}
{\scshape D0} collaboration, S.~Abachi et~al., \emph{{Observation of the top
  quark}}, \href{http://dx.doi.org/10.1103/PhysRevLett.74.2632}{\emph{Phys.
  Rev. Lett.} {\bf 74} (1995) 2632--2637},
  [\href{http://arxiv.org/abs/hep-ex/9503003}{{\tt hep-ex/9503003}}].

\bibitem{Aad:2010ey}
{\scshape ATLAS} collaboration, G.~Aad et~al., \emph{{Measurement of the top
  quark-pair production cross section with ATLAS in pp collisions at
  $\sqrt{s}=7$ TeV}},
  \href{http://dx.doi.org/10.1140/epjc/s10052-011-1577-6}{\emph{Eur. Phys. J.}
  {\bf C71} (2011) 1577}, [\href{http://arxiv.org/abs/1012.1792}{{\tt
  1012.1792}}].

\bibitem{Khachatryan:2010ez}
{\scshape CMS} collaboration, V.~Khachatryan et~al., \emph{{First Measurement
  of the Cross Section for Top-Quark Pair Production in Proton-Proton
  Collisions at $\sqrt{s}=7$ TeV}},
  \href{http://dx.doi.org/10.1016/j.physletb.2010.11.058}{\emph{Phys. Lett.}
  {\bf B695} (2011) 424--443}, [\href{http://arxiv.org/abs/1010.5994}{{\tt
  1010.5994}}].

\bibitem{Aaltonen:2009jj}
{\scshape CDF} collaboration, T.~Aaltonen et~al., \emph{{First Observation of
  Electroweak Single Top Quark Production}},
  \href{http://dx.doi.org/10.1103/PhysRevLett.103.092002}{\emph{Phys. Rev.
  Lett.} {\bf 103} (2009) 092002}, [\href{http://arxiv.org/abs/0903.0885}{{\tt
  0903.0885}}].

\bibitem{Abazov:2009ii}
{\scshape D0} collaboration, V.~M. Abazov et~al., \emph{{Observation of Single
  Top Quark Production}},
  \href{http://dx.doi.org/10.1103/PhysRevLett.103.092001}{\emph{Phys. Rev.
  Lett.} {\bf 103} (2009) 092001}, [\href{http://arxiv.org/abs/0903.0850}{{\tt
  0903.0850}}].

\bibitem{Chatrchyan:2011vp}
{\scshape CMS} collaboration, S.~Chatrchyan et~al., \emph{{Measurement of the
  $t$-channel single top quark production cross section in $pp$ collisions at
  $\sqrt{s}=7$ TeV}},
  \href{http://dx.doi.org/10.1103/PhysRevLett.107.091802}{\emph{Phys. Rev.
  Lett.} {\bf 107} (2011) 091802}, [\href{http://arxiv.org/abs/1106.3052}{{\tt
  1106.3052}}].

\bibitem{Aad:2012ux}
{\scshape ATLAS} collaboration, G.~Aad et~al., \emph{{Measurement of the
  $t$-channel single top-quark production cross section in $pp$ collisions at
  $\sqrt{s}=7$ TeV with the ATLAS detector}},
  \href{http://dx.doi.org/10.1016/j.physletb.2012.09.031}{\emph{Phys. Lett.}
  {\bf B717} (2012) 330--350}, [\href{http://arxiv.org/abs/1205.3130}{{\tt
  1205.3130}}].

\bibitem{Chatrchyan:2014tua}
{\scshape CMS} collaboration, S.~Chatrchyan et~al., \emph{{Observation of the
  associated production of a single top quark and a $W$ boson in $pp$
  collisions at $\sqrt s = $8 TeV}},
  \href{http://dx.doi.org/10.1103/PhysRevLett.112.231802}{\emph{Phys. Rev.
  Lett.} {\bf 112} (2014) 231802}, [\href{http://arxiv.org/abs/1401.2942}{{\tt
  1401.2942}}].

\bibitem{Khachatryan:2015sha}
{\scshape CMS} collaboration, V.~Khachatryan et~al., \emph{{Observation of top
  quark pairs produced in association with a vector boson in pp collisions at $
  \sqrt{s}=8 $ TeV}},
  \href{http://dx.doi.org/10.1007/JHEP01(2016)096}{\emph{JHEP} {\bf 01} (2016)
  096}, [\href{http://arxiv.org/abs/1510.01131}{{\tt 1510.01131}}].

\bibitem{Aad:2015eua}
{\scshape ATLAS} collaboration, G.~Aad et~al., \emph{{Measurement of the $
  t\overline{t}W $ and $ t\overline{t}Z $ production cross sections in pp
  collisions at $ \sqrt{s}=8 $ TeV with the ATLAS detector}},
  \href{http://dx.doi.org/10.1007/JHEP11(2015)172}{\emph{JHEP} {\bf 11} (2015)
  172}, [\href{http://arxiv.org/abs/1509.05276}{{\tt 1509.05276}}].

\bibitem{Aaboud:2016xve}
{\scshape ATLAS} collaboration, M.~Aaboud et~al., \emph{{Measurement of the
  $t\bar{t}Z$ and $t\bar{t}W$ production cross sections in multilepton final
  states using 3.2 fb$^{-1}$ of $pp$ collisions at $\sqrt{s}$ = 13 TeV with the
  ATLAS detector}},
  \href{http://dx.doi.org/10.1140/epjc/s10052-016-4574-y}{\emph{Eur. Phys. J.}
  {\bf C77} (2017) 40}, [\href{http://arxiv.org/abs/1609.01599}{{\tt
  1609.01599}}].

\bibitem{Sirunyan:2017uzs}
{\scshape CMS} collaboration, A.~M. Sirunyan et~al., \emph{{Measurement of the
  cross section for top quark pair production in association with a W or Z
  boson in proton-proton collisions at $\sqrt{s} =$ 13 TeV}},
  \href{http://dx.doi.org/10.1007/JHEP08(2018)011}{\emph{JHEP} {\bf 08} (2018)
  011}, [\href{http://arxiv.org/abs/1711.02547}{{\tt 1711.02547}}].

\bibitem{Sirunyan:2017nbr}
{\scshape CMS} collaboration, A.~M. Sirunyan et~al., \emph{{Measurement of the
  associated production of a single top quark and a Z boson in pp collisions at
  $\sqrt{s} =$ TeV}},
  \href{http://dx.doi.org/10.1016/j.physletb.2018.02.025}{\emph{Phys. Lett.}
  {\bf B779} (2018) 358--384}, [\href{http://arxiv.org/abs/1712.02825}{{\tt
  1712.02825}}].

\bibitem{Aaboud:2018urx}
{\scshape ATLAS} collaboration, M.~Aaboud et~al., \emph{{Observation of Higgs
  boson production in association with a top quark pair at the LHC with the
  ATLAS detector}},
  \href{http://dx.doi.org/10.1016/j.physletb.2018.07.035}{\emph{Phys. Lett.}
  {\bf B784} (2018) 173--191}, [\href{http://arxiv.org/abs/1806.00425}{{\tt
  1806.00425}}].

\bibitem{Sirunyan:2018hoz}
{\scshape CMS} collaboration, A.~M. Sirunyan et~al., \emph{{Observation of
  $\mathrm{t\overline{t}}$H production}},
  \href{http://dx.doi.org/10.1103/PhysRevLett.120.231801,
  10.1130/PhysRevLett.120.231801}{\emph{Phys. Rev. Lett.} {\bf 120} (2018)
  231801}, [\href{http://arxiv.org/abs/1804.02610}{{\tt 1804.02610}}].

\bibitem{Aaboud:2017jvq}
{\scshape ATLAS} collaboration, M.~Aaboud et~al., \emph{{Evidence for the
  associated production of the Higgs boson and a top quark pair with the ATLAS
  detector}}, \href{http://dx.doi.org/10.1103/PhysRevD.97.072003}{\emph{Phys.
  Rev.} {\bf D97} (2018) 072003}, [\href{http://arxiv.org/abs/1712.08891}{{\tt
  1712.08891}}].

\bibitem{Sirunyan:2018shy}
{\scshape CMS} collaboration, A.~M. Sirunyan et~al., \emph{{Evidence for
  associated production of a Higgs boson with a top quark pair in final states
  with electrons, muons, and hadronically decaying $\tau$ leptons at $\sqrt{s}
  =$ 13 TeV}}, \href{http://dx.doi.org/10.1007/JHEP08(2018)066}{\emph{JHEP}
  {\bf 08} (2018) 066}, [\href{http://arxiv.org/abs/1803.05485}{{\tt
  1803.05485}}].

\bibitem{Maltoni:2015ena}
F.~Maltoni, D.~Pagani and I.~Tsinikos, \emph{{Associated production of a
  top-quark pair with vector bosons at NLO in QCD: impact on $
  \mathrm{t}\overline{\mathrm{t}}\mathrm{H} $ searches at the LHC}},
  \href{http://dx.doi.org/10.1007/JHEP02(2016)113}{\emph{JHEP} {\bf 02} (2016)
  113}, [\href{http://arxiv.org/abs/1507.05640}{{\tt 1507.05640}}].

\bibitem{Bylund:2016phk}
O.~Bessidskaia~Bylund, F.~Maltoni, I.~Tsinikos, E.~Vryonidou and C.~Zhang,
  \emph{{Probing top quark neutral couplings in the Standard Model Effective
  Field Theory at NLO in QCD}},
  \href{http://dx.doi.org/10.1007/JHEP05(2016)052}{\emph{JHEP} {\bf 05} (2016)
  052}, [\href{http://arxiv.org/abs/1601.08193}{{\tt 1601.08193}}].

\bibitem{Sirunyan:2018zgs}
{\scshape CMS} collaboration, A.~M. Sirunyan et~al., \emph{{Observation of
  Single Top Quark Production in Association with a $Z$ Boson in Proton-Proton
  Collisions at $\sqrt {s}$ =13 TeV}},
  \href{http://dx.doi.org/10.1103/PhysRevLett.122.132003}{\emph{Phys. Rev.
  Lett.} {\bf 122} (2019) 132003}, [\href{http://arxiv.org/abs/1812.05900}{{\tt
  1812.05900}}].

\bibitem{Czakon:2013goa}
M.~Czakon, P.~Fiedler and A.~Mitov, \emph{{Total Top-Quark Pair-Production
  Cross Section at Hadron Colliders Through $O(\alpha_s^4)$}},
  \href{http://dx.doi.org/10.1103/PhysRevLett.110.252004}{\emph{Phys. Rev.
  Lett.} {\bf 110} (2013) 252004}, [\href{http://arxiv.org/abs/1303.6254}{{\tt
  1303.6254}}].

\bibitem{Brucherseifer:2014ama}
M.~Brucherseifer, F.~Caola and K.~Melnikov, \emph{{On the NNLO QCD corrections
  to single-top production at the LHC}},
  \href{http://dx.doi.org/10.1016/j.physletb.2014.06.075}{\emph{Phys. Lett.}
  {\bf B736} (2014) 58--63}, [\href{http://arxiv.org/abs/1404.7116}{{\tt
  1404.7116}}].

\bibitem{Berger:2016oht}
E.~L. Berger, J.~Gao, C.~P. Yuan and H.~X. Zhu, \emph{{NNLO QCD Corrections to
  t-channel Single Top-Quark Production and Decay}},
  \href{http://dx.doi.org/10.1103/PhysRevD.94.071501}{\emph{Phys. Rev.} {\bf
  D94} (2016) 071501}, [\href{http://arxiv.org/abs/1606.08463}{{\tt
  1606.08463}}].

\bibitem{Liu:2018gxa}
Z.~L. Liu and J.~Gao, \emph{{s-channel Single Top Quark Production and Decay at
  NNLO in QCD}},  \href{http://arxiv.org/abs/1807.03835}{{\tt 1807.03835}}.

\bibitem{Catani:2019hip}
S.~Catani, S.~Devoto, M.~Grazzini, S.~Kallweit and J.~Mazzitelli,
  \emph{{Top-quark pair production at the LHC: Fully differential QCD
  predictions at NNLO}},  \href{http://arxiv.org/abs/1906.06535}{{\tt
  1906.06535}}.

\bibitem{Czakon:2017wor}
M.~Czakon, D.~Heymes, A.~Mitov, D.~Pagani, I.~Tsinikos and M.~Zaro,
  \emph{{Top-pair production at the LHC through NNLO QCD and NLO EW}},
  \href{http://dx.doi.org/10.1007/JHEP10(2017)186}{\emph{JHEP} {\bf 10} (2017)
  186}, [\href{http://arxiv.org/abs/1705.04105}{{\tt 1705.04105}}].

\bibitem{Czakon:2017lgo}
M.~Czakon, D.~Heymes, A.~Mitov, D.~Pagani, I.~Tsinikos and M.~Zaro,
  \emph{{Top-quark charge asymmetry at the LHC and Tevatron through NNLO QCD
  and NLO EW}}, \href{http://dx.doi.org/10.1103/PhysRevD.98.014003}{\emph{Phys.
  Rev.} {\bf D98} (2018) 014003}, [\href{http://arxiv.org/abs/1711.03945}{{\tt
  1711.03945}}].

\bibitem{Czakon:2018nun}
M.~Czakon, A.~Ferroglia, D.~Heymes, A.~Mitov, B.~D. Pecjak, D.~J. Scott et~al.,
  \emph{{Resummation for (boosted) top-quark pair production at NNLO+NNLL' in
  QCD}}, \href{http://dx.doi.org/10.1007/JHEP05(2018)149}{\emph{JHEP} {\bf 05}
  (2018) 149}, [\href{http://arxiv.org/abs/1803.07623}{{\tt 1803.07623}}].

\bibitem{Czakon:2019txp}
M.~L. Czakon et~al., \emph{{Top quark pair production at NNLO+NNLL$'$ in QCD
  combined with electroweak corrections}},  in \emph{{11th International
  Workshop on Top Quark Physics (TOP2018) Bad Neuenahr, Germany, September
  16-21, 2018}}, 2019.
\newblock \href{http://arxiv.org/abs/1901.08281}{{\tt 1901.08281}}.

\bibitem{Frederix:2017wme}
R.~Frederix, D.~Pagani and M.~Zaro, \emph{{Large NLO corrections in
  $t\bar{t}W^{\pm}$ and $t\bar{t}t\bar{t}$ hadroproduction from supposedly
  subleading EW contributions}},
  \href{http://dx.doi.org/10.1007/JHEP02(2018)031}{\emph{JHEP} {\bf 02} (2018)
  031}, [\href{http://arxiv.org/abs/1711.02116}{{\tt 1711.02116}}].

\bibitem{Frederix:2018nkq}
R.~Frederix, S.~Frixione, V.~Hirschi, D.~Pagani, H.~S. Shao and M.~Zaro,
  \emph{{The automation of next-to-leading order electroweak calculations}},
  \href{http://dx.doi.org/10.1007/JHEP07(2018)185}{\emph{JHEP} {\bf 07} (2018)
  185}, [\href{http://arxiv.org/abs/1804.10017}{{\tt 1804.10017}}].

\bibitem{Broggio:2015lya}
A.~Broggio, A.~Ferroglia, B.~D. Pecjak, A.~Signer and L.~L. Yang,
  \emph{{Associated production of a top pair and a Higgs boson beyond NLO}},
  \href{http://dx.doi.org/10.1007/JHEP03(2016)124}{\emph{JHEP} {\bf 03} (2016)
  124}, [\href{http://arxiv.org/abs/1510.01914}{{\tt 1510.01914}}].

\bibitem{Broggio:2016zgg}
A.~Broggio, A.~Ferroglia, G.~Ossola and B.~D. Pecjak, \emph{{Associated
  production of a top pair and a W boson at next-to-next-to-leading logarithmic
  accuracy}}, \href{http://dx.doi.org/10.1007/JHEP09(2016)089}{\emph{JHEP} {\bf
  09} (2016) 089}, [\href{http://arxiv.org/abs/1607.05303}{{\tt 1607.05303}}].

\bibitem{Broggio:2016lfj}
A.~Broggio, A.~Ferroglia, B.~D. Pecjak and L.~L. Yang, \emph{{NNLL resummation
  for the associated production of a top pair and a Higgs boson at the LHC}},
  \href{http://dx.doi.org/10.1007/JHEP02(2017)126}{\emph{JHEP} {\bf 02} (2017)
  126}, [\href{http://arxiv.org/abs/1611.00049}{{\tt 1611.00049}}].

\bibitem{Broggio:2017kzi}
A.~Broggio, A.~Ferroglia, G.~Ossola, B.~D. Pecjak and R.~D. Sameshima,
  \emph{{Associated production of a top pair and a Z boson at the LHC to NNLL
  accuracy}}, \href{http://dx.doi.org/10.1007/JHEP04(2017)105}{\emph{JHEP} {\bf
  04} (2017) 105}, [\href{http://arxiv.org/abs/1702.00800}{{\tt 1702.00800}}].

\bibitem{Broggio:2017oyu}
A.~Broggio, A.~Ferroglia, M.~C.~N. Fiolhais and A.~Onofre, \emph{{Pseudoscalar
  couplings in $t \bar{t} H$ production at NLO+NLL accuracy}},
  \href{http://dx.doi.org/10.1103/PhysRevD.96.073005}{\emph{Phys. Rev.} {\bf
  D96} (2017) 073005}, [\href{http://arxiv.org/abs/1707.01803}{{\tt
  1707.01803}}].  

\bibitem{Kulesza:2017ukk}
A.~Kulesza, L.~Motyka, T.~Stebel and V.~Theeuwes, \emph{{Associated $t \bar{t}
  H$ production at the LHC: Theoretical predictions at NLO+NNLL accuracy}},
  \href{http://dx.doi.org/10.1103/PhysRevD.97.114007}{\emph{Phys. Rev.} {\bf
  D97} (2018) 114007}, [\href{http://arxiv.org/abs/1704.03363}{{\tt
  1704.03363}}].

\bibitem{Kulesza:2018tqz}
A.~Kulesza, L.~Motyka, D.~Schwartl\"{a}nder, T.~Stebel and V.~Theeuwes,
  \emph{{Associated production of a top quark pair with a heavy electroweak
  gauge boson at NLO$+$NNLL accuracy}},
  \href{http://dx.doi.org/10.1140/epjc/s10052-019-6746-z}{\emph{Eur. Phys. J.}
  {\bf C79} (2019) 249}, [\href{http://arxiv.org/abs/1812.08622}{{\tt
  1812.08622}}].

\bibitem{Ju:2019lwp}
W.-L. Ju and L.~L. Yang, \emph{{Resummation of soft and Coulomb corrections for
  $t\bar{t}h$ production at the LHC}},
  \href{http://arxiv.org/abs/1904.08744}{{\tt 1904.08744}}.

\bibitem{Frixione:2014qaa}
S.~Frixione, V.~Hirschi, D.~Pagani, H.~S. Shao and M.~Zaro, \emph{{Weak
  corrections to Higgs hadroproduction in association with a top-quark pair}},
  \href{http://dx.doi.org/10.1007/JHEP09(2014)065}{\emph{JHEP} {\bf 09} (2014)
  065}, [\href{http://arxiv.org/abs/1407.0823}{{\tt 1407.0823}}].

\bibitem{Frixione:2015zaa}
S.~Frixione, V.~Hirschi, D.~Pagani, H.~S. Shao and M.~Zaro, \emph{{Electroweak
  and QCD corrections to top-pair hadroproduction in association with heavy
  bosons}}, \href{http://dx.doi.org/10.1007/JHEP06(2015)184}{\emph{JHEP} {\bf
  06} (2015) 184}, [\href{http://arxiv.org/abs/1504.03446}{{\tt 1504.03446}}].

\bibitem{Pagani:2016caq}
D.~Pagani, I.~Tsinikos and M.~Zaro, \emph{{The impact of the photon PDF and
  electroweak corrections on $t \bar{t}$ distributions}},
  \href{http://dx.doi.org/10.1140/epjc/s10052-016-4318-z}{\emph{Eur. Phys. J.}
  {\bf C76} (2016) 479}, [\href{http://arxiv.org/abs/1606.01915}{{\tt
  1606.01915}}].

\bibitem{Frederix:2016ost}
R.~Frederix, S.~Frixione, V.~Hirschi, D.~Pagani, H.-S. Shao and M.~Zaro,
  \emph{{The complete NLO corrections to dijet hadroproduction}},
  \href{http://dx.doi.org/10.1007/JHEP04(2017)076}{\emph{JHEP} {\bf 04} (2017)
  076}, [\href{http://arxiv.org/abs/1612.06548}{{\tt 1612.06548}}].

\bibitem{Frixione:1995ms}
S.~Frixione, Z.~Kunszt and A.~Signer, \emph{{Three jet cross-sections to
  next-to-leading order}},
  \href{http://dx.doi.org/10.1016/0550-3213(96)00110-1}{\emph{Nucl. Phys.} {\bf
  B467} (1996) 399--442}, [\href{http://arxiv.org/abs/hep-ph/9512328}{{\tt
  hep-ph/9512328}}].

\bibitem{Frixione:1997np}
S.~Frixione, \emph{{A General approach to jet cross-sections in QCD}},
  \href{http://dx.doi.org/10.1016/S0550-3213(97)00574-9}{\emph{Nucl. Phys.}
  {\bf B507} (1997) 295--314}, [\href{http://arxiv.org/abs/hep-ph/9706545}{{\tt
  hep-ph/9706545}}].

\bibitem{Frederix:2009yq}
R.~Frederix, S.~Frixione, F.~Maltoni and T.~Stelzer, \emph{{Automation of
  next-to-leading order computations in QCD: The FKS subtraction}},
  \href{http://dx.doi.org/10.1088/1126-6708/2009/10/003}{\emph{JHEP} {\bf 10}
  (2009) 003}, [\href{http://arxiv.org/abs/0908.4272}{{\tt 0908.4272}}].

\bibitem{Frederix:2016rdc}
R.~Frederix, S.~Frixione, A.~S. Papanastasiou, S.~Prestel and P.~Torrielli,
  \emph{{Off-shell single-top production at NLO matched to parton showers}},
  \href{http://dx.doi.org/10.1007/JHEP06(2016)027}{\emph{JHEP} {\bf 06} (2016)
  027}, [\href{http://arxiv.org/abs/1603.01178}{{\tt 1603.01178}}].

\bibitem{Ossola:2006us}
G.~Ossola, C.~G. Papadopoulos and R.~Pittau, \emph{{Reducing full one-loop
  amplitudes to scalar integrals at the integrand level}},
  \href{http://dx.doi.org/10.1016/j.nuclphysb.2006.11.012}{\emph{Nucl. Phys.}
  {\bf B763} (2007) 147--169}, [\href{http://arxiv.org/abs/hep-ph/0609007}{{\tt
  hep-ph/0609007}}].

\bibitem{Mastrolia:2012bu}
P.~Mastrolia, E.~Mirabella and T.~Peraro, \emph{{Integrand reduction of
  one-loop scattering amplitudes through Laurent series expansion}},
  \href{http://dx.doi.org/10.1007/JHEP11(2012)128,
  10.1007/JHEP06(2012)095}{\emph{JHEP} {\bf 06} (2012) 095},
  [\href{http://arxiv.org/abs/1203.0291}{{\tt 1203.0291}}].

\bibitem{Passarino:1978jh}
G.~Passarino and M.~J.~G. Veltman, \emph{{One Loop Corrections for $e^+ e^-$
  Annihilation Into $\mu^+ \mu^-$ in the Weinberg Model}},
  \href{http://dx.doi.org/10.1016/0550-3213(79)90234-7}{\emph{Nucl. Phys.} {\bf
  B160} (1979) 151--207}.

\bibitem{Davydychev:1991va}
A.~I. Davydychev, \emph{{A Simple formula for reducing Feynman diagrams to
  scalar integrals}},
  \href{http://dx.doi.org/10.1016/0370-2693(91)91715-8}{\emph{Phys. Lett.} {\bf
  B263} (1991) 107--111}.

\bibitem{Denner:2005nn}
A.~Denner and S.~Dittmaier, \emph{{Reduction schemes for one-loop tensor
  integrals}},
  \href{http://dx.doi.org/10.1016/j.nuclphysb.2005.11.007}{\emph{Nucl. Phys.}
  {\bf B734} (2006) 62--115}, [\href{http://arxiv.org/abs/hep-ph/0509141}{{\tt
  hep-ph/0509141}}].

\bibitem{Hirschi:2011pa}
V.~Hirschi, R.~Frederix, S.~Frixione, M.~V. Garzelli, F.~Maltoni and R.~Pittau,
  \emph{{Automation of one-loop QCD corrections}},
  \href{http://dx.doi.org/10.1007/JHEP05(2011)044}{\emph{JHEP} {\bf 05} (2011)
  044}, [\href{http://arxiv.org/abs/1103.0621}{{\tt 1103.0621}}].

\bibitem{Ossola:2007ax}
G.~Ossola, C.~G. Papadopoulos and R.~Pittau, \emph{{CutTools: A Program
  implementing the OPP reduction method to compute one-loop amplitudes}},
  \href{http://dx.doi.org/10.1088/1126-6708/2008/03/042}{\emph{JHEP} {\bf 03}
  (2008) 042}, [\href{http://arxiv.org/abs/0711.3596}{{\tt 0711.3596}}].

\bibitem{Peraro:2014cba}
T.~Peraro, \emph{{Ninja: Automated Integrand Reduction via Laurent Expansion
  for One-Loop Amplitudes}},
  \href{http://dx.doi.org/10.1016/j.cpc.2014.06.017}{\emph{Comput. Phys.
  Commun.} {\bf 185} (2014) 2771--2797},
  [\href{http://arxiv.org/abs/1403.1229}{{\tt 1403.1229}}].

\bibitem{Hirschi:2016mdz}
V.~Hirschi and T.~Peraro, \emph{{Tensor integrand reduction via Laurent
  expansion}}, \href{http://dx.doi.org/10.1007/JHEP06(2016)060}{\emph{JHEP}
  {\bf 06} (2016) 060}, [\href{http://arxiv.org/abs/1604.01363}{{\tt
  1604.01363}}].

\bibitem{Denner:2016kdg}
A.~Denner, S.~Dittmaier and L.~Hofer, \emph{{Collier: a fortran-based Complex
  One-Loop LIbrary in Extended Regularizations}},
  \href{http://dx.doi.org/10.1016/j.cpc.2016.10.013}{\emph{Comput. Phys.
  Commun.} {\bf 212} (2017) 220--238},
  [\href{http://arxiv.org/abs/1604.06792}{{\tt 1604.06792}}].

\bibitem{Cascioli:2011va}
F.~Cascioli, P.~Maierhofer and S.~Pozzorini, \emph{{Scattering Amplitudes with
  Open Loops}},
  \href{http://dx.doi.org/10.1103/PhysRevLett.108.111601}{\emph{Phys. Rev.
  Lett.} {\bf 108} (2012) 111601}, [\href{http://arxiv.org/abs/1111.5206}{{\tt
  1111.5206}}].

\bibitem{Becher:2014oda}
T.~Becher, A.~Broggio and A.~Ferroglia, \emph{{Introduction to Soft-Collinear
  Effective Theory}},
  \href{http://dx.doi.org/10.1007/978-3-319-14848-9}{\emph{Lect. Notes Phys.}
  {\bf 896} (2015) pp.1--206}, [\href{http://arxiv.org/abs/1410.1892}{{\tt
  1410.1892}}].

\bibitem{Bauer:2000yr}
C.~W. Bauer, S.~Fleming, D.~Pirjol and I.~W. Stewart, \emph{{An Effective field
  theory for collinear and soft gluons: Heavy to light decays}},
  \href{http://dx.doi.org/10.1103/PhysRevD.63.114020}{\emph{Phys. Rev.} {\bf
  D63} (2001) 114020}, [\href{http://arxiv.org/abs/hep-ph/0011336}{{\tt
  hep-ph/0011336}}].

\bibitem{Bauer:2001yt}
C.~W. Bauer, D.~Pirjol and I.~W. Stewart, \emph{{Soft collinear factorization
  in effective field theory}},
  \href{http://dx.doi.org/10.1103/PhysRevD.65.054022}{\emph{Phys. Rev.} {\bf
  D65} (2002) 054022}, [\href{http://arxiv.org/abs/hep-ph/0109045}{{\tt
  hep-ph/0109045}}].

\bibitem{Beneke:2002ph}
M.~Beneke, A.~P. Chapovsky, M.~Diehl and T.~Feldmann, \emph{{Soft collinear
  effective theory and heavy to light currents beyond leading power}},
  \href{http://dx.doi.org/10.1016/S0550-3213(02)00687-9}{\emph{Nucl. Phys.}
  {\bf B643} (2002) 431--476}, [\href{http://arxiv.org/abs/hep-ph/0206152}{{\tt
  hep-ph/0206152}}].

\bibitem{Ahrens:2009uz}
V.~Ahrens, A.~Ferroglia, M.~Neubert, B.~D. Pecjak and L.~L. Yang,
  \emph{{Threshold expansion at order $\alpha_s^4$ for the t-tbar invariant
  mass distribution at hadron colliders}},
  \href{http://dx.doi.org/10.1016/j.physletb.2010.03.048}{\emph{Phys. Lett.}
  {\bf B687} (2010) 331--337}, [\href{http://arxiv.org/abs/0912.3375}{{\tt
  0912.3375}}].

\bibitem{Ahrens:2010zv}
V.~Ahrens, A.~Ferroglia, M.~Neubert, B.~D. Pecjak and L.~L. Yang,
  \emph{{Renormalization-Group Improved Predictions for Top-Quark Pair
  Production at Hadron Colliders}},
  \href{http://dx.doi.org/10.1007/JHEP09(2010)097}{\emph{JHEP} {\bf 09} (2010)
  097}, [\href{http://arxiv.org/abs/1003.5827}{{\tt 1003.5827}}].

\bibitem{Broggio:2013uba}
A.~Broggio, A.~Ferroglia, M.~Neubert, L.~Vernazza and L.~L. Yang,
  \emph{{Approximate NNLO Predictions for the Stop-Pair Production Cross
  Section at the LHC}},
  \href{http://dx.doi.org/10.1007/JHEP07(2013)042}{\emph{JHEP} {\bf 07} (2013)
  042}, [\href{http://arxiv.org/abs/1304.2411}{{\tt 1304.2411}}].

\bibitem{Broggio:2014yca}
A.~Broggio, A.~S. Papanastasiou and A.~Signer, \emph{{Renormalization-group
  improved fully differential cross sections for top pair production}},
  \href{http://dx.doi.org/10.1007/JHEP10(2014)098}{\emph{JHEP} {\bf 10} (2014)
  98}, [\href{http://arxiv.org/abs/1407.2532}{{\tt 1407.2532}}].

\bibitem{Ferroglia:2009ep}
A.~Ferroglia, M.~Neubert, B.~D. Pecjak and L.~L. Yang, \emph{{Two-loop
  divergences of scattering amplitudes with massive partons}},
  \href{http://dx.doi.org/10.1103/PhysRevLett.103.201601}{\emph{Phys. Rev.
  Lett.} {\bf 103} (2009) 201601}, [\href{http://arxiv.org/abs/0907.4791}{{\tt
  0907.4791}}].

\bibitem{Ferroglia:2009ii}
A.~Ferroglia, M.~Neubert, B.~D. Pecjak and L.~L. Yang, \emph{{Two-loop
  divergences of massive scattering amplitudes in non-abelian gauge theories}},
  \href{http://dx.doi.org/10.1088/1126-6708/2009/11/062}{\emph{JHEP} {\bf 11}
  (2009) 062}, [\href{http://arxiv.org/abs/0908.3676}{{\tt 0908.3676}}].

\bibitem{Catani:1996yz}
S.~Catani, M.~L. Mangano, P.~Nason and L.~Trentadue, \emph{{The Resummation of
  soft gluons in hadronic collisions}},
  \href{http://dx.doi.org/10.1016/0550-3213(96)00399-9}{\emph{Nucl. Phys.} {\bf
  B478} (1996) 273--310}, [\href{http://arxiv.org/abs/hep-ph/9604351}{{\tt
  hep-ph/9604351}}].

\bibitem{Becher:2007ty}
T.~Becher, M.~Neubert and G.~Xu, \emph{{Dynamical Threshold Enhancement and
  Resummation in Drell-Yan Production}},
  \href{http://dx.doi.org/10.1088/1126-6708/2008/07/030}{\emph{JHEP} {\bf 07}
  (2008) 030}, [\href{http://arxiv.org/abs/0710.0680}{{\tt 0710.0680}}].

\bibitem{Ahrens:2008nc}
V.~Ahrens, T.~Becher, M.~Neubert and L.~L. Yang, \emph{{Renormalization-Group
  Improved Prediction for Higgs Production at Hadron Colliders}},
  \href{http://dx.doi.org/10.1140/epjc/s10052-009-1030-2}{\emph{Eur. Phys. J.}
  {\bf C62} (2009) 333--353}, [\href{http://arxiv.org/abs/0809.4283}{{\tt
  0809.4283}}].

\bibitem{Broggio:2011bd}
A.~Broggio, M.~Neubert and L.~Vernazza, \emph{{Soft-gluon resummation for
  slepton-pair production at hadron colliders}},
  \href{http://dx.doi.org/10.1007/JHEP05(2012)151}{\emph{JHEP} {\bf 05} (2012)
  151}, [\href{http://arxiv.org/abs/1111.6624}{{\tt 1111.6624}}].

\bibitem{Broggio:2013cia}
A.~Broggio, A.~Ferroglia, M.~Neubert, L.~Vernazza and L.~L. Yang, \emph{{NNLL
  Momentum-Space Resummation for Stop-Pair Production at the LHC}},
  \href{http://dx.doi.org/10.1007/JHEP03(2014)066}{\emph{JHEP} {\bf 03} (2014)
  066}, [\href{http://arxiv.org/abs/1312.4540}{{\tt 1312.4540}}].

\bibitem{Bonvini:2012sh}
M.~Bonvini, \emph{{Resummation of soft and hard gluon radiation in perturbative
  QCD}}.
\newblock PhD thesis, Genoa U., 2012.
\newblock \href{http://arxiv.org/abs/1212.0480}{{\tt 1212.0480}}.

\bibitem{Bonvini:2014joa}
M.~Bonvini and S.~Marzani, \emph{{Resummed Higgs cross section at N$^{3}$LL}},
  \href{http://dx.doi.org/10.1007/JHEP09(2014)007}{\emph{JHEP} {\bf 09} (2014)
  007}, [\href{http://arxiv.org/abs/1405.3654}{{\tt 1405.3654}}].

\bibitem{Pecjak:2018lif}
B.~D. Pecjak, D.~J. Scott, X.~Wang and L.~L. Yang, \emph{{Resummation for
  rapidity distributions in top-quark pair production}},
  \href{http://dx.doi.org/10.1007/JHEP03(2019)060}{\emph{JHEP} {\bf 03} (2019)
  060}, [\href{http://arxiv.org/abs/1811.10527}{{\tt 1811.10527}}].

\bibitem{Garzelli:2012bn}
M.~V. Garzelli, A.~Kardos, C.~G. Papadopoulos and Z.~Trocsanyi, \emph{{t
  $\bar{t}$ $W^{+-}$ and t $\bar{t}$ Z Hadroproduction at NLO accuracy in QCD
  with Parton Shower and Hadronization effects}},
  \href{http://dx.doi.org/10.1007/JHEP11(2012)056}{\emph{JHEP} {\bf 11} (2012)
  056}, [\href{http://arxiv.org/abs/1208.2665}{{\tt 1208.2665}}].

\bibitem{Campbell:2012dh}
J.~M. Campbell and R.~K. Ellis, \emph{{$t \bar{t} W^{+-}$ production and decay
  at NLO}}, \href{http://dx.doi.org/10.1007/JHEP07(2012)052}{\emph{JHEP} {\bf
  07} (2012) 052}, [\href{http://arxiv.org/abs/1204.5678}{{\tt 1204.5678}}].

\bibitem{Maltoni:2014zpa}
F.~Maltoni, M.~L. Mangano, I.~Tsinikos and M.~Zaro, \emph{{Top-quark charge
  asymmetry and polarization in $t\overline{t}W^±$ production at the LHC}},
  \href{http://dx.doi.org/10.1016/j.physletb.2014.07.033}{\emph{Phys. Lett.}
  {\bf B736} (2014) 252--260}, [\href{http://arxiv.org/abs/1406.3262}{{\tt
  1406.3262}}].

\bibitem{Li:2014ula}
H.~T. Li, C.~S. Li and S.~A. Li, \emph{{Renormalization group improved
  predictions for $t\bar{t}W^\pm$ production at hadron colliders}},
  \href{http://dx.doi.org/10.1103/PhysRevD.90.094009}{\emph{Phys. Rev.} {\bf
  D90} (2014) 094009}, [\href{http://arxiv.org/abs/1409.1460}{{\tt
  1409.1460}}].

\bibitem{deFlorian:2016spz}
{\scshape LHC Higgs Cross Section Working Group} collaboration, D.~de~Florian
  et~al., \emph{{Handbook of LHC Higgs Cross Sections: 4. Deciphering the
  Nature of the Higgs Sector}},  \href{http://arxiv.org/abs/1610.07922}{{\tt
  1610.07922}}.

\bibitem{Dror:2015nkp}
J.~A. Dror, M.~Farina, E.~Salvioni and J.~Serra, \emph{{Strong tW Scattering at
  the LHC}}, \href{http://dx.doi.org/10.1007/JHEP01(2016)071}{\emph{JHEP} {\bf
  01} (2016) 071}, [\href{http://arxiv.org/abs/1511.03674}{{\tt 1511.03674}}].

\bibitem{Beenakker:2001rj}
W.~Beenakker, S.~Dittmaier, M.~Kramer, B.~Plumper, M.~Spira and P.~M. Zerwas,
  \emph{{Higgs radiation off top quarks at the Tevatron and the LHC}},
  \href{http://dx.doi.org/10.1103/PhysRevLett.87.201805}{\emph{Phys. Rev.
  Lett.} {\bf 87} (2001) 201805},
  [\href{http://arxiv.org/abs/hep-ph/0107081}{{\tt hep-ph/0107081}}].

\bibitem{Beenakker:2002nc}
W.~Beenakker, S.~Dittmaier, M.~Kramer, B.~Plumper, M.~Spira and P.~M. Zerwas,
  \emph{{NLO QCD corrections to t anti-t H production in hadron collisions}},
  \href{http://dx.doi.org/10.1016/S0550-3213(03)00044-0}{\emph{Nucl. Phys.}
  {\bf B653} (2003) 151--203}, [\href{http://arxiv.org/abs/hep-ph/0211352}{{\tt
  hep-ph/0211352}}].

\bibitem{Dawson:2002tg}
S.~Dawson, L.~H. Orr, L.~Reina and D.~Wackeroth, \emph{{Associated top quark
  Higgs boson production at the LHC}},
  \href{http://dx.doi.org/10.1103/PhysRevD.67.071503}{\emph{Phys. Rev.} {\bf
  D67} (2003) 071503}, [\href{http://arxiv.org/abs/hep-ph/0211438}{{\tt
  hep-ph/0211438}}].

\bibitem{Dawson:2003zu}
S.~Dawson, C.~Jackson, L.~H. Orr, L.~Reina and D.~Wackeroth, \emph{{Associated
  Higgs production with top quarks at the large hadron collider: NLO QCD
  corrections}},
  \href{http://dx.doi.org/10.1103/PhysRevD.68.034022}{\emph{Phys. Rev.} {\bf
  D68} (2003) 034022}, [\href{http://arxiv.org/abs/hep-ph/0305087}{{\tt
  hep-ph/0305087}}].

\bibitem{Lazopoulos:2008de}
A.~Lazopoulos, T.~McElmurry, K.~Melnikov and F.~Petriello,
  \emph{{Next-to-leading order QCD corrections to $t \bar{t} Z$ production at
  the LHC}},
  \href{http://dx.doi.org/10.1016/j.physletb.2008.06.073}{\emph{Phys. Lett.}
  {\bf B666} (2008) 62--65}, [\href{http://arxiv.org/abs/0804.2220}{{\tt
  0804.2220}}].

\bibitem{Garzelli:2011is}
M.~V. Garzelli, A.~Kardos, C.~G. Papadopoulos and Z.~Trocsanyi, \emph{{Z0 -
  boson production in association with a top anti-top pair at NLO accuracy with
  parton shower effects}},
  \href{http://dx.doi.org/10.1103/PhysRevD.85.074022}{\emph{Phys. Rev.} {\bf
  D85} (2012) 074022}, [\href{http://arxiv.org/abs/1111.1444}{{\tt
  1111.1444}}].

\bibitem{Kardos:2011na}
A.~Kardos, Z.~Trocsanyi and C.~Papadopoulos, \emph{{Top quark pair production
  in association with a Z-boson at NLO accuracy}},
  \href{http://dx.doi.org/10.1103/PhysRevD.85.054015}{\emph{Phys. Rev.} {\bf
  D85} (2012) 054015}, [\href{http://arxiv.org/abs/1111.0610}{{\tt
  1111.0610}}].

\bibitem{Plehn:2015cta}
M.~L. Mangano, T.~Plehn, P.~Reimitz, T.~Schell and H.-S. Shao, \emph{{Measuring
  the Top Yukawa Coupling at 100 TeV}},
  \href{http://dx.doi.org/10.1088/0954-3899/43/3/035001}{\emph{J. Phys.} {\bf
  G43} (2016) 035001}, [\href{http://arxiv.org/abs/1507.08169}{{\tt
  1507.08169}}].

\bibitem{Manohar:2016nzj}
A.~Manohar, P.~Nason, G.~P. Salam and G.~Zanderighi, \emph{{How bright is the
  proton? A precise determination of the photon parton distribution function}},
  \href{http://dx.doi.org/10.1103/PhysRevLett.117.242002}{\emph{Phys. Rev.
  Lett.} {\bf 117} (2016) 242002}, [\href{http://arxiv.org/abs/1607.04266}{{\tt
  1607.04266}}].

\bibitem{Manohar:2017eqh}
A.~V. Manohar, P.~Nason, G.~P. Salam and G.~Zanderighi, \emph{{The Photon
  Content of the Proton}},  \href{http://arxiv.org/abs/1708.01256}{{\tt
  1708.01256}}.

\bibitem{Butterworth:2015oua}
J.~Butterworth et~al., \emph{{PDF4LHC recommendations for LHC Run II}},
  \href{http://dx.doi.org/10.1088/0954-3899/43/2/023001}{\emph{J. Phys.} {\bf
  G43} (2016) 023001}, [\href{http://arxiv.org/abs/1510.03865}{{\tt
  1510.03865}}].

\bibitem{Ball:2014uwa}
{\scshape NNPDF} collaboration, R.~D. Ball et~al., \emph{{Parton distributions
  for the LHC Run II}},
  \href{http://dx.doi.org/10.1007/JHEP04(2015)040}{\emph{JHEP} {\bf 04} (2015)
  040}, [\href{http://arxiv.org/abs/1410.8849}{{\tt 1410.8849}}].

\bibitem{Harland-Lang:2014zoa}
L.~A. Harland-Lang, A.~D. Martin, P.~Motylinski and R.~S. Thorne, \emph{{Parton
  distributions in the LHC era: MMHT 2014 PDFs}},
  \href{http://dx.doi.org/10.1140/epjc/s10052-015-3397-6}{\emph{Eur. Phys. J.}
  {\bf C75} (2015) 204}, [\href{http://arxiv.org/abs/1412.3989}{{\tt
  1412.3989}}].

\bibitem{Dulat:2015mca}
S.~Dulat, T.-J. Hou, J.~Gao, M.~Guzzi, J.~Huston, P.~Nadolsky et~al.,
  \emph{{New parton distribution functions from a global analysis of quantum
  chromodynamics}},
  \href{http://dx.doi.org/10.1103/PhysRevD.93.033006}{\emph{Phys. Rev.} {\bf
  D93} (2016) 033006}, [\href{http://arxiv.org/abs/1506.07443}{{\tt
  1506.07443}}].

\bibitem{deFlorian:2015ujt}
D.~de~Florian, G.~F.~R. Sborlini and G.~Rodrigo, \emph{{QED corrections to the
  Altarelli?Parisi splitting functions}},
  \href{http://dx.doi.org/10.1140/epjc/s10052-016-4131-8}{\emph{Eur. Phys. J.}
  {\bf C76} (2016) 282}, [\href{http://arxiv.org/abs/1512.00612}{{\tt
  1512.00612}}].

\bibitem{deFlorian:2016gvk}
D.~de~Florian, G.~F.~R. Sborlini and G.~Rodrigo, \emph{{Two-loop QED
  corrections to the Altarelli-Parisi splitting functions}},
  \href{http://dx.doi.org/10.1007/JHEP10(2016)056}{\emph{JHEP} {\bf 10} (2016)
  056}, [\href{http://arxiv.org/abs/1606.02887}{{\tt 1606.02887}}].

\bibitem{Frederix:2011ss}
R.~Frederix, S.~Frixione, V.~Hirschi, F.~Maltoni, R.~Pittau et~al.,
  \emph{{Four-lepton production at hadron colliders: aMC@NLO predictions with
  theoretical uncertainties}},
  \href{http://dx.doi.org/10.1007/JHEP02(2012)099}{\emph{JHEP} {\bf 1202}
  (2012) 099}, [\href{http://arxiv.org/abs/1110.4738}{{\tt 1110.4738}}].

\bibitem{Aaltonen:2011kc}
{\scshape CDF} collaboration, T.~Aaltonen et~al., \emph{{Evidence for a Mass
  Dependent Forward-Backward Asymmetry in Top Quark Pair Production}},
  \href{http://dx.doi.org/10.1103/PhysRevD.83.112003}{\emph{Phys. Rev.} {\bf
  D83} (2011) 112003}, [\href{http://arxiv.org/abs/1101.0034}{{\tt
  1101.0034}}].

\bibitem{Kuhn:1998kw}
J.~H. Kuhn and G.~Rodrigo, \emph{{Charge asymmetry of heavy quarks at hadron
  colliders}}, \href{http://dx.doi.org/10.1103/PhysRevD.59.054017}{\emph{Phys.
  Rev.} {\bf D59} (1999) 054017},
  [\href{http://arxiv.org/abs/hep-ph/9807420}{{\tt hep-ph/9807420}}].

\bibitem{Hollik:2011ps}
W.~Hollik and D.~Pagani, \emph{{The electroweak contribution to the top quark
  forward-backward asymmetry at the Tevatron}},
  \href{http://dx.doi.org/10.1103/PhysRevD.84.093003}{\emph{Phys. Rev.} {\bf
  D84} (2011) 093003}, [\href{http://arxiv.org/abs/1107.2606}{{\tt
  1107.2606}}].

\bibitem{Czakon:2014xsa}
M.~Czakon, P.~Fiedler and A.~Mitov, \emph{{Resolving the Tevatron Top Quark
  Forward-Backward Asymmetry Puzzle: Fully Differential
  Next-to-Next-to-Leading-Order Calculation}},
  \href{http://dx.doi.org/10.1103/PhysRevLett.115.052001}{\emph{Phys. Rev.
  Lett.} {\bf 115} (2015) 052001}, [\href{http://arxiv.org/abs/1411.3007}{{\tt
  1411.3007}}].

\bibitem{Degrassi:2016wml}
G.~Degrassi, P.~P. Giardino, F.~Maltoni and D.~Pagani, \emph{{Probing the Higgs
  self coupling via single Higgs production at the LHC}},
  \href{http://dx.doi.org/10.1007/JHEP12(2016)080}{\emph{JHEP} {\bf 12} (2016)
  080}, [\href{http://arxiv.org/abs/1607.04251}{{\tt 1607.04251}}].

\bibitem{Maltoni:2017ims}
F.~Maltoni, D.~Pagani, A.~Shivaji and X.~Zhao, \emph{{Trilinear Higgs coupling
  determination via single-Higgs differential measurements at the LHC}},
  \href{http://dx.doi.org/10.1140/epjc/s10052-017-5410-8}{\emph{Eur. Phys. J.}
  {\bf C77} (2017) 887}, [\href{http://arxiv.org/abs/1709.08649}{{\tt
  1709.08649}}].

\end{thebibliography}\endgroup

\end{document}